%% file: main.tex
\documentclass[twocolumn,aps,superscriptaddress,floatfix,longbibliography]{revtex4-1}

\usepackage{graphicx}
\usepackage{subfig}
\usepackage{bm}
\usepackage{amsmath,amsfonts} 

 \usepackage{cancel} 

\usepackage{xspace}  

\usepackage[font=,labelfont={normalsize},labelsep=period, 
           justification=centerlast]{caption}

\usepackage{floatrow}
\usepackage{xcolor}
\usepackage[colorlinks=true,citecolor=blue,linkcolor=magenta]{hyperref}

\newcommand{\la}{\langle}
\newcommand{\ra}{\rangle}

\newcommand{\ben}{\begin{eqnarray}}
\newcommand{\een}{\end{eqnarray}}

\newcommand{\be}{\begin{equation}}
\newcommand{\ee}{\end{equation}}

\newcommand{\bea}{\begin{eqnarray}}
\newcommand{\eea}{\end{eqnarray}}

\input{shortcuts.tex}

\hypersetup{
  colorlinks=true,
  linkcolor=blue,
  anchorcolor = blue,
  citecolor = blue,
  filecolor = blue,
  urlcolor = blue
}

  \def\YX{\ensuremath{\mathit{YX}}\xspace} 
  \def\XY{\ensuremath{\mathit{XY}}\xspace}
  \def\YZ{\ensuremath{\mathit{YZ}}\xspace}
  \def\ZY{\ensuremath{\mathit{ZY}}\xspace}

  \def\XX{\ensuremath{\mathit{XX}}\xspace}
  \def\YY{\ensuremath{\mathit{YY}}\xspace}
  \def\ZZ{\ensuremath{\mathit{ZZ}}\xspace}

\begin{document}
	
	 \title{Feedback-based Quantum Algorithm Inspired by Counterdiabatic Driving}

\author{Rajesh K. Malla}
\email{rmalla@bnl.gov}
\affiliation{Condensed Matter Physics and Materials Science Division, Brookhaven National Laboratory, Upton, New York 11973, USA}

\author{Hiroki Sukeno}
\affiliation{C.N. Yang Institute for Theoretical Physics \& Department of Physics and Astronomy, Stony Brook University, Stony Brook, New York 11794, USA}

\author{Hongye Yu}
\affiliation{C.N. Yang Institute for Theoretical Physics \& Department of Physics and Astronomy, Stony Brook University, Stony Brook, New York 11794, USA}

\author{Tzu-Chieh Wei}
\affiliation{C.N. Yang Institute for Theoretical Physics \& Department of Physics and Astronomy, Stony Brook University, Stony Brook, New York 11794, USA}

\author{Andreas Weichselbaum}
\affiliation{Condensed Matter Physics and Materials Science Division, Brookhaven National Laboratory, Upton, New York 11973, USA}

\author{Robert M. Konik}
\affiliation{Condensed Matter Physics and Materials Science Division, Brookhaven National Laboratory, Upton, New York 11973, USA}
	
\date{\today}
	
\begin{abstract}
In recent quantum algorithmic developments, a feedback-based approach has shown promise for preparing quantum many-body
system ground states and solving combinatorial optimization
problems. This method utilizes quantum Lyapunov control to
iteratively construct quantum circuits. Here, we propose a
substantial enhancement by implementing a protocol that uses ideas from quantum Lyapunov control
and the counterdiabatic driving protocol, a key concept from
quantum adiabaticity. Our approach introduces an additional
control field inspired by counterdiabatic driving. We apply our
algorithm to prepare ground states in one-dimensional quantum
Ising spin chains. Comprehensive simulations demonstrate a
remarkable acceleration in population transfer to low-energy
states within a significantly reduced time frame compared to
conventional feedback-based quantum algorithms. This
acceleration translates to a reduced quantum circuit depth, a
critical metric for potential quantum computer implementation.
We validate our algorithm on the IBM cloud computer,
highlighting its efficacy in expediting quantum computations for
many-body systems and combinatorial optimization problems. 
\end{abstract}


\maketitle

\section{Introduction}
\label{sec:intro}

The quest for efficient quantum algorithms for many-body ground
state preparation has been a central focus in quantum simulation
research \cite{Wocjan2009,tubman2018postponing,ge2019faster,
JBL:lin,wang2023ground}, marked by the evolution from early
adiabatic approaches to recent quantum-classical hybrid
structures. Early approaches leveraged the concept of
adiabaticity utilizing an effective time-dependent Hamiltonian
to undergo a gradual time evolution from an initial state to the
ground state at large times. These types of algorithms can be categorized by either quantum adiabatic algorithm~\cite{farhi2000quantum,
GordonScience2005,FarhiScience2001, RMDadiabatic2018} or
quantum annealing~\cite{finnila1994quantum, kadowaki1998quantum,
brooke1999quantum}. In contrast, recent advancements in quantum
algorithms have shifted towards quantum-classical hybrid
structures, harnessing the combined effect of both quantum and
classical computing. This approach is particularly apt for the
era of noisy intermediate-scale quantum (NISQ) devices
\cite{JBL:nisq}. These algorithms are known as variational
quantum algorithms (VQA) \cite{mcclean2016theory,
bravo2019variational, grimsley2019adaptive,
cerezo2021variational, PRXQuantum.2.010101, RMDKBharti2022},
with some notable examples including the Quantum Approximate
Optimization Algorithm (QAOA) \cite{QAOA, lloyd2018quantum,
dalzell2020many} and the Variational Quantum Eigensolver (VQE)
\cite{peruzzo2014variational, kandala2017hardware,
colless2018computation, tilly2022variational}. These algorithms
have demonstrated superior performance compared to classical
counterparts, particularly in addressing combinatorial
problems~\cite{zhou2020quantum, liu2022layer} and challenges in
quantum chemistry~\cite{peruzzo2014variational, cao2019quantum}.
We note, however, that classical optimization within variational
quantum algorithms often faces numerical challenges due to
optimization landscapes that contain false local  minima and
barren plateaus~\cite{chakrabarti2007quantum,
russell2017control, mcclean2018barren, wiersema2020exploring,
bittel2021training, larocca2022diagnosing}.

Recently, a novel quantum algorithm named the Feedback-Based
Quantum Algorithm (FQA) has been
introduced~\cite{FeedbackPRL,FeedbackPRA}. This approach draws
inspiration from the principles of Quantum Lyapunov Control
(QLC)~\cite{grivopoulos2003lyapunov,
QLC_Wang_2013}, negating the necessity for predetermined time
evolution or classical optimization. Instead, FQA constructs the
quantum circuit iteratively, introducing new layers where the
parameters are meticulously determined through 
feedback derived from qubit measurements in the preceding layer. A fundamental
unit, called a layer, in the FQA architecture comprises two
unitaries, echoing the structure found in QAOA and quantum
annealing. These unitaries represent the parent Hamiltonian
governing the desired ground state and a mixer Hamiltonian, the
ground state of which serves as the algorithm's initial state.
This innovative methodology represents a departure from
conventional approaches, exemplifying the potential of QLC in
shaping the landscape of quantum algorithms.

\begin{figure*}
\centering
\includegraphics[scale=0.4]{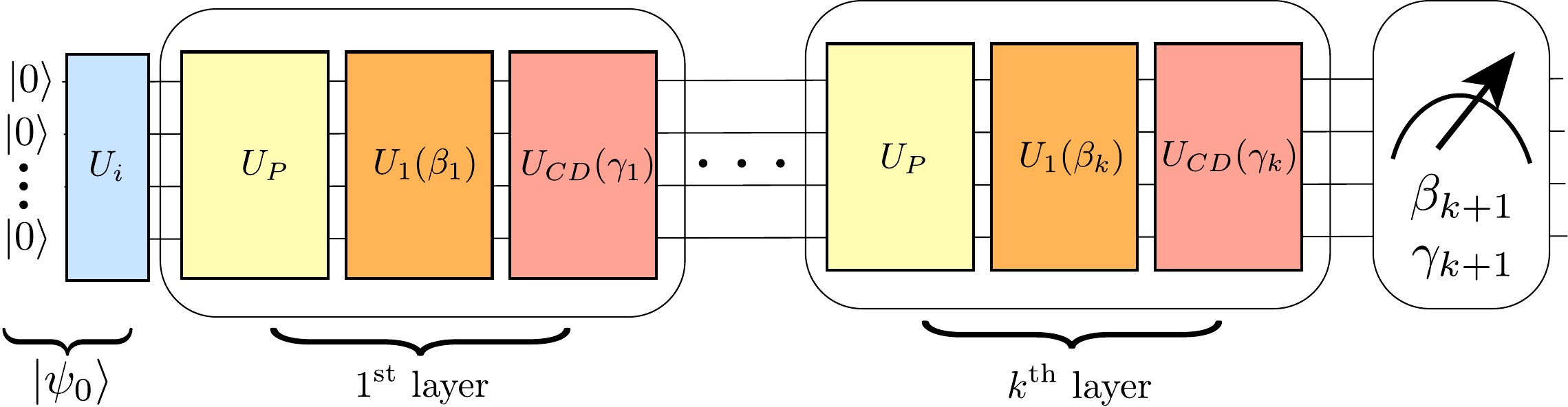}
\caption{
   The schematic diagram of the CD-FQA quantum circuit up to
   $k^{\text{th}}$ layer is shown. The  $(k+1)^{\text{th}}$
   layer is parameterized by $\beta_{k+1}$ and $\gamma_{k+1}$.
   These are obtained by measuring the respective 
   commutator expectation values 
   $\beta_{k+1} = i\langle \psi_{k}|[H_{P}, H_{1} ]|\psi_{k}\rangle$ and
   $\gamma_{k+1} = i\langle \psi_{k}|[H_{P}, H_{\rm CD}]|\psi_{k}\rangle$.}
\label{fig:1}
\end{figure*}

This work presents a significant enhancement to the FQA through
the integration of the counterdiabatic driving
protocol~\cite{demirplak2003adiabatic}, formally termed the
Counterdiabatic Feedback-Based Quantum Algorithm (CD-FQA). While
FQA draws upon QLC principles, counterdiabaticity, a concept
derived from adiabaticity, is employed to effect rapid changes
in the time-dependent Hamiltonian without inducing nonadiabatic
transitions. The utilization of counterdiabaticity in quantum
circuits has been previously applied within the framework of VQA~\cite{yao2021reinforcement,chandarana2022digitized,keever2023adiabatic}. Here, we
explore the dynamic interplay between QLC and
counterdiabaticity, applying these principles to the design of
quantum circuits for the ground-state preparation of
Hamiltonians representing one-dimensional (1D) Ising model
Hamiltonians. Distinct from FQA, each layer in CD-FQA includes a
third unitary inspired by the counterdiabatic driving protocol, see Fig.~\ref{fig:1}.
This addition results in a notable reduction in depth compared
to the standard FQA. The selection of the third unitary is
performed from a pool of counterdiabatic operators. It is
demonstrated that an improper choice from this pool can lead to
convergence issues in the dynamics. The implications of these
findings are discussed in the context of advancing quantum
algorithms for ground-state preparation.




The article is structured as follows:
\Sec{sec:QLC} 
provides a review of QLC, while 
\Sec{sec:CD} 
establishes a connection
between the counterdiabatic driving protocol and QLC, exploring
the selection of the second control Hamiltonian for the CD-FQA.
In \Sec{sec:CD-FQA} 
we present the CD-FQA, comparing complexities
with the standard FQA. \SEC{sec:results:Ising} 
applies the CD-FQA to diverse
Ising models and discusses the outcomes. 
\SEC{sec:experiment} 
demonstrates the implementation of CD-FQA on cloud quantum computers.
\SEC{sec:discussion} 
discusses various implications of CD-FQA, and the conclusion is presented
in \Sec{sec:conclusion}. 

\section{Quantum Lyapunov control}
\label{sec:QLC}

Quantum Lyapunov Control (QLC) represents a form of quantum
control engineered to guide a quantum system from an arbitrary
initial state, denoted as $\vert \psi_i \rangle$, to a specified
final state, $\vert \psi_f \rangle$. This steering process is
facilitated by a target-specific control function $V(t)$,
referred to as the Lyapunov function. The design of such
controlled dynamics often involves placing constraints on the
Lyapunov function.

Consider, for instance, the task of preparing the ground state
of a many-body system for a given  
physical or {\it problem} Hamiltonian $H_P$.
In cases where
the ground state is initially unknown, the system's energy
$E_P(t) = \langle \psi(t) \vert H_P \vert \psi(t) \rangle
\equiv \langle H_P \rangle_t$
naturally emerges as a suitable Lyapunov function. The
controlled dynamics is formulated to adhere to the constraint
$\frac{d}{dt}E_P \equiv \dot{E}_P \leq 0$, 
ensuring that at each time step, the system's
energy experiences a decrement. This condition guarantees a
systematic reduction in the system's energy as the controlled
evolution unfolds.

Let us begin with a driven quantum system, where the dynamics
is governed by the Schr{\" o}dinger equation
\begin{equation}
    i\tfrac{d}{dt}|\psi(t)\rangle
  = \left(H_{P}+H_{C}(t)\right)|\psi(t)\rangle,
\label{Schrodinger}
\end{equation}
where $H_P$ is the problem Hamiltonian as above, and
$H_C(t)$ is the {\it control Hamiltonian}. For convenience, we
set $\hbar=1$, throughout. 
The control Hamiltonian $H_C(t)$ 
can be expressed in the general form
\begin{equation}
H_C(t)=\sum_{m=1}^{M} \beta_{m}(t)H_{m}.
    \label{ctrl-ham}
\end{equation}
In this formulation, the $H_m$'s represent $m=1,\ldots,M$ 
time-independent Hermitian {\it mixing} operators, 
with 
the time-dependence 
embedded in the
real-valued control fields $\beta_m(t)$. 
These control parameters are chosen such that they
ensure a negative rate of change of the 
energy $E_P(t) \equiv \langle H_P \rangle_t$
of the problem Hamiltonian, 
\begin{eqnarray}
   \tfrac{d}{dt}E_P(t)
 &=& i\langle [H_{C},H_{P}] \rangle _t
 \label{EP-dot} \\
 &=& \sum_{m}\beta_m(t) \underbrace{
   i\langle [H_{m},H_{P}] \rangle _t}_{\equiv A_m(t) \ \in\ \mathbb{R}}
   \equiv \boldsymbol{\beta}(t) \cdot {\bf A}(t)
 \ \leq\ 0
\text{ ,}\notag
\end{eqnarray}
with real-valued $M$-dimensional vectors $\boldsymbol{\beta}$ and $\bf A$
in the last expression, and expectation values are obtained
with respect to the wavefunction $\psi(t)$.
To ensure negative
$\dot{E}_P(t) \leq 0$, the conventional choice for the control field is used
$\boldsymbol{\beta}(t) = - \alpha {\bf A}(t)$
with $\alpha>0$. We note that when the system size, $N$, increases the expectation value of commutators $A_{m}(t)$ increases linearly with system size. Therefore, to keep the protocol system independent, we choose 
\be
     \boldsymbol{\beta}(t)
  = -\alpha \,  \tfrac{1}{NJ^2} {\bf A}(t)
\text{ ,}\label{alpha}
\ee
where we applied a factor $1/J^2$
on the r.h.s. to make both, $\alpha$ and $\beta$ dimensionless, and $J$ is the energy scale of the system.
Then, each protocol can be defined by a fixed $\alpha$ that is
independent of the system size.

While usually $M=1$,
the QLC method seamlessly extends to scenarios involving
multiple control fields, $M>1$. 
The inclusion of additional parameters therefore emerges as an
intuitive approach to expedite the preparation of the target
state.
The Lyapunov function, derived from the solution
of \Eq{Schrodinger}, converges to the minimum of $E_P(t)$
under specific sufficient conditions \cite{grivopoulos2003lyapunov,
QLC_Beauchard_2007, QLC_Zhao_2012, QLC_Wang_2013}. Furthermore,
the state converges to a set of states characterized by La
Salle’s invariance principle \cite{la1976stability}.


In the subsequent section, we demonstrate that these additional
control fields can be derived from a pool of operators commonly
employed in the context of the counterdiabatic driving protocol.
This methodology presents a promising avenue for
enhancing the efficiency of state preparation using a QLC protocol.

\section{Counterdiabatic drive inspired control Hamiltonians}
\label{sec:CD}

The counterdiabatic driving protocol is a pivotal concept in
non-equilibrium physics, employed to induce
rapid changes in the time-dependent Hamiltonian without inducing
transitions \cite{demirplak2003adiabatic,sels2017minimizing}
across instantaneous eigenstates.
This phenomenon is also recognized as a ``shortcut to
adiabaticity" \cite{chen2010fast, del2013shortcuts,
guery2019shortcuts,takahashi2019hamiltonian}.

Consider a time-dependent Hamiltonian $H(\beta(t))$, where
$\beta$ represents an arbitrary function of time. When a quantum system, initially prepared in an eigenstate of the initial Hamiltonian, evolves
under $H(\beta(t))$, it undergoes nonadiabatic excitations,
causing it to deviate from the instantaneous eigenstate. To
eliminate such transitions, a velocity-dependent term
proportional to $\dot{\beta}$ is introduced to the original
Hamiltonian, yielding $H(\beta)+\dot{\beta} A_{\beta}$.
Here, $A_{\beta}$ is defined as
\begin{equation}
\langle m|A_{\beta}|n\rangle=i \langle m|\partial_{\beta} n\rangle=-i\frac{\langle m|\partial_{\beta} H | n\rangle}{\epsilon_{m}-\epsilon_n},
    \label{gaugepotential}
\end{equation}
is known as adiabatic gauge potential
with $|m\rangle$ and $|n\rangle$ being two instantaneous eigenstates of $H(\beta(t))$ and $\epsilon_m$ and $\epsilon_n$ are the corresponding time-dependent energies.

Finding an exact form of $A_{\beta}$ for many-body quantum
Hamiltonians is impractical since it requires diagonalizing the
time-dependent many-body Hamiltonian. However, recently it has been shown in Ref.~\cite{PolkovnikovPRL2019}
that an approximate gauge potential can be obtained without the
need for diagonalization. This approximation is constructed
using 
nested commutators, 
\begin{equation}
  A_{\beta}^{l}=i\sum_{k=1}^{l} \gamma_{k}(t) \underbrace{[H(\beta),[H(\beta),...[H(\beta)}_{2k-1~{\text{times}}},
  \underbrace{\partial_{\beta} H(\beta)}_{
    \equiv H_1
  }]]
\,, \label{gauge}
\end{equation}
determined by a set of coefficients
$\{\gamma_1,\gamma_2,...,\gamma_l \}$, where $l$ is the order of
the expansion. By properly tuning these coefficients one can suppress the nonadiabatic excitations in the system. As seen from \Eq{gaugepotential},
the matrix elements of the counterdiabatic operator
$\hat{A}$ change sign when taking the transpose.
For real matrix elements $\langle m|\partial_{\beta} H | n\rangle$, 
the operator $\hat{A}$ is antisymmetric and purely imaginary.
 In this case (relevant for our work here) only odd nested commutators are \cite{PolkovnikovPRL2019}
included in \Eq{gauge}.
For large values of $l$, the gauge
potential incorporates long-range interacting terms. In cases
where the physical Hamiltonian encompasses terms up to
nearest-neighbor interactions, it may be possible 
to approximate the
adiabatic gauge potential solely with local and two-body
interaction terms \cite{vcepaite2023counterdiabatic}.


Having reviewed the counterdiabatic driving protocol, we now apply it in the context of QLC, where we introduce additional control fields inspired by counterdiabatic driving protocols.  This inspiration is only in spirit and stems from the fact that the imaginary operators computed from Eq.~(\ref{gauge}) can generate fast mixing between different eigenstates. However, we note that our goal is not to make the system adiabatic. Rather, we propose to find the coefficient $\gamma_{k}(t)$ using multi-control QLC. Then, the resulting $A_{\beta}^{l}$ may not be truly a gauge potential. The main idea here is to use the operators from the nested commutators and the coefficients from the QLC and integrate them into the quantum circuit. 


The time-dependent Hamiltonian including the first control field 
\begin{equation}
   H(\beta(t)) 
   = H_P+\beta(t) \,H_1 \ ,
\label{Ham-one}
\end{equation}
where $\beta \equiv \beta_1$ is the control field that takes the
role of $\beta$ earlier. The time evolution mixes the
eigenstates of $H_P$, with couplings $\langle n|H_1|m\rangle$
between the $n^{\text{th}}$ and $m^{\text{th}}$ levels. 
The first control Hamiltonian, $H_1$, is chosen
heuristically. It 
is inspired from quantum annealing or QAOA. 
{It is} 
also called a mixer {Hamiltonian, in 
that it} mixes the eigenstates of the
problem Hamiltonian. Here we will consider the $H_1$ to be a sum
of operators that act only on local qubits.
For example, for the applications on the Ising model below, we will choose $H_1$
as a sum over Pauli-$x$ operators
which mixes across different $S_z$ sectors.

To enhance population transfer, it is essential to devise a
dynamic process that facilitates swift mixing between
instantaneous eigenstates and surpasses 
the efficiency of the QLC
with a single control parameter. To address this, we incorporate
an additional control Hamiltonian into the feedback algorithm.
While the potential addition of any number of control
Hamiltonians are feasible in principle, our preference is to
limit it to one due to the practical considerations associated
with quantum circuit implementations.

To speed up transitions across wider energy
intervals, one can weight transitions based on the
energy differences between instantaneous eigenstates.
Hence
\Eq{Schrodinger} with an additional control Hamiltonian $H_{\rm CD}$, i.e.,
\begin{equation}
    H_P + H_C(t) = H(\beta(t))+\gamma(t)H_{CD},
\end{equation}
reads
\begin{equation}
i\dot{a}_n=\epsilon_n a_n+ \gamma(t) \sum_m 
    \la n|H_{\rm CD} |m \ra a_m,
    \label{Schrodinger-amp1}
\end{equation}
where we have written $|\psi(t)\rangle$ as
$$
|\psi(t)\rangle = \sum_n a_n(t)|n(t)\rangle,
$$
where the $|n\ra$'s are the instantaneous eigenstates of
$H(\beta(t))$.  The matrix element $\la n|H_{\rm CD} |m \ra$
must be dependent on the energy differences between level $n$
and level $m$. We select $H_{\rm CD}$ from a pool constructed by the nested commutator
$A_{\beta}^{l}$ in Eq.~(\ref{gauge}). 
The control field
$\gamma(t)$ is determined by the QLC protocol described in
\Sec{sec:QLC}. 



The energy differences in \Eq{Schrodinger-amp1} are based on the
eigenstates of the Hamiltonian $H(\beta(t))$.  The control
Hamiltonian $H_{\rm CD}$ can also be constructed from
\Eq{gauge}, by replacing $H(\beta(t))$ with the problem
Hamiltonian $H_P$.  The operator pool generated by the nested
commutator constructed from $H_P$ is a subset of the operator
pool generated by Eq.~(\ref{gauge}). For practical purposes, one
can use both sets of operator pools. Since we strongly truncate
the series in \Eq{gauge} anyway, we expect both pools to have
comparable performance.

\section{Counterdiabatic feedback-based quantum algorithm (CD-FQA)}
\label{sec:CD-FQA}

We enhance the FQA by introducing an additional control field inspired by counterdiabatic driving protocol. The resulting digital
quantum circuit for counterdiabatic FQA (CD-FQA) 
discretizes the
time evolution of the Schr\"odinger equation (\ref{Schrodinger}) with two control fields,
\begin{equation}
  i\tfrac{d}{dt}|\psi(t)\rangle
  = \bigl(
       H_P + \beta(t) H_1 + \gamma(t) H_{\rm CD}
    \bigr)|\psi(t\rangle)
\text{ ,}\label{Schrodinger-CD}
\end{equation}
where $H_{\rm CD}$ is an operator selected from the pool of
operators inspired by counterdiabatic driving protocol. 
\EQ{Schrodinger-CD} can be seen as specialization
of \Eqs{Schrodinger}-\eqref{EP-dot} for $M=2$,
with $\gamma \equiv \beta_2$ and $H_{\rm CD}$
a particular choice for $H_2$ motivated from 
counterdiabatic driving.

The CD-FQA quantum circuit is assembled by successively applying
three unitaries,
\begin{equation}
   |\psi_{l}\rangle
   =\prod_{k=1}^l {\cal U}_k|\psi_{0}\rangle
   =\prod_{k=1}^l U_{\rm CD}(\gamma_k) U_1(\beta_k)  U_{P}|\psi_{0}\rangle
\,,\label{Falqon-1}
\end{equation}
Here, $|\psi_{0}\rangle$ represents the arbitrary initial state,
$|\psi_{l}\rangle$ is the quantum state after applying $l$
layers of unitaries, and each layer is parameterized by
$\{\beta_k,\gamma_k\}$. The unitaries are defined as $U_{P}\equiv
e^{-iH_{P}\Delta t}$, $U_{1}(\beta_{k})\equiv
e^{-i\beta_{k}H_{1}\Delta t}$, and $U_{\rm CD}(\gamma_{k})\equiv
e^{-i\gamma_{k}H_{\rm CD}\Delta t}$. For small $\Delta t$, this
evolution closely resembles the
continuous-time evolution of the system. The parameter $\Delta t$
must be small enough so that the first-order reduction in
energy must exceed all the higher-order terms
\cite{FeedbackPRA}. 

The quantum circuit in CD-FQA is constructed iteratively.
The 
unitaries $U_1$ and $U_{\rm CD}$ for the $(k+1)^{\text{th}}$ 
layer depend on the respective parameters
$\beta_{k+1}$ and $\gamma_{k+1}$. 
To determine these parameters, 
we compute the commutators $i\langle [H_{1},
H_{P}]\rangle$ and $i \langle [H_{\rm CD}, H_{P}]\rangle $
using a quantum circuit for the 
state $|\psi_{k}\rangle$, 
i.e., a state that is built up to the $k^{\text{th}}$ layer. 
Following the conventional choice for the application of QLC,
we set the control fields
to the following expectation values 
(cf. \Fig{1}):
\begin{eqnarray}
   \beta _{k+1} &=& \tfrac{i\alpha}{N}\,\langle \psi_{k}|[H_{P}, H_{1}\ \ ]|\psi_{k}\rangle, \notag \\
   \gamma_{k+1} &=& \tfrac{i\alpha}{N}\,\langle \psi_{k}|[H_{P}, H_{\rm CD}]|\psi_{k}\rangle
\text{ .}\label{eq:params:k+1}
\end{eqnarray}
Once these parameters are determined, the 
procedure is repeated iteratively
to construct the 
{next layer $k+2$}. 
Since the parameters $\beta$ and $\gamma$ enter as
prefactors to Hamiltonians, they need to scale independent
of system size. This necessitates the $1/N$ scale factor
in \Eq{eq:params:k+1}.  

The expectation value of an operator $\mathcal{O}$
is obtained by expanding it 
as a linear combination of Pauli 
operators,
{$\hat{O} = \sum_{i}^{\tilde{N}}\! \gamma_i\, \hat{P}_{i}$, such that}
$\langle \psi_k| {\hat O} |\psi_k\rangle=\sum_{i}^{\tilde{N}}\!
\gamma_i\, \la {\hat P}_{i}\ra$ where $\gamma_{i}$ are scalar
coefficients and $P_{i}$ are {compact finite-size strings}
of Pauli basis operators. The
measurement of the Pauli operators $P_i$ is repeated to collect
statistics. 
The resulting 
expectation values are combined to find 
the expectation value of $\mathcal{O}$
in $|\psi_k\rangle$. 
The number of
Pauli operators and the number of measurements will depend on
the structure of $H_P$, $H_1$, and $H_{\rm CD}$. Here, we
consider Hamiltonians with nearest-neighbor hopping. 
Therefore the number of Pauli operators is 
${\tilde{N} \propto} N$. 

The measurement of {$\tilde{N} \propto N$} 
Pauli operators can be efficiently parallelized
\cite{gokhale2019minimizing,
verteletskyi2020measurement, reggio2023fast, anastasiou2023really,
 zhu2023optimizing}. 
Consider, for example, the measurement of a spin Hamiltonian \(\sum_{i}^{N}
\sigma_{i}^a \sigma_{i+1}^b\) for $a\neq b$. The total number of
two-qubit Pauli strings required to be measured per layer is
{$\tilde{N} = N$}. The terms \(\sigma_{i}^a \sigma_{i+1}^b\) and
\(\sigma_{j}^a \sigma_{j+1}^b\) can be measured simultaneously
if they commute.
This holds trivially if the Pauli strings do not overlap,
i.e., for $|i-j|>1$. On the other hand, overlapping,
yet commuting Pauli strings share the Pauli basis
operators.
In the example above the number of Pauli strings can be
divided into two sets containing commuting Pauli strings 
that act either on even or odd bonds.
%
All Pauli strings within a set can be measured simultaneously.
Consequently, the number of parallel measurements required
per layer is~2. This number corresponds to the two noncommuting
terms in the Hamiltonian that act on any given spin. For any
counterdiabatic Hamiltonian $H_{\rm CD}$, the number
measurements can be obtained from the number of noncommuting 
terms in the commutator $[H_P, H_{\rm CD}]$ that act on any
given spin. 
We discuss 
in detail below
the number of
measurements needed for the LFI model with different
counterdiabatic Hamiltonians and compare it with the standard FQA.


 

Each layer in CD-FQA comprises two parameters. In comparison
with standard FQA, 
CD-FQA demands twice
the number of measurements per layer. The circuit depth in
CD-FQA is $3l$, while the circuit depth in FQA is only $2l$,
where $l$ is the number of layers. This extended depth per layer
reflects the added complexity introduced by the counterdiabatic
control field, emphasizing the need for enhanced computational
resources per each layer in CD-FQA. Nevertheless, the
incorporation of an additional control field in CD-FQA leads to
a reduced number of layers in CD-FQA. For a small number of
layers our algorithm shows tremendous improvement over standard
FQA.


\begin{figure}
\begin{center}
  \sidesubfloat[]{\label{:a}%
  \includegraphics[scale=0.179,trim = 0in 1.5in 0in 0in, clip=true]{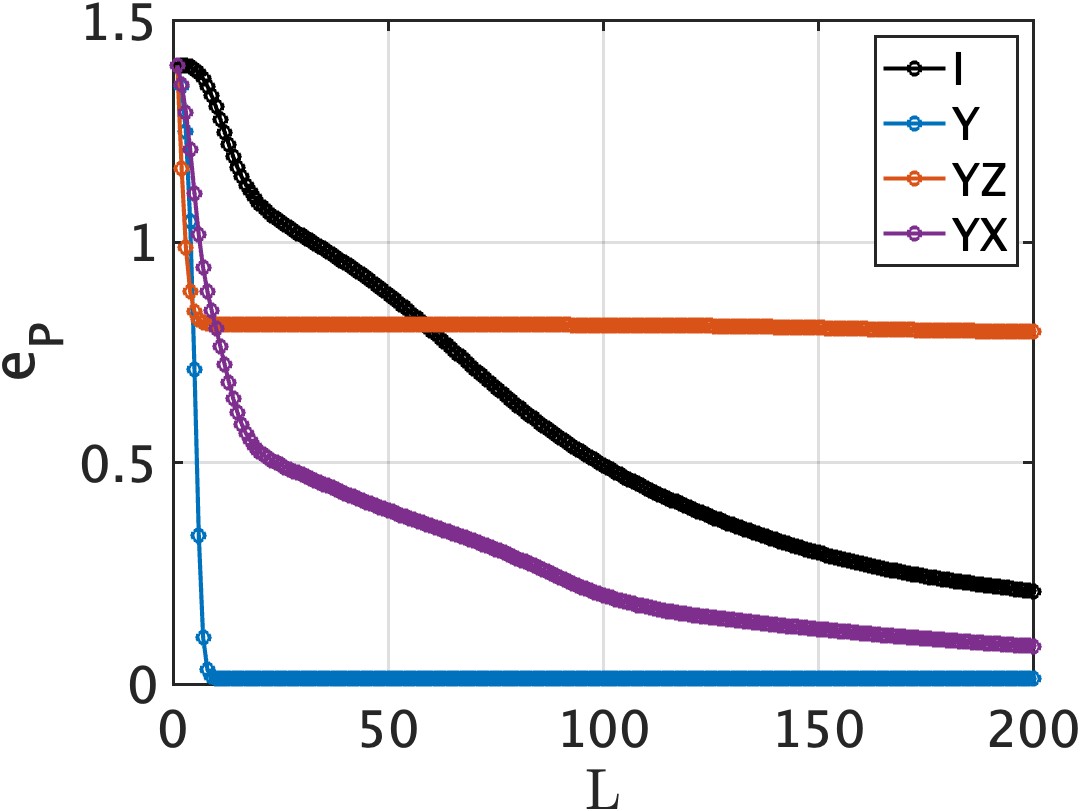}} \\
  \sidesubfloat[]{\label{:b}\ \ %
  \includegraphics[scale=0.175,trim = 0in 1.4in 0in 0in, clip=true]{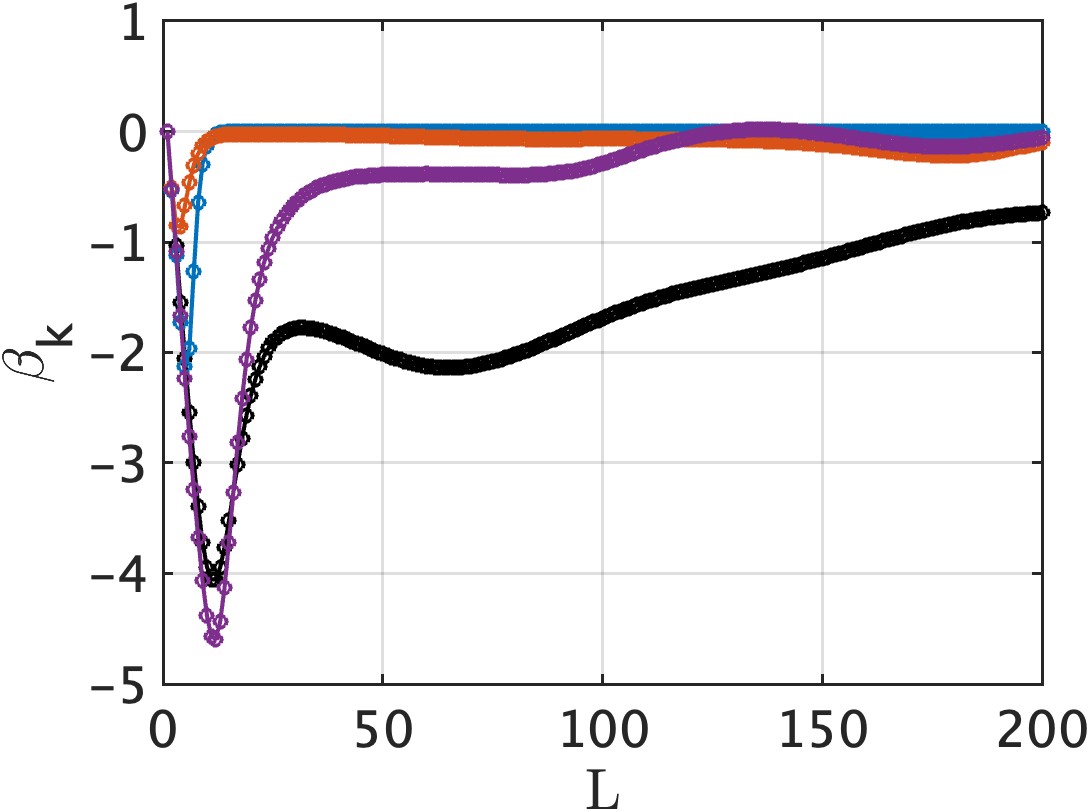}} \\
  \sidesubfloat[]{\label{:c}%
  \includegraphics[scale=0.179]{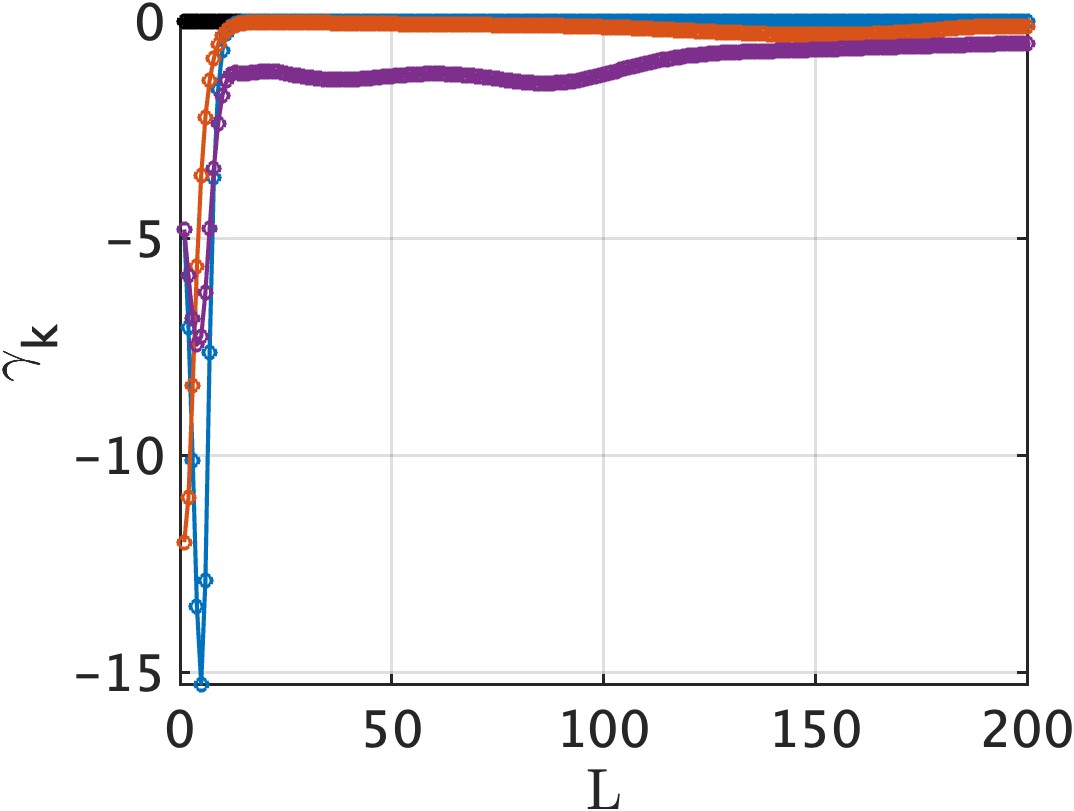}}\\
\end{center}
\caption{
(a) The average energy per site is
    shown as a function of the number of layers for LFI 
    $(h_z=0.4$, $h_x=0)$ with $N=6$ spins for various CD-FQA protocols. The black color represents
    the standard FQA which is equivalent to taking
    the identity for $H_{\rm CD}$, i.e., $H_{\rm CD}=I$
    as indicated in the legend.
    The other colors represent CD-FQA with a
    particular operator  denoted by $H_{\rm CD}$ selected from the pool (\ref{pool}).
(b) The first and (c) the second control fields, $\beta_k$ and
    $\gamma_k$, respectively, are shown as a function of the
    number of layers for different $H_{\rm CD}$. Parameters $\alpha=6$
    and $\Delta t = 0.01 / J$. 
}
\label{fig:LFI}
\end{figure}

\begin{figure}
\centering
\includegraphics[scale=0.18]{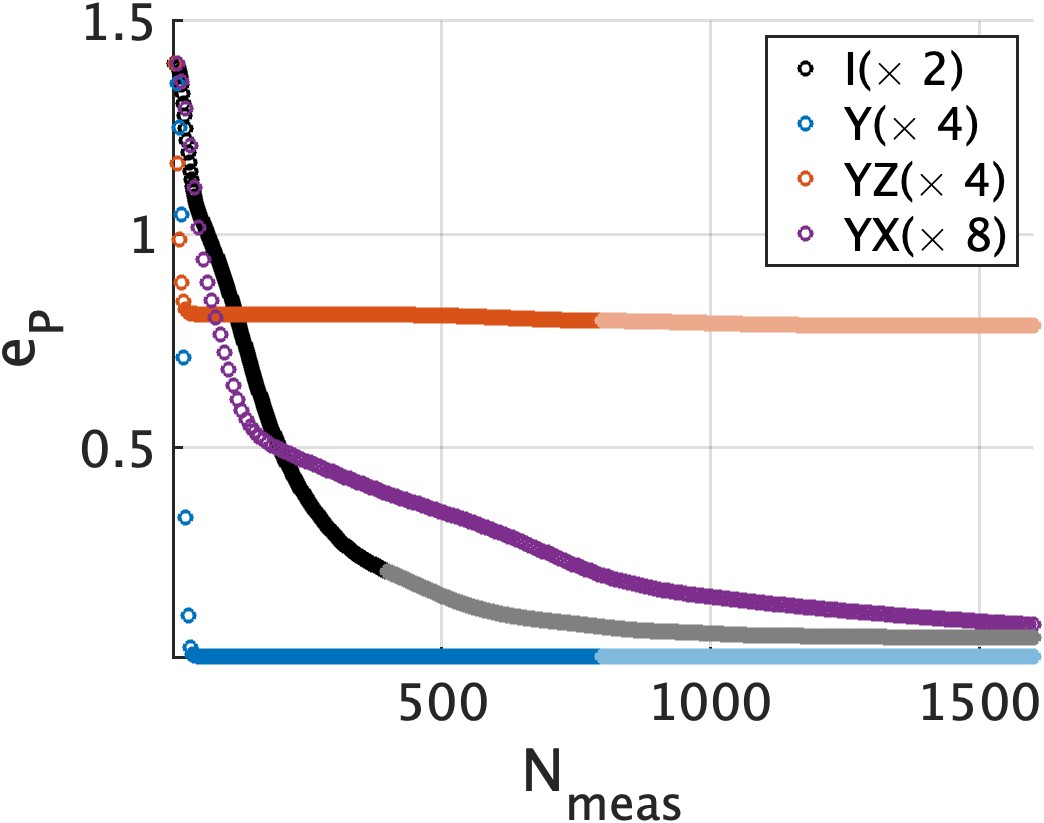}
\caption{
The average energy per site for LFI 
  vs. 
  number of parallel measurements
  up to repeats.
  Same data as in \Fig{LFI}, yet plotted vs. the number 
  of measurement layers $N_{\rm meas}$. 
Because the number of parallel measurements per layer varies
across different CD-FQA protocols, 
this applies different horizontal scale factors 
to the data in \Fig{LFI}(a) as indicated with the legend here. The darker sections 
of each curve correspond to
the $L=200$ layers in \Fig{LFI}.
}\label{fig:measurements}
\end{figure}


\begin{figure}
\centering
\includegraphics[scale=0.18]{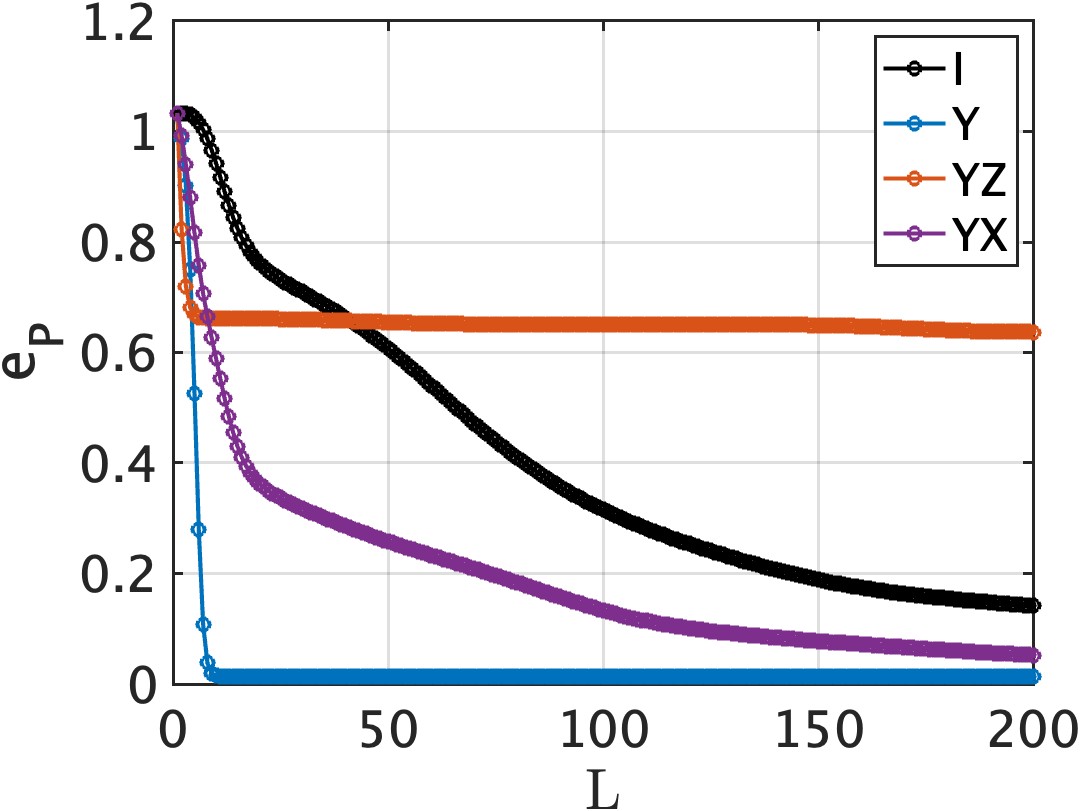}
\caption{
   The average energy per site is shown as a function of the number of
   layers for MFI ($h_z=h_x=0.4$). 
   All other parameters are same as Fig. \ref{fig:LFI}.
}
\label{fig:MFI}
\end{figure}


\begin{figure}
    \centering
    \sidesubfloat[]{\label{a}\!\!\!\includegraphics[scale=0.1]{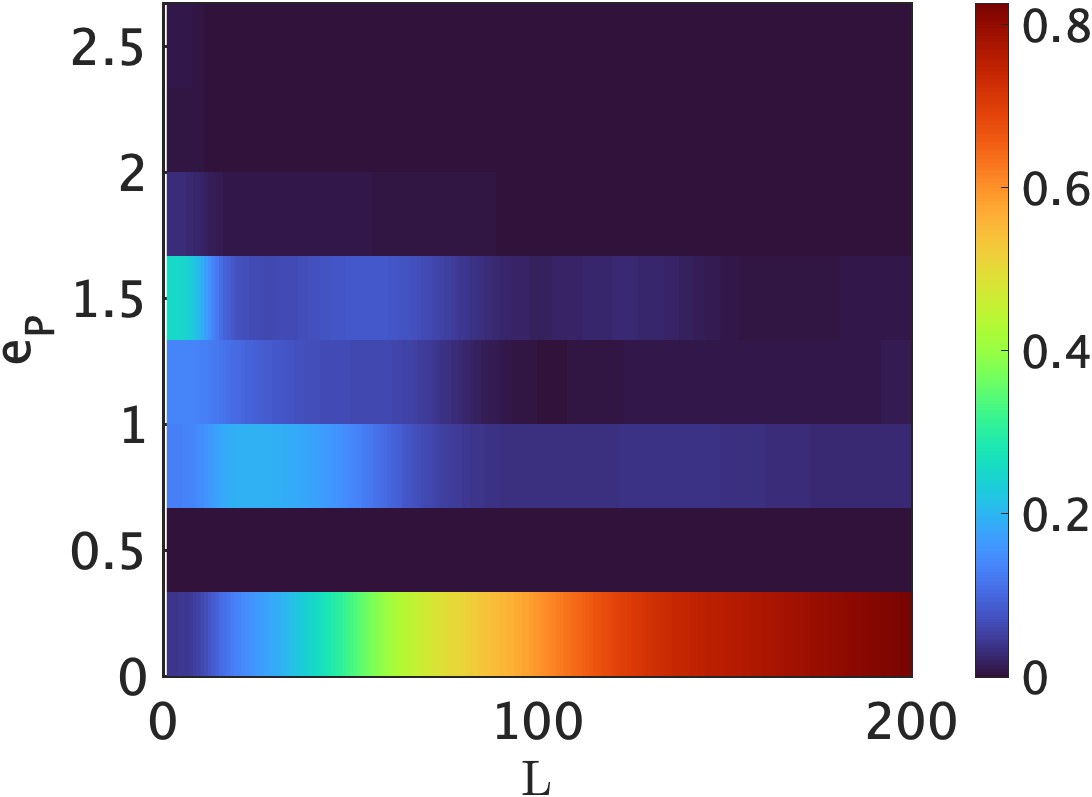}\ } 
    \sidesubfloat[]{\label{b}\!\!\!\includegraphics[scale=0.1]{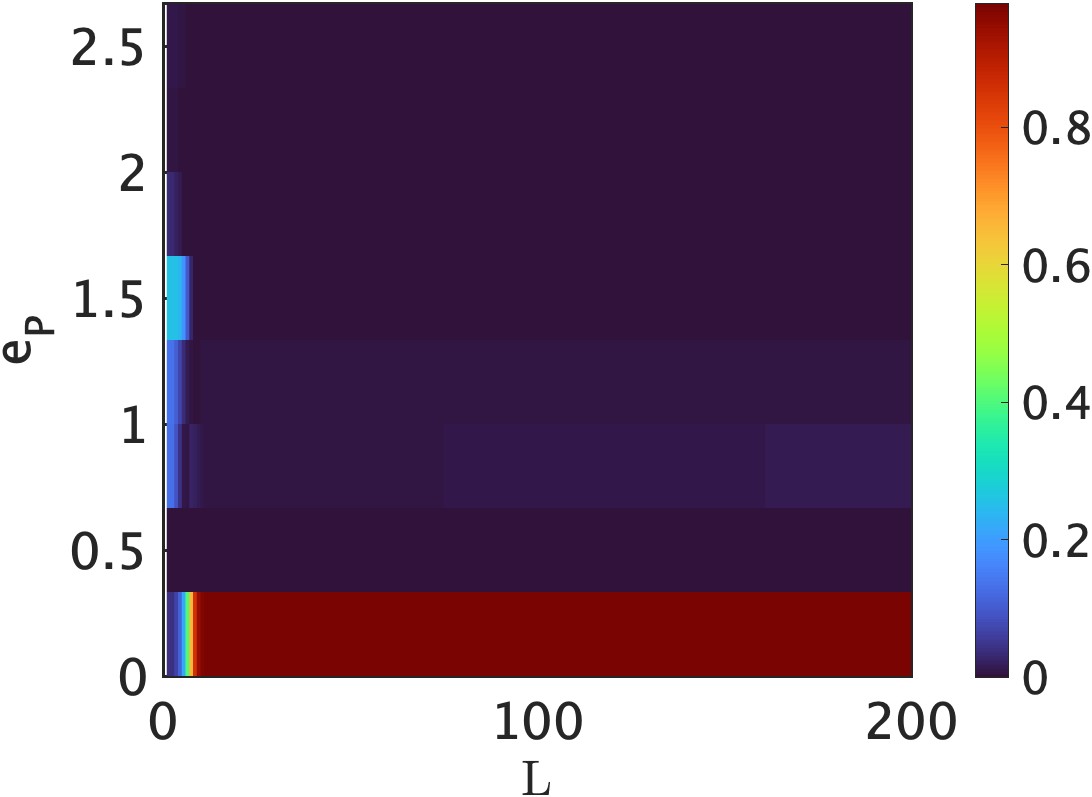}}\\
    \sidesubfloat[]{\label{c}\!\!\!\includegraphics[scale=0.1]{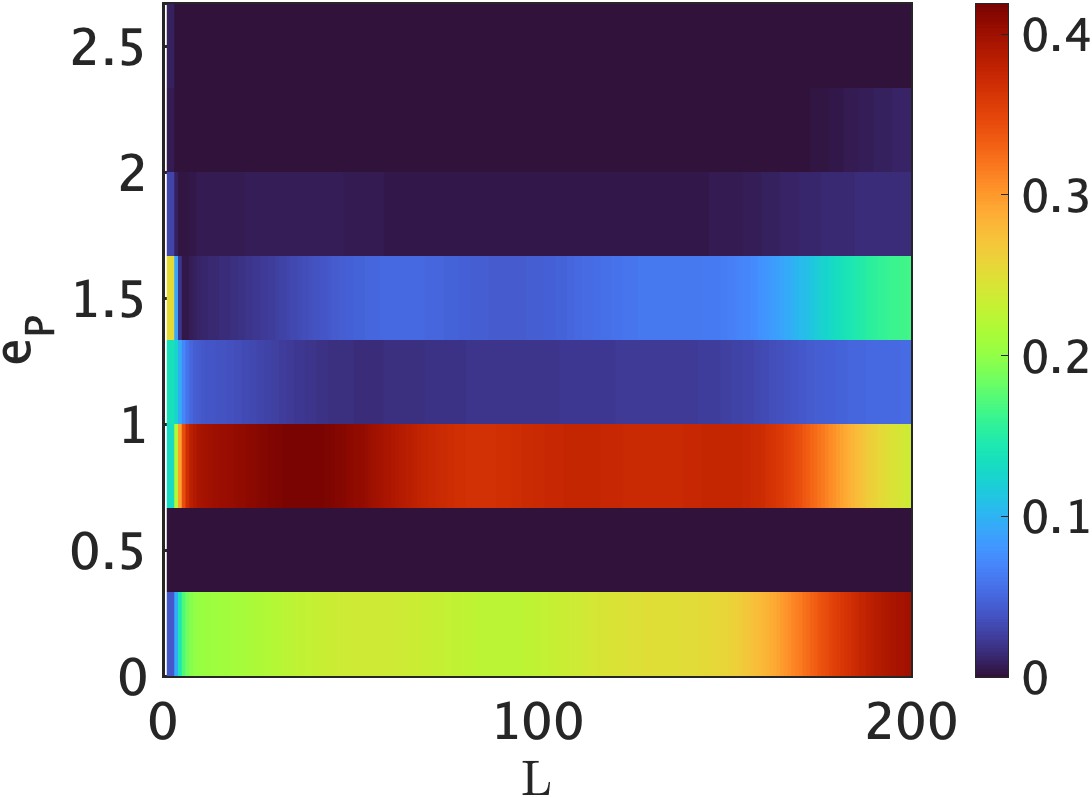}\ }
    \sidesubfloat[]{\label{d}\!\!\!\includegraphics[scale=0.1]{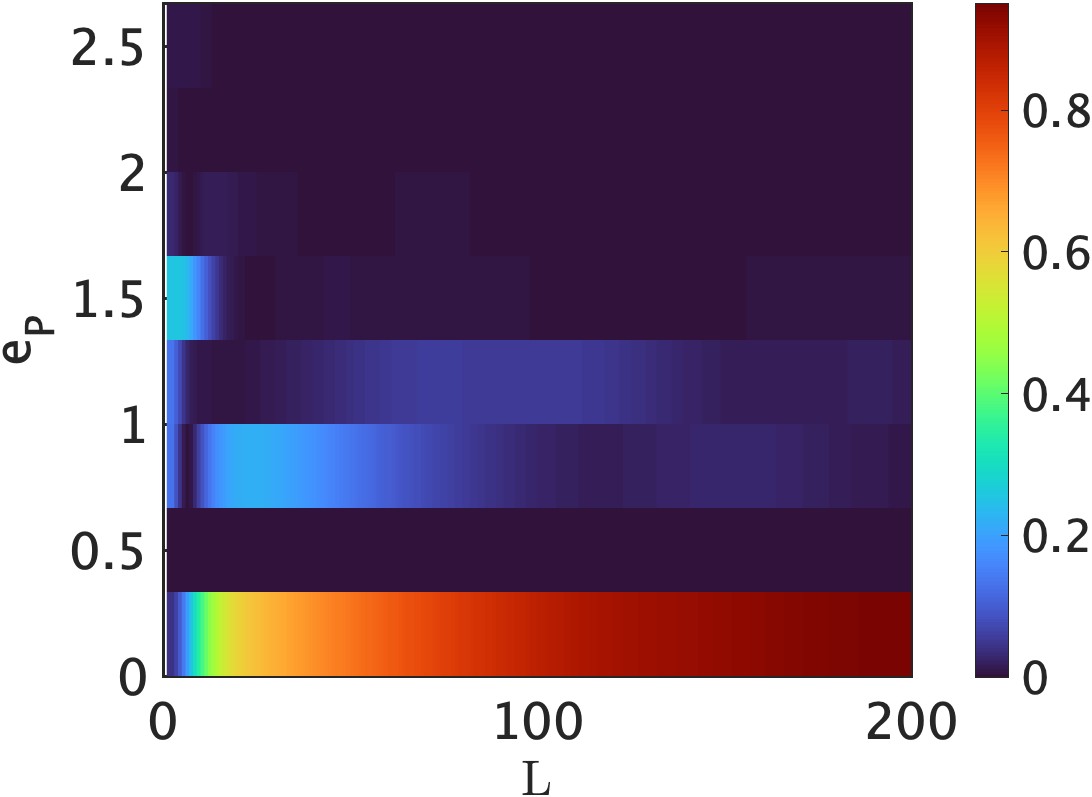}}
\caption{
   Binned energy distribution 
   of the state $\psi$  vs. circuit depth for the MFI
   for the simulation of \Fig{MFI}.
   The panels 
   represent CD-FQA protocols with operators
      (a) $I$ (i.e., plain FQA),
      (b) $Y$,
      (c) \YZ, and
      (d) \YX, respectively.
   The average energies per site $e_P$
   are coarse-grained 
   into 8 bins of equal width 
   $2J$. For each circuit depth,
   the energy densities integrate to $1$ vertically
   over all energies.}
\label{fig:MFIcolor}
\end{figure}


\section{Results: Preparing the ground-state of Ising spin models}
\label{sec:results:Ising}

We apply the CD-FQA to Ising chains
of length $N$  
\begin{eqnarray}
   H_{I} 
   &=& \sum_{i}^N \Bigl(
        -J \sigma_{i}^z\sigma_{i+1}^z
     - h_z \sigma_{i}^z
     - h_x \sigma_{i}^x
   \Bigr)
\label{eq:IsingModel} \\
  &\equiv& - \ZZ - h_z \, Z - h_x \, X
\text{ ,}\notag
\end{eqnarray}
and for various parameter settings.
The nearest-neighbor interaction $J:=1$ 
sets the unit of energy, throughout,
while $h_z$ and $h_x$ specify 
 the longitudinal and transverse fields, respectively. The operators
$\sigma_{i}^a$ with $a \in \{x,y,z\}$ are the standard Pauli
operators acting on site $i$.
We use periodic boundary
conditions (PBC) in all classical simulations except for Fig. \ref{fig:experiment}
where we use open boundary conditions  (OBC) for quantum simulation.  
For convenience, we employ shorthand notations to describe the
sums of Pauli operators. The sum of local operators
is denoted by $A \equiv \sum_{i}^N \sigma_{i}^a$,
with $A \in \{X,Y,Z\}$ corresponding to $a\in \{x,y,z\}$, respectively.
Similarly, two-body terms are denoted by
$AB \equiv \sum_{i}^N \sigma_{i}^a\sigma_{i+1}^b$.
With this, the Hamiltonian in~\Eq{eq:IsingModel}
can be written as shown in the line below it.

By varying the parameters
$h_z$ and $h_x$,
we investigate four distinct types of Ising models:
(i)   longitudinal field Ising (LFI) when $h_x$ = 0,
(ii)  transverse field Ising (TFI) when $h_z$ = 0,
(iii) mixed-Field Ising (MFI) with non-zero values for all 
parameters, and
(iv)  A special case $h_x=h_z=0$. 
This study showcases the versatility of the CD-FQA approach in
tackling various Ising models with different field
configurations. 

In all considered cases, the first control Hamiltonian is
defined as $H_1=X$. This Hamiltonian, commonly referred to as a
mixer Hamiltonian, is a standard choice in QAOA and quantum
annealing protocols, particularly when the problem Hamiltonian
consists of Z 
terms. Here, we set the initial state
to be the ground state of $-H_1$ 
$|\psi_{0}\rangle = |X\rangle \equiv
|\rightarrow\rightarrow \ldots \rightarrow\rangle$,
where all spins
are aligned along the $x$-axis. This initial state, being a
product state, can be readily prepared on a quantum circuit
using only Hadamard gates. Equivalently,
$|X\rangle = e^{-i\frac{\pi}{4} Y} |Z\rangle$
with $|Z\rangle \equiv |{+}Z\rangle
\equiv |\uparrow \uparrow \ldots \uparrow \rangle$.

The second control Hamiltonian in our approach draws inspiration
from the counterdiabatic protocol as discussed in 
\Sec{sec:CD}, 
and is selected from an operator pool generated by the nested
commutator Eq.~(\ref{gauge}). Importantly, we restrict
the operator pool to include only local and two-body terms. When
the problem Hamiltonian and the first control Hamiltonian are
real then the operator pool comprises solely operators with imaginary matrix elements.
The counterdiabatic operator pool $A_{\text{pool}}$
is a subset of the operator pool consisting of all the local and
two-body operators and is given by
\begin{equation}
   A_{\rm pool}\subseteq 
   H_{\rm pool} = \{Y, \ZY, \YZ, \XY, \YX \}
\text{ .}\label{pool}
\end{equation}
The terms in \Eq{pool} are generated by commuting individual
terms of $H_P = H_I$ with $H_1=X$. As it turns out, 
\YZ and \ZY exhibit identical behavior,
similar to \YX vs. \XY. 
With this, we eliminate \ZY and \XY from the pool above.
Yet for the sake of the presentation, we include the
identity $I$ to the pool, which then simply represents
the standard FQA.

\subsection{$h_z\neq 0$ (LFI and MFI)}
First, we consider the case of non-zero
longitudinal field 
$h_z = 0.4$ where we find that both LFI ($h_x=0$, \Fig{LFI})
and MFI ($h_x=h_z$, \Fig{MFI}) yield
similar results.
The system consists of $N=6$ spins, 
and the simulation is performed up to $200$
circuit layers using 
$\Delta t=0.01$. The parameter $\alpha$ in the QLC protocol is assumed to be $6$, so that the prefactor $\alpha/N=1$. 

We depict three distinct CD-FQA protocols, each associated with
a different operator selected from the pool, characterized by
$H_{\rm CD} \in 
\{ Y, \YZ, \YX \}$. These 
are derived from terms that arise from
the nested commutators (\ref{gauge}),
$Y$ and \YZ at 
first order, and 
\YX at 
second-order. 
The performance of CD-FQA, for each choice of $H_{\rm CD}$,
is compared against the standard FQA represented by the black
curves. 

In \Fig{LFI}a we plot
the average energy per site relative to the ground state energy
$E_0^P$ of $H_P$,
\begin{eqnarray}
   e_P &\equiv& 
   \tfrac{1}{N}\, (\langle H_P\rangle_t -E_0^P)
\label{eq:ep}
\end{eqnarray}
against the number of circuit layers. While, by construction,
all four curves show monotonic decay,
there are significant qualitative differences. 
The 
CD-FQA approaches demonstrate a strongly accelerated 
reduction of the energy at early times, i.e.,
small number of layers. However, with
an increasing number of layers, the CD-FQA protocol associated
with \YZ shows early plateau-like behavior, 
thus failing to decrease the energy to $E_0$. 
We find that the CD-FQA with $Y$ achieves the
most favorable results, followed by the \YX protocol.
These findings highlight the effectiveness of CD-FQA in the
ground-state preparation, yet also reveal clear differences
depending on the choice of the
operator for $H_{\rm CD}$.

The control fields $\beta(t)$ and $\gamma(t)$ are presented in
\Fig{LFI}b and \Fig{LFI}c. 
Starting from 
zero, these fields decrease rapidly towards a minimum, before returning to zero at large times
with an irregular oscillatory intermediate behavior.
The initial changes in $\gamma(t)$ [\Fig{LFI}c] strongly
surpass those in $\beta(t)$ [\Fig{LFI}c], thus
contributing to a significantly more rapid decay in average
energy as
compared to standard FQA. 
The control fields in the CD-FQA with
$Y$ and \YZ reach zero within a short time, while in the
standard FQA and CD-FQA with \YX
they have sizeable value over a significantly longer
times.

The number of measurements is a key resource in our protocol. As
we have established, the number of measurements per layer 
{without repeats} equals
the number of non-commuting Pauli strings acting on any site. For
the LFI model the commutator $[H_P, H_1]$ yields terms $YZ+ZY$
and $Y$. The term $\sigma_{i}^y\sigma_{i+1}^z$ commutes with
$\sigma_{j}^z\sigma_{j+1}^y$ for $j = i \pm 1$.
Therefore they can be measured
simultaneously. With this, 
the number of parallel
measurements needed to find expectation values of $YZ+ZY$ and
$Y$ is 2, since measuring $YZ{+ZY}$ 
is enough to extract information about the $Y$-measurement.
Similarly {for the CD-FQA protocol, with} $H_{\rm CD}=Y$,
the commutator $[H_P, H_{\rm CD}]$ yields terms $\{X, XZ, ZX\}$.
Again this requires 2 parallel measurements, {adding up to} 
a total of 4 measurements per layer. For $H_{\rm CD}=YZ$, the
commutator $[H_P, H_{\rm CD}]$ yields $\{X, ZXZ, XZ\}$.
{This also} 
requires 2 parallel measurements per layer, {adding up to
a total of 4 parallel measurements for that protocol}. For $H_{\rm
CD}=YX$, the commutator $[H_P, H_{\rm CD}]$ yields $\{XY, YYZ,
ZXX, XX, YY\}$. 
Since the {expectation values for} 
$YZ+ZY$ and $Y$ {out of $[H_P,H_1]$}
can be {obtained from the data for this set},
the protocol requires a total of~8 parallel measurements per layer. In \Fig{measurements}, we plot
the average energy per site vs number of measurements required for
different CD-FQA protocols. The number of measurements in the
protocol with $Y$ as a CD operator converges much faster than
other protocols. The protocol with $YX$ as a CD operator
requires more measurements per layer as compared to standard FQA.
Therefore the measurement cost between different protocols
depends on the degree of locality in the system.
{We emphasize, however,} 
that the number of {parallel} measurements is {\it independent}
of the system size {for any protocol}.

We repeat the same simulation as in \Fig{LFI}
but this time also with $h_x = 0.4\, (= h_z)$ turned on (MFI).
The data presented in \Fig{MFI} 
is very similar to \Fig{LFI}a,
except that the system starts out a somewhat lower
energy in the system when adding the transverse term.

The key distinction between the LFI and MFI models lies in the
additional $X$ term present in the problem Hamiltonian.
Since $H_1=X$, the introduction of the $X$ term in $H_P$
has minimal influence on $\beta \sim \langle [H_P,H_1]\rangle$.
On the other hand, 
$\gamma\sim \langle [H_P,H_{\rm CD}]\rangle $ 
does acquire 
additional terms. 
However,
given that the initial state is the ground state of $X$,
these 
contribute insignificantly to the rates $\gamma$ at early times
since for the transverse term $X$ in $H_P$,
$\langle[X, H_{\rm CD}]\rangle \sim 0$.
Further discussions on the rate of energy change and its
dependence on various commutators and the initial state are
elaborated in \Sec{sec:discussion} 
in the context of different CD-FQA
protocols.

The emergence of plateaus in \Fig{LFI}a and
\Fig{MFI} raises questions on the nature of the `steady'
state reached. In the worst case, the system might converge to an
excited eigenstate of $H_P$, in which case also
$\beta,\gamma \to 0$. Hence in order
to gain deeper insights into the impact of CD-FQA on ground
state preparation, \Fig{MFIcolor} 
tracks the energy distribution in the system
vs. circuit depth with respect to the eigenspectrum of $H_P$.
For this purpose, we partitioned 
the full many-body energy window 
into eight bins. The general behavior 
in \Fig{MFIcolor} largely aligns with those in \Fig{MFI}.
In the standard FQA [\Fig{MFIcolor}a], the
population gradually transfers to the ground state, reaching
approximately a weight of $p_0 \sim 0.825$ in the lowest bin, where $p_i$ is the overlap of the wavefunction with all the eigenstates in the $(i-1)^{\text{th}}$ bin. 
In contrast, the Y-FQA protocol achieves 
$p_0 \sim 0.976$ within just a few layers.
Similarly, 
the YX-FQA protocol, in agreement with the results in
\Fig{MFI}a reaches 
$p_0 \sim 0.948$.

The YZ-FQA protocol [\Fig{MFIcolor}c]
notably fails to reach the
ground state with 
$p_0 \sim 0.398$.
This is consistent with the plateau observed in \Fig{MFI}a
for YZ-FQA. However, as seen with the energy resolution here,
by having considerable weight at low
energy, the energy distribution remains broad, overall.
Therefore the CD-FQA protocol does not drive
the state into an excited eigenstate. Instead, several
eigenstates of $H_P$ over a wider energy window conspire
to form an approximate steady state for the FQA protocol.
Despite the variations vs. circuit depth in the
energy distribution, the average energy in the system
barely changes. For example toward the largest times
(circuit depth), there are three bins
with pronounced weight ($p_0$, $p_2$, and $p_4$).
While $p_0$ gains weight, for $l\gtrsim 175$,
so does $p_4$, at the cost of the intermediate 
energy bin $p_2$. Therefore overall, the energy
expectation value remains nearly the same.

%

So far, we have illustrated the behavior of CD-FQA protocols for different second-control Hamiltonians while keeping all the other parameters fixed. In the following paragraphs we present the effect of CD-FQA protocols for different values of $\alpha$, systems size $N$, and the time-step $\Delta t$.

\subsubsection{CD-FQA for different values of $\alpha$}
\begin{figure}
    \centering
    \includegraphics[scale=0.11]{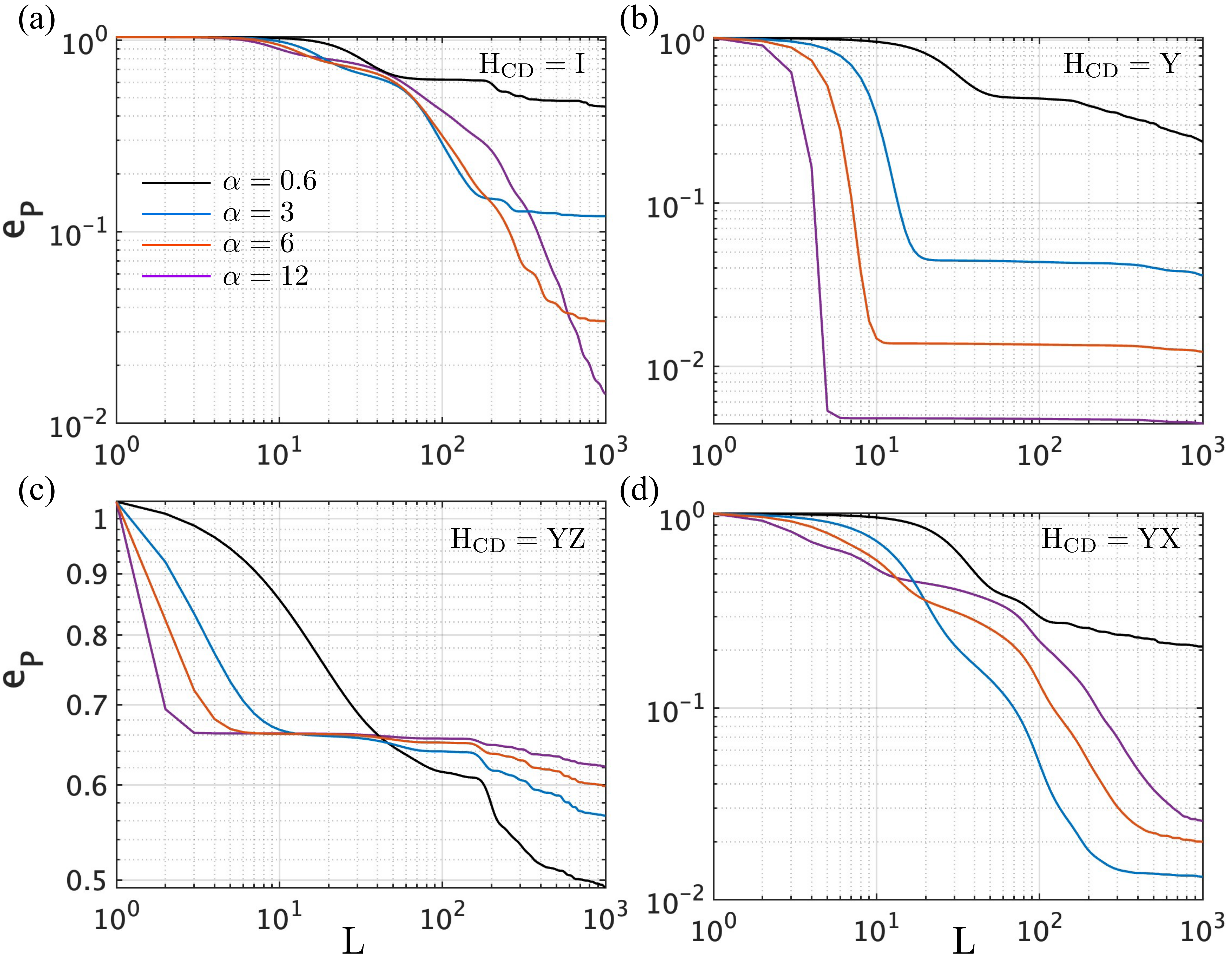}
    \caption{Average energy difference per site shown as a function of circuit depth in log-scale for the MFI model
    for 
    various $\alpha$ as specified in the legend
    in (a) for all panels, having $N=6$,
    $\Delta t = 0.01 / J$. 
    Each panel 
    corresponds to a different CD-FQA protocol
    for the $H_{\rm CD}$ as specified.
}\label{fig:alpha_final}
\end{figure}

The parameter $\alpha$ assumes a pivotal role in the CD-FQA protocol, as shown by the proportional relationship between the rate of change of average energy and $\alpha$ in Eq.~\ref{alpha}. The dependency of $\alpha$ on the protocol for arbitrary circuit depth, however, is non-trivial. Our findings are presented in Fig. \ref{fig:alpha_final}, where the average energy is plotted against circuit depth for four distinct values of $\alpha$.
In the context of standard FQA
[\Fig{alpha_final}a] 
a larger $\alpha$ induces a rapid reduction in energy for shallow circuit depths, while for large circuit depth the protocol with a smaller $\alpha$ demonstrates superior convergence. Notably, in the CD-FQA protocol with $Y$
[\Fig{alpha_final}b],
performance improves with increasing $\alpha$. However, beyond a certain value 
of $\alpha$, the average energy exhibits small oscillations, failing to decrease after a few layers. 
Similar behaviors are observed in CD-FQA with $YZ$ and $YX$
[\Figs{alpha_final}c and d, respectively].
The monotonic decrease of the energy
is guaranteed in \Eq{EP-dot} only for the differential
setting with infinitesimally small $\Delta t$.
Hence the onset of oscillations in the energy, where
the energy also intermittently increase, and is necessarily
due to the finite $\Delta t$ chosen.
\FIG{alpha_final} thus suggests an upper limit
$\alpha \, \Delta t \lesssim 0.1/J$, above which
\Eq{EP-dot} fails to decrease energy (see \Fig{deltat_final}
for a more detailed analysis still). This
underscores the importance of $\alpha$ 
together with $\Delta t$
in governing the performance dynamics of CD-FQA protocols, providing valuable considerations for the optimized CD-FQA protocol for a given quantum circuit depth.

\subsubsection{CD-FQA for different system sizes}
\begin{figure}
    \centering
    \includegraphics[scale=0.11]{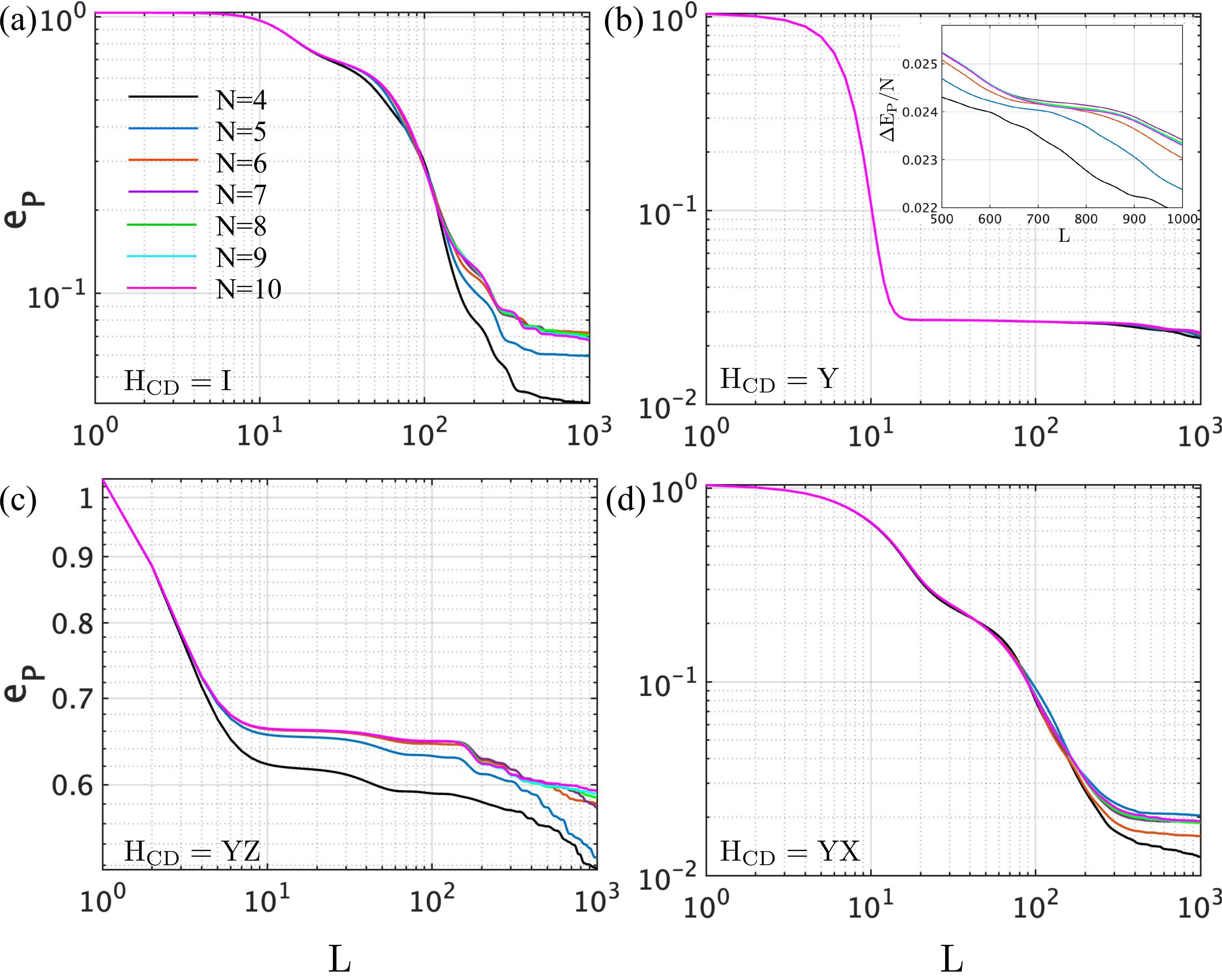}
    \caption{Average energy difference shown as a function of circuit depth in log-scale for the MFI model
    for 
    various values of $N$ as specified in the
    legend in (a) for all panels,
    having $\alpha=4$, $\Delta t=0.01/J$. 
    Each panel 
    corresponds to a different CD-FQA protocol
    for the $H_{\rm CD}$ as specified.
    The inset in (b) demonstrates the large-time behavior.
}\label{fig:N_final}
\end{figure}

In Fig.~\ref{fig:N_final}, we present a comprehensive analysis of energy reduction versus circuit depth across a range of system sizes ($N=4$ to $N=10$) using various CD-FQA protocols, maintaining a fixed $\alpha=4$. The results depicted in all four figures underscore the \textit{robustness} of the CD-FQA protocol, demonstrating its independence from system sizes up to $100$ layers where $L\Delta t\sim 1$. Deviations in the curves for $N=4$ and $N=5$ are attributed to the influence of small system size. A noteworthy comparison can be drawn with the findings in Ref.~\cite{FeedbackPRL}, where the authors establish a linear relationship between the number of layers and system sizes. It is crucial to highlight a key distinction: unlike the approach in Ref.~\cite{FeedbackPRL}, our methodology involves normalizing the prefactor, as illustrated in \Eq{alpha}. Here, the parameters $\beta$ and $\gamma$ are normalized by a factor of $N$. 
For large circuit depth, 
on the other hand,
we find across all panels in \Fig{N_final}
that the smaller system sizes show a somewhat
improved performance. This may be attributed
to the larger finite-size level spacing within
the excited states.

The observed  
independence of the circuit depth for the above case is due the finite correlation length of the system
bearing in mind that the system is gapped. This 
has the advantage that
one can use the results of small sizes as an insight to design the protocol for large sizes.
Another relevant
comparison can be drawn with the findings in Ref. \cite{yao2021reinforcement}, where the authors, employing a
Reinforcement Learning method, observed similar independence of the number of layers on system sizes in a QAOA-type
architecture. In their study, unitaries composed of the MFI
Hamiltonian, $X$, and $Y$ were strategically ordered using a
policy derived from Reinforcement Learning techniques. The signature that the number of layers in a CD-FQA circuit is
nearly independent of the system size for the MFI model highlights the practical usage of such a protocol for large-system sizes.


\subsubsection{CD-FQA for different time-steps $\Delta t$}

\begin{figure}
    \centering
    \includegraphics[scale=0.11]{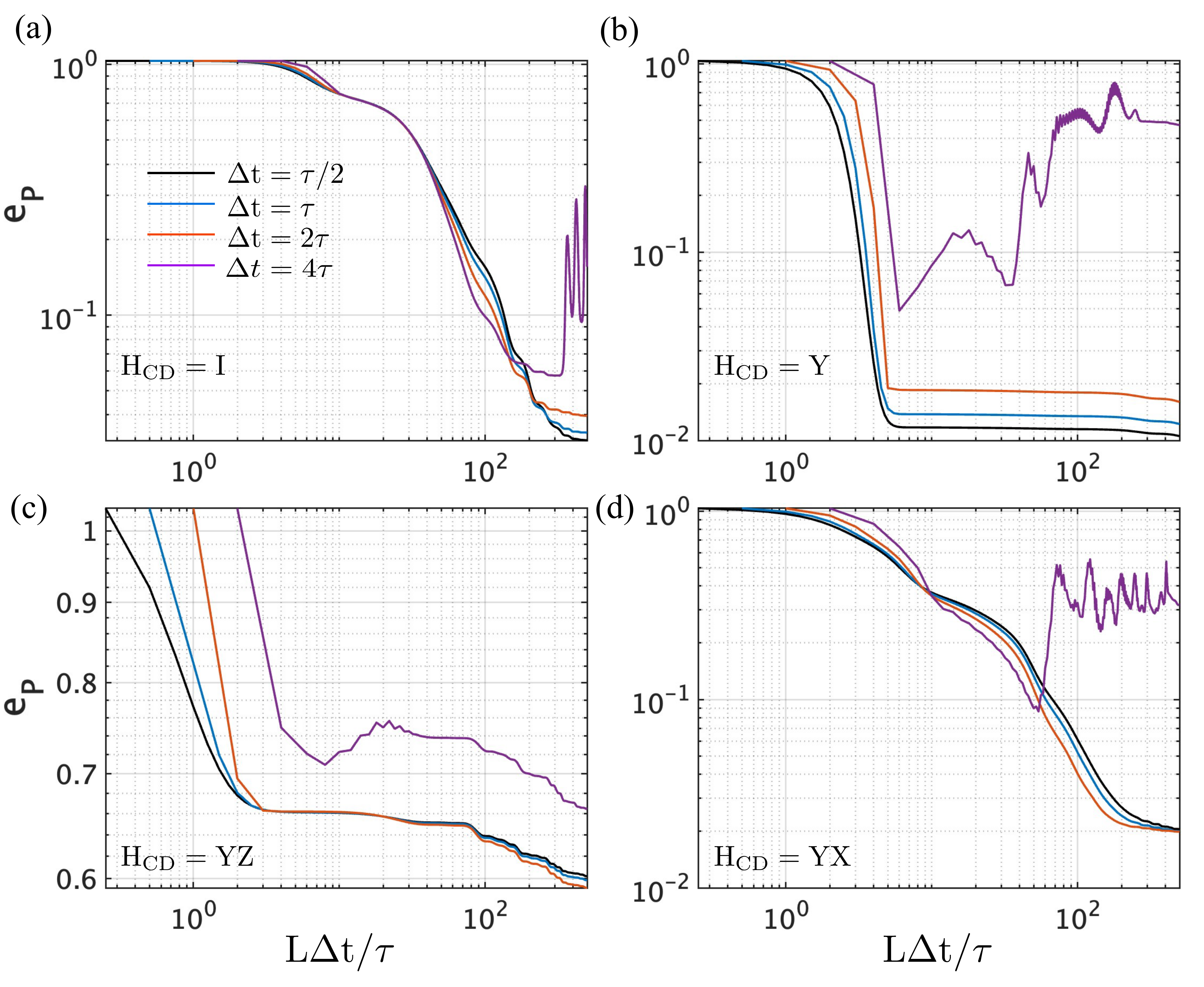}
    \caption{
    Average energy difference shown as a function of circuit depth in log-scale for the MFI model
    for 
    various $\Delta t$ relative to the constant
    $\tau = 0.01/J$ 
    used previously,
    with values specified in the legend of (a) for
    all panels, having $N=6$, and $\alpha=6$.
    Each panel 
    corresponds to a different CD-FQA protocol
    for the $H_{\rm CD}$ as specified.
    Panel (c) demonstrates the early plateau behavior.
}\label{fig:deltat_final}
\end{figure}

Finally, 
we also 
analyze the dependence 
of the CD-FQA protocols on 
the value of $\Delta t$.
Larger 
$\Delta t$ leads to a shallower 
circuit depth which is   
desirable 
for the implementation in a quantum circuit.
Too large a 
$\Delta t$, however, can lead to a deviation from the QLC protocol that results in an energy increase
in \Eq{EP-dot}. In \Fig{deltat_final} we present the average energy decay as a function of 
the total simulated time
$T\equiv  L\Delta t$ for four values of $\Delta t$. For
$\Delta t \gtrsim 2\tau$,
and therefore $\alpha \Delta t \gtrsim 0.1/J$,
the CD-FQA protocol starts to oscillate after a few layers and the average energy does not decrease with circuit depth. 
The limit on $\Delta t$ to describe the differential
setting in \Eq{EP-dot} is thus comparable
to what one may use in a Trotterized setting,
bearing in mind that $\alpha$ enters as a scale
factor to the full Hamiltonian in \Eq{Schrodinger}.
Nevertheless, since we are not interested in the
the trajectory of the prepared state per se, but only
in the final result, in principle, this opens the
possibility to use adaptive $\Delta t$ along
the circuit starting from larger values.
We leave this additional fine-tuning as an
outlook for future studies.
For the present paper, however,
we keep $\Delta t$ constant throughout the circuit.

\begin{figure}[tbh!]
\centering
\includegraphics[scale=0.185]{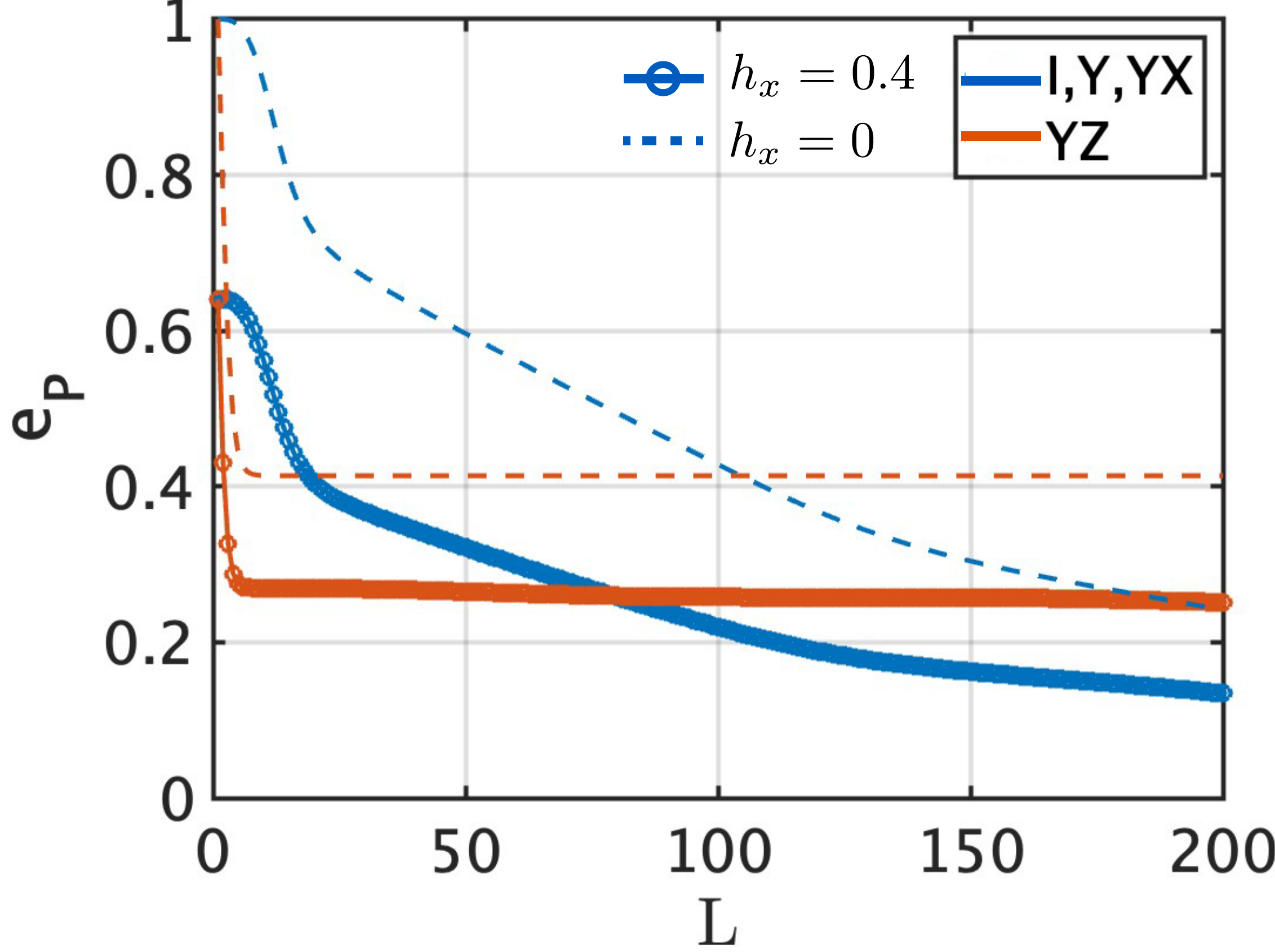}
\caption{
   Average energy 
   vs. circuit depth for TFI with  $h_x=0.4$ (solid lines)
   and $h_x=0$ (dashed lines). The standard FQA and CD-FQA with $H_{\rm CD}=Y$ and $YX$ yield the same result. 
}\label{fig:TFI}
\end{figure}

\begin{figure}
    \centering
    \includegraphics[scale=0.185]{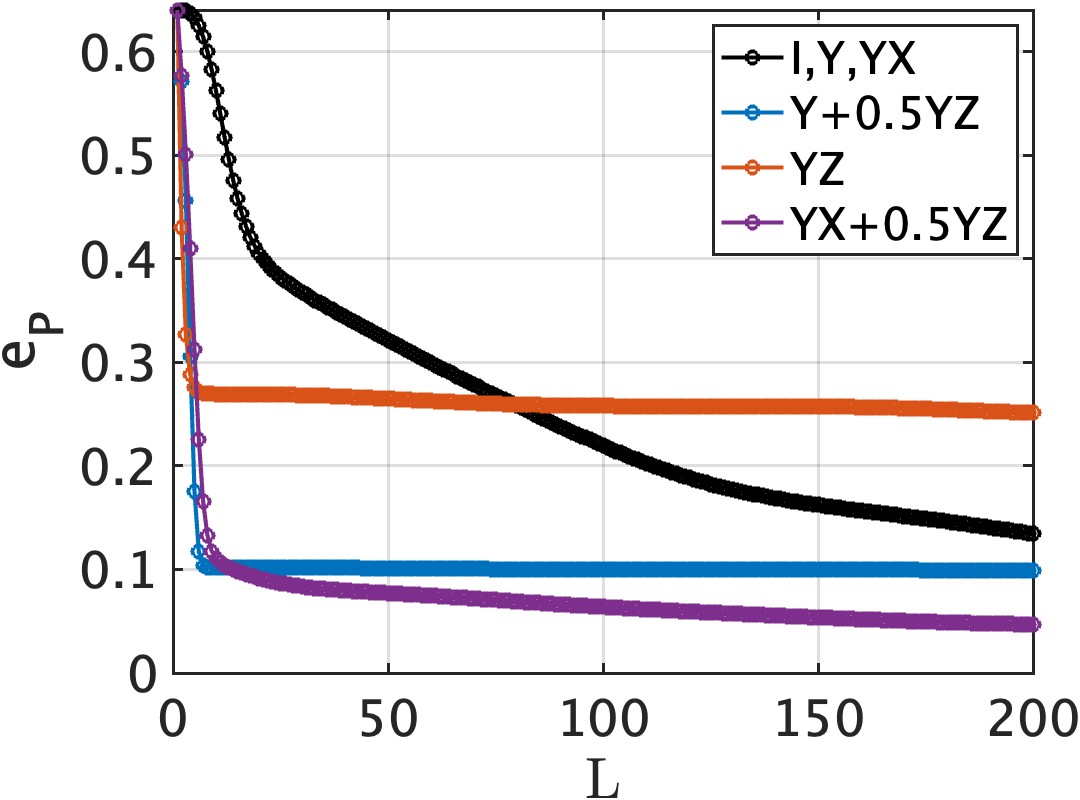}
    \caption{Average energy
   vs. circuit depth for TFI with  $h_x=0.4$ with CD-FQA operator is a linear combination of two operators from the pool.
   }
    \label{fig:Lcomb}
\end{figure}

\begin{figure}[tbh!]
\centering
\includegraphics[scale=0.185]{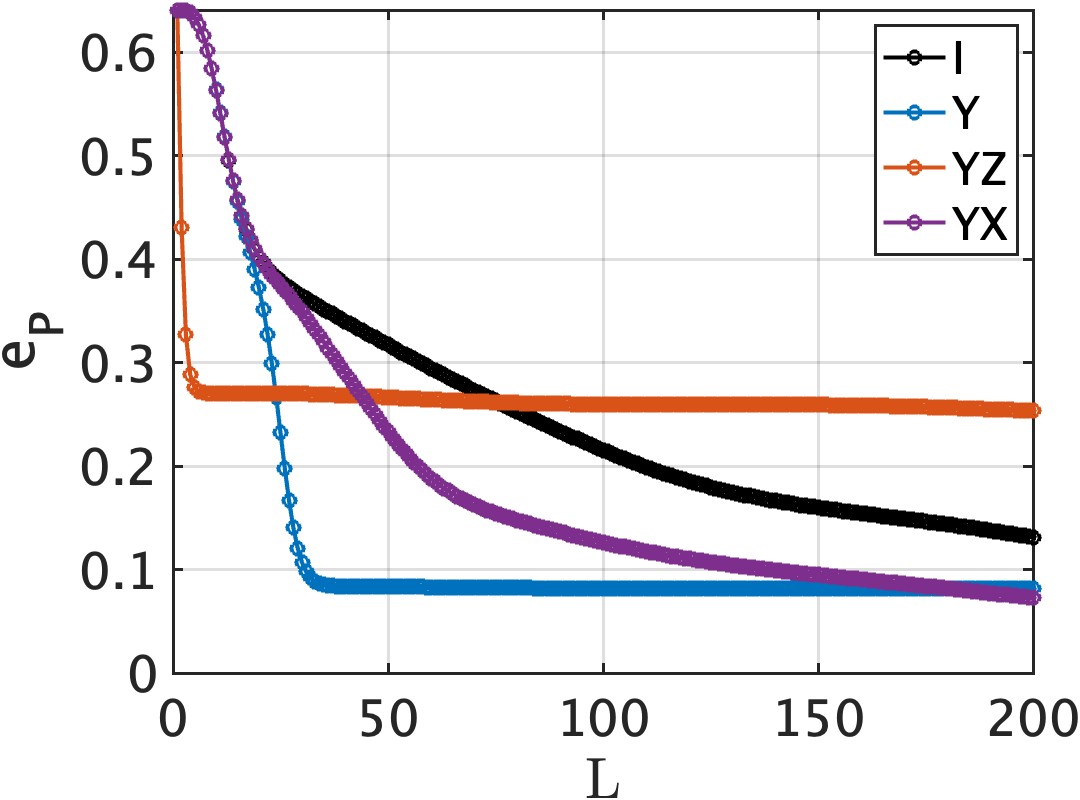}
\caption{ 
   Average energy 
   vs. circuit depth for the TFI at $h_x=0$,
   but now including the  term
   $H_{\rm add}(k) 
   =e^{-(k-1)/5} \, Z$ for layer $k$. 
}\label{fig:GHZpert}
\end{figure}



\subsection{TFI (including $h_x=0$)}
Now we apply the CD-FQA protocols to Ising chains with the longitudinal field turned off, i.e., $h_z=0$,
with the results presented in 
\Fig{TFI} for $h_x=0.4$, and  $h_x=0$.
The ground state of
the TFI is in the ferromagnetic phase and at $h_x=0$ there are
two degenerate ground states
$|\psi_{g1}\ra=|\uparrow\uparrow...\uparrow\ra$ and
$|\psi_{g2}\ra=|\downarrow\downarrow...\downarrow\ra$. 

The CD-FQA protocol employing the $H_{\rm CD}=\YZ$ operator
demonstrates a rapid energy reduction initially but becomes
exceedingly slow as we increase the number of layers, exhibiting
a plateau similar to the ones seen for LFI and MFI. This
protocol fails to reach the ground state, rendering it
undesirable. Notably, CD-FQA with $H_{\rm CD}=Y$ and \YX
mirrors the outcomes of the standard FQA, since $\gamma=0$ for all time steps.

In situations where the CD-FQA faces challenges in
convergence such as in \Fig{TFI} one may consider various approaches to improve the performance of the protocol. For instance, one may choose $H_{\rm CD}$ as a linear combination of operators from the pool of operators as the second control Hamiltonian, or add a small time-dependent Hamiltonian that vanishes at the end of the protocol.

\subsubsection{Improving CD-FQA with linear combination of operators}

To address the limitations of the \YZ operator in the TFI model,
we first introduce the counterdiabatic operator as a linear
combination of two operators from the operator pool. In
\Fig{Lcomb}, we plot the CD-FQA protocol with $H_{\rm CD}=Y+
\frac{1}{2} 
\YZ$ and $\YX + \frac{1}{2} 
\YZ$ together with $H_{\rm CD}=\YZ$ and
the standard FQA.  The combination of these imaginary operators
promotes a better mixing between instantaneous eigenstates
compared to individual operators from the pool. As depicted in
\Fig{Lcomb} the CD-FQA protocol with linear combination achieves
a rapid decrease in average energy, surpassing the energy
obtained with the \YZ term alone. Although employing a linear
combination of multiple counterdiabatic operators circumvents
the plateau observed with a single counterdiabatic operator, the
implementation on a quantum circuit requires more than a single
layer. A similar linear combination of operators have been utilized as a counterdiabatic term in Ref. \cite{Chen2022} in the digitalized counterdiabatic quantum optimization for MFI model. 


\subsubsection{Improving CD-FQA with an additional time-dependent Hamiltonian}

Another viable strategy to improve the CD-FQA is to introduce a time-dependent term
into the dynamics. It is crucial to configure the time
dependence of the additional term in a manner such that the magnitude 
of the term gradually diminishes throughout the protocol,
ultimately reaching zero upon the protocol's completion.
Furthermore, the additional term must commute
with the problem Hamiltonian so that the protocol does not
introduce additional measurement for $\beta$ and $\gamma$.

Let us illustrate this approach with a specific example,
considering the case where $H_P=-\ZZ$ and $H_1=X$. As previously
demonstrated, the counterdiabatic protocol with \YZ fails to
converge to the ground state, resulting in the average energy
plateauing at a value higher than the ground state. To address
this, 
we introduce a commuting time-dependent term,
denoted as $H_{\text{add}}=f(t_{k})Z$, where $t_k$ corresponds
to the time at the $k^{\text{th}}$ layer. The modified total
time-dependent Hamiltonian is now expressed as
$H(t)=H_{P}+H_{\text{add}}(t)+\beta(t)H_1+\gamma(t)H_{\rm CD}.$
The QLC protocol continues to be determined by the condition
$d\la H_P \ra/dt \leq 0$. However, the introduction of
$H_{\text{add}}(t)$ alters the effectiveness of counterdiabatic
operators, allowing the utilization of $Y$ as $H_{\rm CD}$ for
constructing the CD-FQA protocol. In Fig. \ref{fig:GHZpert}, we
present the average energy as a function of the number of
layers, incorporating the additional term and selecting
$H_{\rm CD}=Y$. The CD-FQA protocol with the additional term
exhibits better performance compared to the protocol with the
\YZ operator.

It is important to note that the introduction of the $Z$ term
breaks the degeneracy of the ground state of $H_{P}=-\ZZ$ as well as some of the excited states. 
Depending on the sign of the $Z$, the CD-FQA converges
to a state closer to one of the degenerate states within the
ground state manifold. Indeed, in our protocol, if it were not for the
$Z$, the state would be forced to flow to the GHZ
state, which is known to have a linear circuit
complexity~\cite{bravyi2006lieb, yu2023learning}, since both the initial state and the unitaries in CD-FQA are symmetric under $\prod_j X_j$.  

To see this, recall that we start with an initial state  $|{+}X\rangle$, and both $H_P$ and $H_1$ commute with $\prod_j X_j$. Operators from nested commutators (i.e., the \YZ operator) also commute with $\prod_j X_j$. Since we minimize the energy of $H_P=\ZZ$, when we are restricted to the $\prod_j X_j=1$ subspace, we are forced to flow to GHZ. An additional $Z$ term in the Hamiltonian biases the system towards either $|{+}Z\rangle$ or $|{-}Z\rangle$ and breaks away from the symmetry subspace. This addition of $Z$ further allows us to utilize $Y$ as a second control Hamiltonian. So one possible strategy to expedite conversion to a ground state is to select a
symmetry-breaking operator as the additional term when an obstruction due to long-range order is expected.


Both strategies  employed above highlight the importance of alternate approaches to tweak the CD-FQA protocol to navigate
the quantum landscape effectively. However, one must take into
account that the introduction of an additional Hamiltonian
always increases the number of gates and thereby the depth of
the quantum circuit. 

Furthermore, in Appendix A, we extend our investigation for the TFI by exploring an alternative first control Hamiltonian, specifically $H_1=Z$, and the corresponding initial state $|\psi_0\ra=|\uparrow \uparrow ...\uparrow\ra$. The CD-FQA under this configuration exhibits superior performance compared to the scenario shown in Fig. \ref{fig:TFI}. All three counterdiabatic operators demonstrate better performance in achieving ground-state compared to the standard FQA Consequently, it is evident that the choice of $H_1$ significantly impacts the CD-FQA's performance in the context of the TFI. In this comparison, $Z$ emerges as a more favorable first control Hamiltonian than $X$ for the TFI, highlighting the importance of the specific control Hamiltonian selection in the CD-FQA.

\section{Demonstration on cloud quantum computers}
\label{sec:experiment}


\begin{figure}
\centering
\includegraphics[scale=.2]{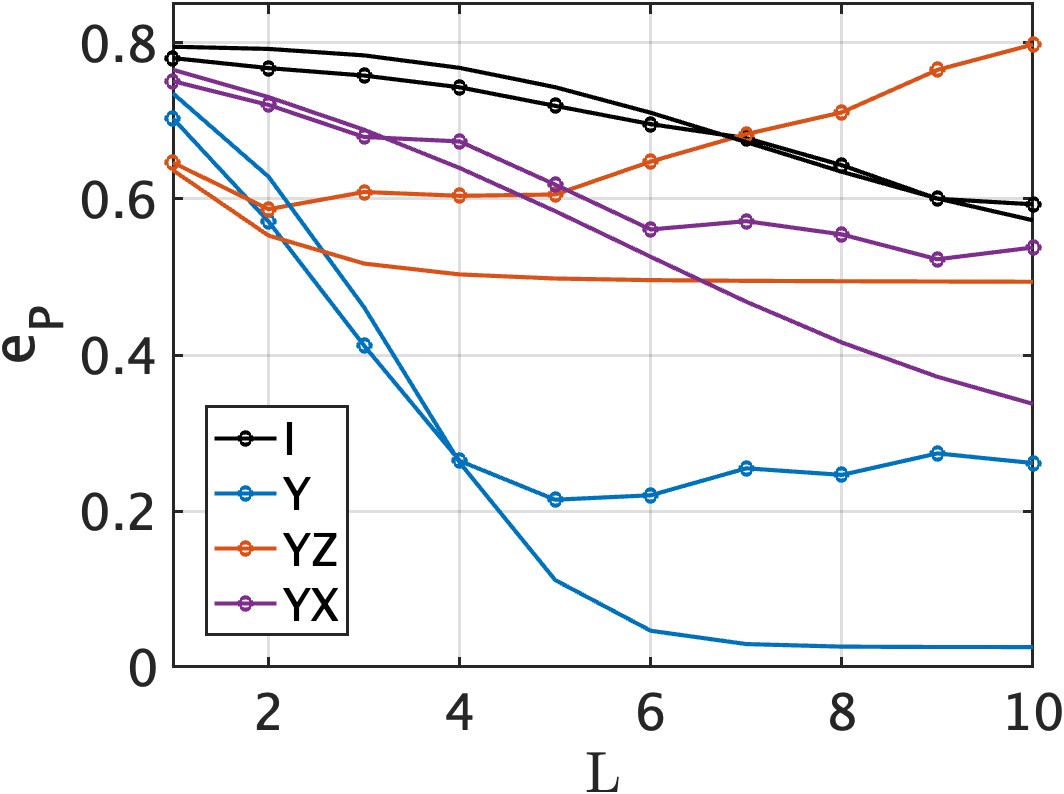}
\caption{
    The average energy is shown as a function of the number of
    layers for 4-qubit MFI with $h_z=h_x=0.4$ and $\Delta t = 0.02$.
    The simulation is performed on ibm\_torino with 28\,192
    repetitions for each measurement. The curves with markers
    represent data from quantum computers while the solid lines represent
    classical simulations.
}\label{fig:experiment}
\end{figure}

\begin{figure}[h!]
    \centering
    \includegraphics[scale=0.2]{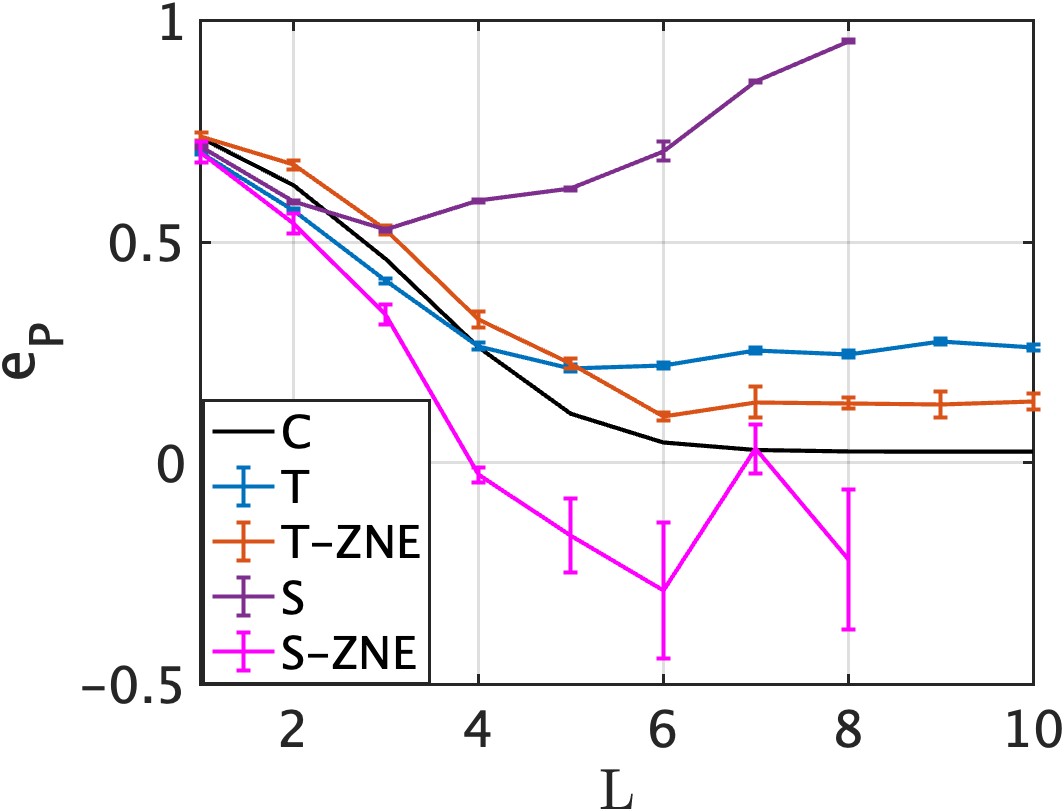}
\caption{
   The experimental results from IBM\_Torino (T), IBM\_Sherbrooke (S), and IBM\_Sherbrooke with zero-noise extrapolations (S-ZNE) is compared with the classical simulation (C) for the CD-FQA with Y as the CD-operator. The average gate error per layer for IBM\_Torino is 0.8\%, and 1.7\% for IBM\_Sherbrooke,thus resulting in the deviations observed for $L=1$.
   All other parameters are same as in Fig.~\ref{fig:experiment}. The deviations of the energies for the four curves at $L=1$ are from below 1\% to around 5\%, with T-ZNE having the best accuracy and S the worst.}

\label{fig:exprtZNE}
\end{figure}

%

To showcase the enhanced performance of CD-FQA over FQA on an
actual quantum platform, we conducted demonstrations
on IBM's
superconducting quantum computer through cloud-based
simulations, employing CD-FQA for the MFI with a four-spin system with OBC for parameters $h_x=h_z=0.4$. As we set a small time step interval $\Delta t=0.02$, we find that one-layer Trotterization (e.g. $e^{i (ZZ+X)}\approx e^{i ZZ}e^{iX}$) suffices for an accurate simulation. The experiment is done recursively. We use the $k$-step circuit to measure the commutators (i.e. expectation values of each Pauli term in the commutators), and get $\beta_{k+1},\gamma_{k+1}$ for the circuit of next step.  The outcomes,
illustrated in Fig. \ref{fig:experiment}, reveal a consistent
monotonic decay of energy in FQA and YX-FQA up to the
$9^{\text{th}}$ layer. Conversely, in the $Y$ case, we observed a
decrease only up to the $5^{\text{th}}$ layer, and \YZ failed to
exhibit any decay even in the second layer.

Considering the impact of gate noise inherent in quantum
circuits, deploying a substantial number of 2-qubit gates
 resulted in undesirable outcomes. Therefore, in conventional FQA 
and CD-FQA utilizing the Y operator, where local operators are
employed as $H_1$ and $H_{\rm CD}$, we found such circuits demonstrated a decrease in energy performance up to the $6^{\text{th}}$ layer. Moreover, the CD-FQA with $Y$ performs the best.
The state generated after the application of numerous layers
exhibited an undesirable growth in energy, a phenomenon
unsuitable for any protocols. The influence of noise became
evident in measurements, where all Pauli operator expectation
values approached zero. Nevertheless, for a limited number of
layers, we observed an enhancement of CD-FQA over conventional
FQA in terms of both energy accuracy and convergence speed,
particularly in the case of Y-FQA. Below we discuss the effect of gate noise and methods to improve experimental results. The discussion on statistical errors that may appear due to the measurements is discussed in the Appendix C.



\subsection{{Improving results from IBM's cloud quantum computers}} 

The experimental data in \Fig{experiment}
deviates from the classical simulations.  The primary
reason for such discrepancy stems from the limitations of the
current NISQ devices. 
Gate errors accumulate 
{exponentially with circuit depth $L$}. 
In Fig.~\ref{fig:exprtZNE} we {compare} 
the quantum simulations performed on 
IBM\_Sherbrooke with 
{the newer} IBM\_Torino {system}. 
The energy curve obtained from the IBM\_Sherbrooke
for $Y$-FQA deviates significantly from the theoretical curve
already for $L>3$. To improve this discrepancy one can use
either a less noisy machine, or error mitigation schemes such as zero-noise
extrapolation (ZNE) \cite{temme2017error,giurgica2020digital}.
As shown in Fig.~\ref{fig:exprtZNE}, curves obtained from
both machines get closer to the theoretical one after applying
ZNE. The average gate
error per layer for IBM\_Sherbrooke is around 1.7\%, while the
rate for IBM\_Torino is 0.8\%. Thus the data 
from IBM\_Torino {is more accurate and} 
consistent with the theoretical curve up to $L=5$. In the
ZNE 
approach of Ref. \cite{giurgica2020digital},
the circuit, created by the overall unitary $U$, is `folded' by
looking at a sequence of equivalent unitaries,
$UU^\dagger U$, $UU^\dagger UU^\dagger U$, $\ldots
= U (U^\dagger U)^n$.
Since the noise gets amplified with increasing $n$,
this permits an extrapolation to a zero-noise limit.
While this 
can achieve better accuracy {for} 
expectation values, 
at the same time {it may also} 
increase the uncertainty of the results,
as seen 
in Fig.~\ref{fig:exprtZNE}.

\section{Discussion}
\label{sec:discussion}


An important aspect of 
CD-FQA lies in the strategic selection of the
additional 
control Hamiltonian, 
$H_{\rm CD}$,
within the given context of the provided $H_P$ and $H_1$.
This choice significantly influences the evolution of the control
field $\gamma(t)\propto \la [H_P,H_{\rm CD}] \ra$  
which depends on 
both, the commutator and the state $\psi(t)$. 
In our investigation with $H_1=X$,
the initial state is chosen as the ground state
of $-H_1$, 
$|\psi_{0}\rangle = |X\rangle \equiv |\rightarrow\rightarrow \ldots \rightarrow\rangle $.
As this extremizes $\la X \ra$,
a strategy to enhance the energy
reduction via $\gamma \sim \langle [H_P, H_{\rm CD}] \rangle$
at early times is to look for operators $H_{\rm CD}$ that yield
$ [H_P, H_{\rm CD}] \sim X$. Given our initial state, 
therefore the commutators resulting in $X$ or \XX lead
to the largest $\gamma$'s at early times. This leaves only
a few choices for $H_{\rm CD}$:
\begin{subequations}\label{eq:HCD:dominant} 
\begin{eqnarray}
   H_{\rm CD} &=& Y                   \label{eq:HCD:Y}  \\
   H_{\rm CD} &=& \YX \text{ or } \XY \label{eq:HCD:YX} \\
   H_{\rm CD} &=& \YZ \text{ or } \ZY \label{eq:HCD:YZ}
\end{eqnarray}
\end{subequations}
since for \Eq{eq:HCD:Y}, $[(H_P \to Z),Y] \sim X$, 
for \Eq{eq:HCD:YX}, e.g., $[(H_P \to Z), \YX] \sim -\XX + \YY$.
For these to occur, this also shows the importance of the
longitudinal term $Z$ in $H_P$ to be present. 
The last option in \Eq{eq:HCD:YZ} arises since, e.g., 
$[(H_P \to \ZZ), \YZ] \sim - \mathit{XZZ} - \mathit{ZXZ}$
where the diagonal terms in \ZZ lead to $X$.
In all cases, $H_{\rm CD}$ needs to include $Y$.
This makes intuitive sense, since $Y$ is required to rotate
the initial state $|X\rangle$ to $|Z\rangle$ which is close
to the Ising ground state, exactly so for $h_x=0$.


In the presence of the $Z$ term in the Hamiltonian
(LFI and MFI) 
all three counterdiabatic operators in \Eqs{eq:HCD:dominant}
exhibit a rapid reduction in average energy values over a small
number of layers.
As seen in \Figs{LFI} and \ref{fig:MFI},
the best performance for long times is given by $a>b>c$
[based on the subequation numbering in \eqref{eq:HCD:dominant}].
Notably, the one with the worst long-term performance ($c$) 
demonstrates the fastest energy reduction at early times.
This shows that focusing solely on the most rapid energy reduction in the choice of $H_{CD}$
can drive the system into barren plateaus in terms of quasi-steady
states at elevated energy [cf. \Fig{MFIcolor}c].

In stark contrast, in 
the absence of the $Z$ term in the Hamiltonian (TFI,
 including $h_x=0$), only the option in \Eq{eq:HCD:YZ} remains in order to effectively reduce the energy, as also seen in \Fig{TFI}. 
In this case, using \Eq{eq:HCD:Y} has no effect
whatsoever, with the data identical to plain FQA.
Similar to the LFI case, however, the \YZ operator
is prone to getting stuck at finite energy.
As seen in \Fig{TFI},
adding \YX, while irrelevant at early times,
nevertheless does permit to drive the system to
lower energy after the initial stage.

We have also addressed variations of the CD-FQA protocol to overcome the early plateaus seen in \Fig{TFI}. The two approaches involve incorporating a linear combination of two operators from (\ref{eq:HCD:dominant}) or adding a time-dependent term to the $H_{P}$ that diminishes over time. Such variations yield excellent results for the TFI.


 Feedback-based quantum algorithms require a deep circuit with many layers and are beyond the scope of the current NISQ devices. However, a shallow FQA circuit can be utilized to ``warm start" a QAOA-type quantum circuit, i.e., to use the FQA parameters $\beta_{k}$ to initialize the QAOA algorithm. Note that, each layer in QAOA is parametrized by two real numbers, and one of the parameter is $1$ and the second parameter is approximated by $\beta_{k}$. 
 We observe that a similar extension is also possible for the CD-FQA, where one can utilize the parameters of CD-FQA circuit to warm start a VQA, where each layer is parameterized by three layers. This translates to a digitalized-counterdiabtic inspired QAOA \cite{chandarana2022digitized} where the initial parameters are selected from the CD-FQA.    

\section{Conclusion}
\label{sec:conclusion}

In this study, we have extended the FQA by incorporating the
principles of QLC  with the counterdiabatic driving protocol.
Departing from the conventional use of a single control field,
we propose the integration of a second control field inspired by
the counterdiabatic driving protocol. This modification proves
instrumental in accelerating ground state preparation,
showcasing its effectiveness for implementation on digital
quantum circuits. The CD-FQA has been systematically applied to
four distinct variants of the Ising models. Our results showcase
the intricate interplay between various parameters involved in
the CD-FQA and how a proper selection of the second control
Hamiltonian depends on the problem Hamiltonian, first control
Hamiltonian, and the initial state for fast preparation of the
ground state. On the one hand, 
the introduction of a second
control field contributes to an increase in complexity within a single layer. On the other hand, this 
effectively reduces the overall circuit complexity.  
These two system-dependent aspects need to be
balanced for optimal performance.
The parameters of the presented FQA can serve
as an initial seed for further classical optimization.
This opens the
potential for extending the CD-FQA parameters to set up QAOA
circuits, thus allowing one to combine 
quantum and classical
optimization methodologies. This is particularly noteworthy, as
previous research has successfully demonstrated a three-unitary
QAOA utilizing a similar operator pool
\cite{chandarana2022digitized}. Beyond its applications in
quantum optimization algorithms, the CD-FQA unveils novel
possibilities in the realm of fast quantum control methods
employing counterdiabatic driving protocols. This convergence of quantum algorithmic advancements and control theory holds
promise for shaping the future landscape of quantum computing.
In conclusion, our work not only enhances the understanding of
ground state preparation in quantum many-body systems but also
adds insights for quantum control strategies with far-reaching
implications in the field. 

{\it Data availability:} 
The source code is open-source, readily available online, and can be easily modified to address similar problems \cite{malla2024cdfqa}. 

\section{Acknowledgments}

The authors thank Ning Bao and Ananda Roy for useful discussions. All authors of this work were supported in its production by the U.S. Department of Energy, Office of Basic Energy Sciences, under Contract No. DE-SC0012704.
TCW acknowledges the support of an SBU Presidential Innovation and Excellence (PIE)
Fund. This research also used resources from the Oak Ridge Leadership Computing Facility, which is a DOE Office of Science User Facility supported under Contract DE-AC05-00OR22725, and the Brookhaven National Laboratory operated IBM-Q Hub. The results presented in this work do not reflect the views of IBM and its employees.

\appendix

\begin{figure}[b!]
    \centering
    \includegraphics[width=0.85\linewidth]{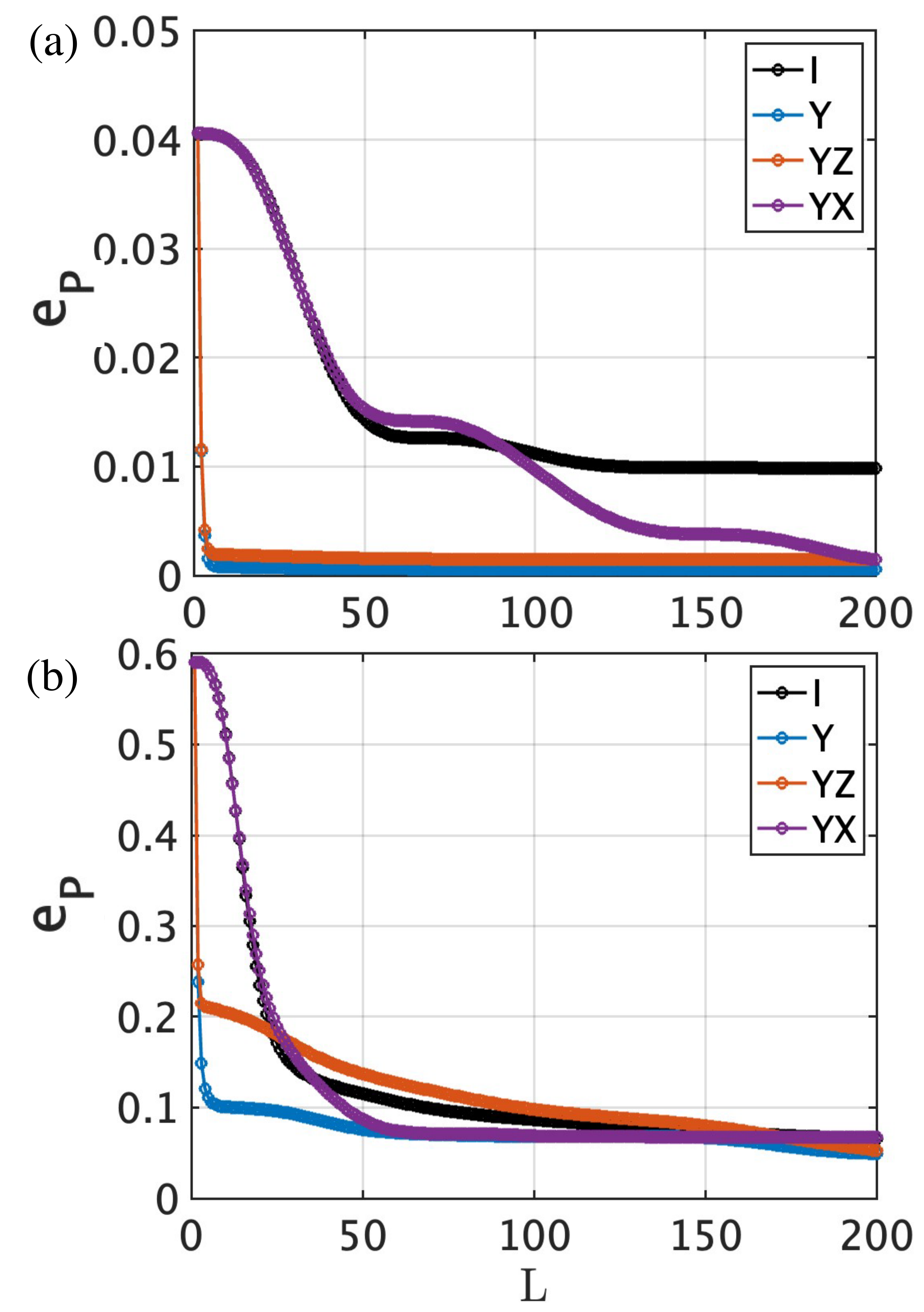} 
    \caption{The CD-FQA protocol for the TFI with alternate first control Hamiltonian $H_1=Z$ 
    for $h_x$ = (a) $0.4$, (b) $1.4$. The initial state is the ground state of $Z$.}
    \label{fig:TFIwithZ}
\end{figure}

\begin{figure}[tb]
    \centering
    \includegraphics[width=0.85\linewidth]{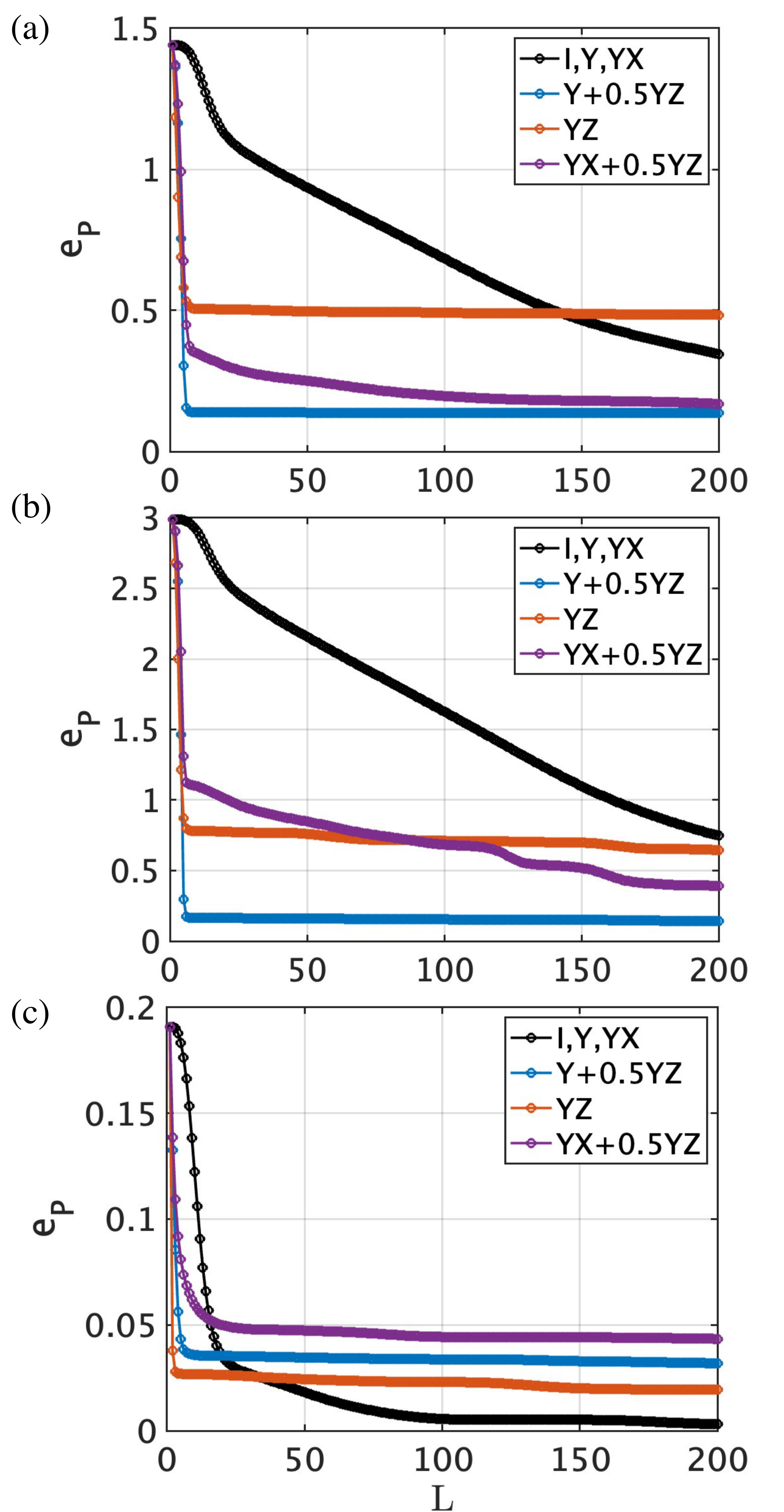} 
\caption{
   The CD-FQA protocol for additional cases for the TFI with
   $h_x$ = (a) $-0.4$, (b) $-1.4$,  (c) $1.4$.
}\label{fig:TFIappendix1}
\end{figure}

\begin{figure}[h!]
    \centering
    \includegraphics[width=0.8\linewidth]{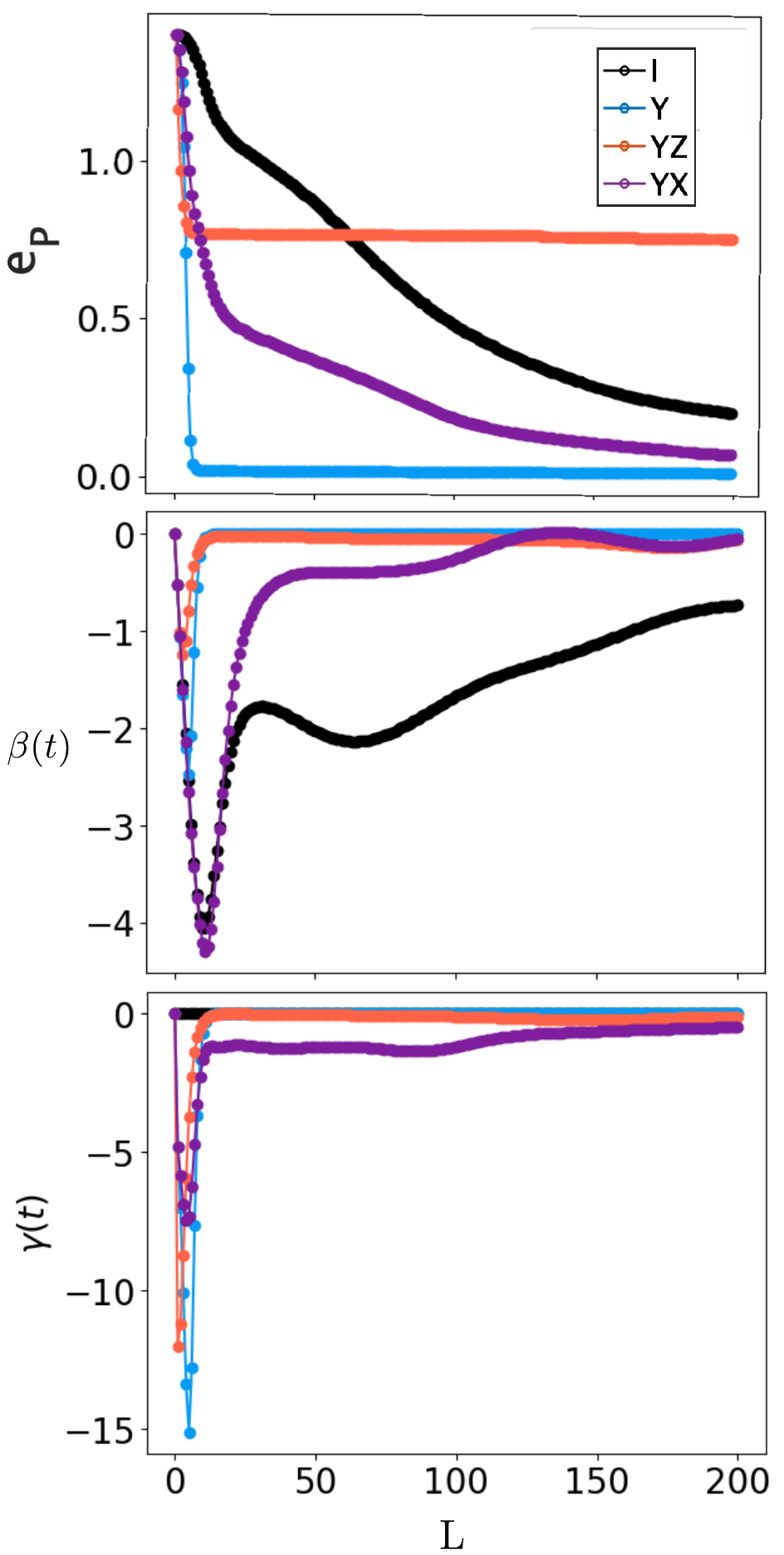} 
    \caption{FQA simulation results for LFI with $N=6$, $\Delta t = 0.01$, $\alpha=6$, and parameters $\{h_z,h_x\} = \{0.4,0.0\}$. The simulation was performed with PennyLane~\cite{bergholm2018pennylane}, where each unitary evolution is implemented with the first-order Trotter approximation. The result is consistent with the quasi-exact simulation without Trotterization presented in Fig.~\ref{fig:LFI}.}
    \label{fig:PLLFI}
\end{figure}

\section{CD-FQA for the TFI: additional plots}

In Fig.~\ref{fig:TFIwithZ}, we apply the CD-FQA protocol to
two TFI models where $H_1=Z$ serves as the first control
Hamiltonian, coupled with the initial state
$|\psi_0\ra=|\uparrow \uparrow ...\uparrow\ra$. Unlike the
results observed in Fig. \ref{fig:TFI}, where the CD-FQA with
$Y$ and \YX aligns with the standard FQA, the CD-FQA protocols
in Fig. \ref{fig:TFIwithZ} yield different curves for the
average energy. The contrast between the two figures underscores
the significant impact of the first control Hamiltonian on the
overall performance of the algorithm. In the context of the TFI
model, the choice of $Z$ as the first control Hamiltonian tends
to be more effective than opting for $X$.

We also introduce additional plots that extend the
analysis of the TFI model with variations in parameters. Fig.~\ref{fig:TFIappendix1} provides insights into the performance of
the CD-FQA across three TFI models characterized by distinct
parameter sets, $h_x=-0.4$, $-1.4$, and $1.4$. For the case of
$h_x=-0.4$, the ground state is in a antiferromagnetic phase.
The performance of CD-FQA protocol is similar to that shown in
Fig. \ref{fig:TFI} in the main text. When the $|h_x|>1$, the
spins in the ground state are more likely to be aligned along
x-axis. Therefore, the \YX operator does not play any
significant role compared to the \YZ operator, see Figs.~\ref{fig:TFIappendix1}b,c. Finally,  when $h_x>1$, the ground
state is close to the initial state and therefore we see a fast
convergence using the standard FQA.

\begin{figure}[t!]
    \centering
    \includegraphics[width=0.85\linewidth]{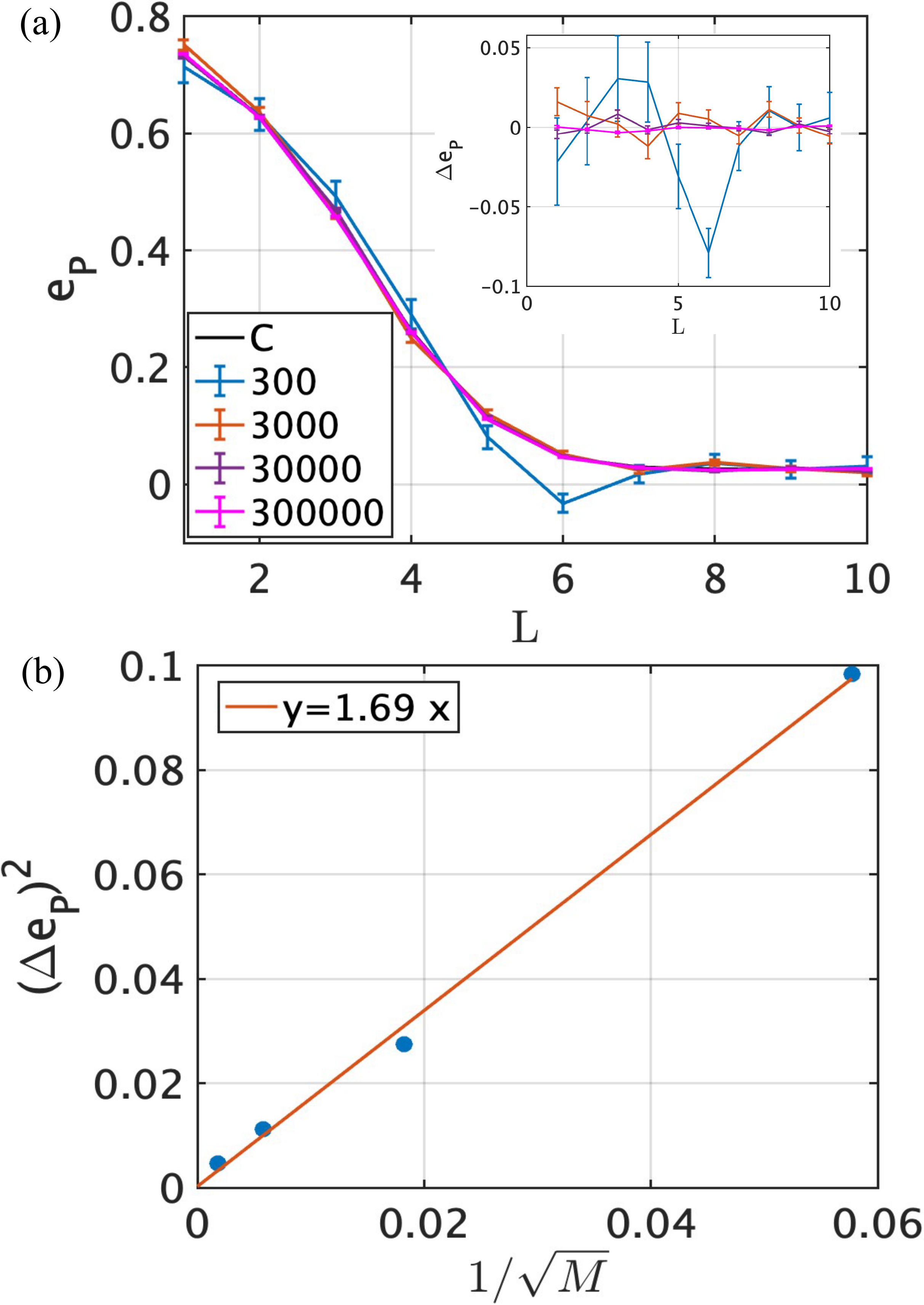} 
    \caption{(a) The average energy vs the circuit depth is plotted for CD-FQA with $Y$ as the CD operator for different number of measurement samples, $M$. Here, ``C" refers to the classical simulation where expectation values are calculated directly. The inset shows the deviation of energy from the classical simulation result. (b) The standard deviations for different sample sizes are shown and fitted linearly with $1/\sqrt{M}$ (red line). The standard deviation is obtained from the energy deviations across all layers.
}\label{fig:staterror}
\end{figure}

\section{Simulation on PennyLane} 

In the main text, simulation results were presented for
unitaries without Trotterization. However, in practical quantum
devices, it is common practice to execute unitary (time)
evolutions using the Trotter approximation.
In Fig.~\ref{fig:PLLFI}, 
we showcase a classical simulation conducted through the
application of Pennylane tools in the first-order Trotter
approximation. 
For
demonstration purposes, we choose the LFI model, and the plots
affirm the consistency of the curves with the results previously
depicted in Fig.~\ref{fig:LFI}.

\section{Effects of statistical errors}

The experimental results are affected by two sources of
error: 
gate errors, and statistical errors 
{when evaluating expectation values}.
We show in \Fig{staterror} 
that the dynamics does not accumulate statistical errors
with circuit depth $L$, and thus can easily be made negligible
compared to the gate errors by increasing the repetition number
for each measurement. The reason for the non-accumulation is
that, if we obtain $\beta$'s and $\gamma$'s with some
uncertainties, it effectively changes the parameter $\alpha$
 {as in \Eq{eq:params:k+1}} slightly for each layer. 
When the uncertainties are small, we
will obtain another parameter trajectory for $\beta$'s and
$\gamma$'s close to the exact ones, {typically} 
leading to the same converged energies. Thus the
statistical error will not accumulate during the feedback-based
protocol. The overall uncertainty due to 
statistical errors is proportional to $\frac{1}{\sqrt{M}}$, where $M$ is
the number of repetitions for each measurement. To showcase the
argument, we did ``dry-run" simulations of our protocol, where
we performed repetitive measurements on a noiseless classical
simulator for quantum circuits to get expectation values,
compared to the results obtained from directly calculating the
expectation values of states. The results 
{in \Fig{staterror}(b)} 
demonstrate that the statistical errors
do not accumulate and converge to the same energy values.

\bibliography{bib}

\end{document}

%% file: shortcuts.tex
  \newcommand{\Sec}[1]{Sec.\,\ref{#1}}
  \newcommand{\SEC}[1]{Section\,\ref{#1}}

  \newcommand{\EQ}[1]{Equation\,\eqref{#1}}
  \newcommand{\Eq}[1]{Eq.\,\eqref{#1}}
  \newcommand{\Eqs}[1]{Eqs.\,\eqref{#1}}

  \newcommand{\FIG}[1]{Figure\,\ref{fig:#1}}
  \newcommand{\Fig}[1]{Fig.\,\ref{fig:#1}}
  \newcommand{\Figs}[1]{Figs.\,\ref{fig:#1}}

  \newcommand{\rsym}[1]{\ensuremath{%
  \mathfrak{r}_{\mathrm{\ifx\@empty{}\else{#1}\fi}}}}

  \definecolor{cEdit}{rgb}{.17,.44,.76}

%% file: main.bbl
\begin{thebibliography}{65}%
\makeatletter
\providecommand \@ifxundefined [1]{%
 \@ifx{#1\undefined}
}%
\providecommand \@ifnum [1]{%
 \ifnum #1\expandafter \@firstoftwo
 \else \expandafter \@secondoftwo
 \fi
}%
\providecommand \@ifx [1]{%
 \ifx #1\expandafter \@firstoftwo
 \else \expandafter \@secondoftwo
 \fi
}%
\providecommand \natexlab [1]{#1}%
\providecommand \enquote  [1]{``#1''}%
\providecommand \bibnamefont  [1]{#1}%
\providecommand \bibfnamefont [1]{#1}%
\providecommand \citenamefont [1]{#1}%
\providecommand \href@noop [0]{\@secondoftwo}%
\providecommand \href [0]{\begingroup \@sanitize@url \@href}%
\providecommand \@href[1]{\@@startlink{#1}\@@href}%
\providecommand \@@href[1]{\endgroup#1\@@endlink}%
\providecommand \@sanitize@url [0]{\catcode `\\12\catcode `\$12\catcode
  `\&12\catcode `\#12\catcode `\^12\catcode `\_12\catcode `\%12\relax}%
\providecommand \@@startlink[1]{}%
\providecommand \@@endlink[0]{}%
\providecommand \url  [0]{\begingroup\@sanitize@url \@url }%
\providecommand \@url [1]{\endgroup\@href {#1}{\urlprefix }}%
\providecommand \urlprefix  [0]{URL }%
\providecommand \Eprint [0]{\href }%
\providecommand \doibase [0]{http://dx.doi.org/}%
\providecommand \selectlanguage [0]{\@gobble}%
\providecommand \bibinfo  [0]{\@secondoftwo}%
\providecommand \bibfield  [0]{\@secondoftwo}%
\providecommand \translation [1]{[#1]}%
\providecommand \BibitemOpen [0]{}%
\providecommand \bibitemStop [0]{}%
\providecommand \bibitemNoStop [0]{.\EOS\space}%
\providecommand \EOS [0]{\spacefactor3000\relax}%
\providecommand \BibitemShut  [1]{\csname bibitem#1\endcsname}%
\let\auto@bib@innerbib\@empty
\bibitem [{\citenamefont {Poulin}\ and\ \citenamefont
  {Wocjan}(2009)}]{Wocjan2009}%
  \BibitemOpen
  \bibfield  {author} {\bibinfo {author} {\bibfnamefont {David}\ \bibnamefont
  {Poulin}}\ and\ \bibinfo {author} {\bibfnamefont {Pawel}\ \bibnamefont
  {Wocjan}},\ }\bibfield  {title} {\enquote {\bibinfo {title} {Preparing ground
  states of quantum many-body systems on a quantum computer},}\ }\href
  {\doibase 10.1103/PhysRevLett.102.130503} {\bibfield  {journal} {\bibinfo
  {journal} {Phys. Rev. Lett.}\ }\textbf {\bibinfo {volume} {102}},\ \bibinfo
  {pages} {130503} (\bibinfo {year} {2009})}\BibitemShut {NoStop}%
\bibitem [{\citenamefont {Tubman}\ \emph {et~al.}(2018)\citenamefont {Tubman},
  \citenamefont {Mejuto-Zaera}, \citenamefont {Epstein}, \citenamefont {Hait},
  \citenamefont {Levine}, \citenamefont {Huggins}, \citenamefont {Jiang},
  \citenamefont {McClean}, \citenamefont {Babbush}, \citenamefont
  {Head-Gordon},\ and\ \citenamefont {Whaley}}]{tubman2018postponing}%
  \BibitemOpen
  \bibfield  {author} {\bibinfo {author} {\bibfnamefont {Norm~M.}\ \bibnamefont
  {Tubman}}, \bibinfo {author} {\bibfnamefont {Carlos}\ \bibnamefont
  {Mejuto-Zaera}}, \bibinfo {author} {\bibfnamefont {Jeffrey~M.}\ \bibnamefont
  {Epstein}}, \bibinfo {author} {\bibfnamefont {Diptarka}\ \bibnamefont
  {Hait}}, \bibinfo {author} {\bibfnamefont {Daniel~S.}\ \bibnamefont
  {Levine}}, \bibinfo {author} {\bibfnamefont {William}\ \bibnamefont
  {Huggins}}, \bibinfo {author} {\bibfnamefont {Zhang}\ \bibnamefont {Jiang}},
  \bibinfo {author} {\bibfnamefont {Jarrod~R.}\ \bibnamefont {McClean}},
  \bibinfo {author} {\bibfnamefont {Ryan}\ \bibnamefont {Babbush}}, \bibinfo
  {author} {\bibfnamefont {Martin}\ \bibnamefont {Head-Gordon}}, \ and\
  \bibinfo {author} {\bibfnamefont {K.~Birgitta}\ \bibnamefont {Whaley}},\
  }\bibfield  {title} {\enquote {\bibinfo {title} {Postponing the orthogonality
  catastrophe: efficient state preparation for electronic structure simulations
  on quantum devices},}\ }\href@noop {} {\  (\bibinfo {year} {2018})},\ \Eprint
  {http://arxiv.org/abs/1809.05523} {arXiv:1809.05523 [quant-ph]} \BibitemShut
  {NoStop}%
\bibitem [{\citenamefont {Ge}\ \emph {et~al.}(2019)\citenamefont {Ge},
  \citenamefont {Tura},\ and\ \citenamefont {Cirac}}]{ge2019faster}%
  \BibitemOpen
  \bibfield  {author} {\bibinfo {author} {\bibfnamefont {Yimin}\ \bibnamefont
  {Ge}}, \bibinfo {author} {\bibfnamefont {Jordi}\ \bibnamefont {Tura}}, \ and\
  \bibinfo {author} {\bibfnamefont {J.~Ignacio}\ \bibnamefont {Cirac}},\
  }\bibfield  {title} {\enquote {\bibinfo {title} {{Faster ground state
  preparation and high-precision ground energy estimation with fewer
  qubits}},}\ }\href {\doibase 10.1063/1.5027484} {\bibfield  {journal}
  {\bibinfo  {journal} {Journal of Mathematical Physics}\ }\textbf {\bibinfo
  {volume} {60}},\ \bibinfo {pages} {022202} (\bibinfo {year}
  {2019})}\BibitemShut {NoStop}%
\bibitem [{\citenamefont {Lin}\ and\ \citenamefont {Tong}(2020)}]{JBL:lin}%
  \BibitemOpen
  \bibfield  {author} {\bibinfo {author} {\bibfnamefont {Lin}\ \bibnamefont
  {Lin}}\ and\ \bibinfo {author} {\bibfnamefont {Yu}~\bibnamefont {Tong}},\
  }\bibfield  {title} {\enquote {\bibinfo {title} {Near-optimal ground state
  preparation},}\ }\href {\doibase 10.22331/q-2020-12-14-372} {\bibfield
  {journal} {\bibinfo  {journal} {{Quantum}}\ }\textbf {\bibinfo {volume}
  {4}},\ \bibinfo {pages} {372} (\bibinfo {year} {2020})}\BibitemShut {NoStop}%
\bibitem [{\citenamefont {Wang}\ \emph {et~al.}(2023)\citenamefont {Wang},
  \citenamefont {Zhu}, \citenamefont {Jing},\ and\ \citenamefont
  {Wang}}]{wang2023ground}%
  \BibitemOpen
  \bibfield  {author} {\bibinfo {author} {\bibfnamefont {Youle}\ \bibnamefont
  {Wang}}, \bibinfo {author} {\bibfnamefont {Chenghong}\ \bibnamefont {Zhu}},
  \bibinfo {author} {\bibfnamefont {Mingrui}\ \bibnamefont {Jing}}, \ and\
  \bibinfo {author} {\bibfnamefont {Xin}\ \bibnamefont {Wang}},\ }\bibfield
  {title} {\enquote {\bibinfo {title} {Ground state preparation with shallow
  variational warm-start},}\ }\href@noop {} {\  (\bibinfo {year} {2023})},\
  \Eprint {http://arxiv.org/abs/2303.11204} {arXiv:2303.11204 [quant-ph]}
  \BibitemShut {NoStop}%
\bibitem [{\citenamefont {Farhi}\ \emph {et~al.}(2000)\citenamefont {Farhi},
  \citenamefont {Goldstone}, \citenamefont {Gutmann},\ and\ \citenamefont
  {Sipser}}]{farhi2000quantum}%
  \BibitemOpen
  \bibfield  {author} {\bibinfo {author} {\bibfnamefont {Edward}\ \bibnamefont
  {Farhi}}, \bibinfo {author} {\bibfnamefont {Jeffrey}\ \bibnamefont
  {Goldstone}}, \bibinfo {author} {\bibfnamefont {Sam}\ \bibnamefont
  {Gutmann}}, \ and\ \bibinfo {author} {\bibfnamefont {Michael}\ \bibnamefont
  {Sipser}},\ }\bibfield  {title} {\enquote {\bibinfo {title} {Quantum
  computation by adiabatic evolution},}\ }\href@noop {} {\  (\bibinfo {year}
  {2000})},\ \Eprint {http://arxiv.org/abs/quant-ph/0001106}
  {arXiv:quant-ph/0001106 [quant-ph]} \BibitemShut {NoStop}%
\bibitem [{\citenamefont {Aspuru-Guzik}\ \emph {et~al.}(2005)\citenamefont
  {Aspuru-Guzik}, \citenamefont {Dutoi}, \citenamefont {Love},\ and\
  \citenamefont {Head-Gordon}}]{GordonScience2005}%
  \BibitemOpen
  \bibfield  {author} {\bibinfo {author} {\bibfnamefont {Alán}\ \bibnamefont
  {Aspuru-Guzik}}, \bibinfo {author} {\bibfnamefont {Anthony~D.}\ \bibnamefont
  {Dutoi}}, \bibinfo {author} {\bibfnamefont {Peter~J.}\ \bibnamefont {Love}},
  \ and\ \bibinfo {author} {\bibfnamefont {Martin}\ \bibnamefont
  {Head-Gordon}},\ }\bibfield  {title} {\enquote {\bibinfo {title} {Simulated
  quantum computation of molecular energies},}\ }\href {\doibase
  10.1126/science.1113479} {\bibfield  {journal} {\bibinfo  {journal}
  {Science}\ }\textbf {\bibinfo {volume} {309}},\ \bibinfo {pages} {1704--1707}
  (\bibinfo {year} {2005})}\BibitemShut {NoStop}%
\bibitem [{\citenamefont {Farhi}\ \emph {et~al.}(2001)\citenamefont {Farhi},
  \citenamefont {Goldstone}, \citenamefont {Gutmann}, \citenamefont {Lapan},
  \citenamefont {Lundgren},\ and\ \citenamefont {Preda}}]{FarhiScience2001}%
  \BibitemOpen
  \bibfield  {author} {\bibinfo {author} {\bibfnamefont {Edward}\ \bibnamefont
  {Farhi}}, \bibinfo {author} {\bibfnamefont {Jeffrey}\ \bibnamefont
  {Goldstone}}, \bibinfo {author} {\bibfnamefont {Sam}\ \bibnamefont
  {Gutmann}}, \bibinfo {author} {\bibfnamefont {Joshua}\ \bibnamefont {Lapan}},
  \bibinfo {author} {\bibfnamefont {Andrew}\ \bibnamefont {Lundgren}}, \ and\
  \bibinfo {author} {\bibfnamefont {Daniel}\ \bibnamefont {Preda}},\ }\bibfield
   {title} {\enquote {\bibinfo {title} {A quantum adiabatic evolution algorithm
  applied to random instances of an np-complete problem},}\ }\href {\doibase
  10.1126/science.1057726} {\bibfield  {journal} {\bibinfo  {journal}
  {Science}\ }\textbf {\bibinfo {volume} {292}},\ \bibinfo {pages} {472--475}
  (\bibinfo {year} {2001})}\BibitemShut {NoStop}%
\bibitem [{\citenamefont {Albash}\ and\ \citenamefont
  {Lidar}(2018)}]{RMDadiabatic2018}%
  \BibitemOpen
  \bibfield  {author} {\bibinfo {author} {\bibfnamefont {Tameem}\ \bibnamefont
  {Albash}}\ and\ \bibinfo {author} {\bibfnamefont {Daniel~A.}\ \bibnamefont
  {Lidar}},\ }\bibfield  {title} {\enquote {\bibinfo {title} {Adiabatic quantum
  computation},}\ }\href {\doibase 10.1103/RevModPhys.90.015002} {\bibfield
  {journal} {\bibinfo  {journal} {Rev. Mod. Phys.}\ }\textbf {\bibinfo {volume}
  {90}},\ \bibinfo {pages} {015002} (\bibinfo {year} {2018})}\BibitemShut
  {NoStop}%
\bibitem [{\citenamefont {Finnila}\ \emph {et~al.}(1994)\citenamefont
  {Finnila}, \citenamefont {Gomez}, \citenamefont {Sebenik}, \citenamefont
  {Stenson},\ and\ \citenamefont {Doll}}]{finnila1994quantum}%
  \BibitemOpen
  \bibfield  {author} {\bibinfo {author} {\bibfnamefont {A.B.}\ \bibnamefont
  {Finnila}}, \bibinfo {author} {\bibfnamefont {M.A.}\ \bibnamefont {Gomez}},
  \bibinfo {author} {\bibfnamefont {C.}~\bibnamefont {Sebenik}}, \bibinfo
  {author} {\bibfnamefont {C.}~\bibnamefont {Stenson}}, \ and\ \bibinfo
  {author} {\bibfnamefont {J.D.}\ \bibnamefont {Doll}},\ }\bibfield  {title}
  {\enquote {\bibinfo {title} {Quantum annealing: A new method for minimizing
  multidimensional functions},}\ }\href {\doibase
  https://doi.org/10.1016/0009-2614(94)00117-0} {\bibfield  {journal} {\bibinfo
   {journal} {Chemical Physics Letters}\ }\textbf {\bibinfo {volume} {219}},\
  \bibinfo {pages} {343--348} (\bibinfo {year} {1994})}\BibitemShut {NoStop}%
\bibitem [{\citenamefont {Kadowaki}\ and\ \citenamefont
  {Nishimori}(1998)}]{kadowaki1998quantum}%
  \BibitemOpen
  \bibfield  {author} {\bibinfo {author} {\bibfnamefont {Tadashi}\ \bibnamefont
  {Kadowaki}}\ and\ \bibinfo {author} {\bibfnamefont {Hidetoshi}\ \bibnamefont
  {Nishimori}},\ }\bibfield  {title} {\enquote {\bibinfo {title} {Quantum
  annealing in the transverse ising model},}\ }\href {\doibase
  10.1103/PhysRevE.58.5355} {\bibfield  {journal} {\bibinfo  {journal} {Phys.
  Rev. E}\ }\textbf {\bibinfo {volume} {58}},\ \bibinfo {pages} {5355--5363}
  (\bibinfo {year} {1998})}\BibitemShut {NoStop}%
\bibitem [{\citenamefont {Brooke}\ \emph {et~al.}(1999)\citenamefont {Brooke},
  \citenamefont {Bitko}, \citenamefont {F.}, \citenamefont {null},\ and\
  \citenamefont {Aeppli}}]{brooke1999quantum}%
  \BibitemOpen
  \bibfield  {author} {\bibinfo {author} {\bibfnamefont {J.}~\bibnamefont
  {Brooke}}, \bibinfo {author} {\bibfnamefont {D.}~\bibnamefont {Bitko}},
  \bibinfo {author} {\bibfnamefont {T.}~\bibnamefont {F.}}, \bibinfo {author}
  {\bibnamefont {null}}, \ and\ \bibinfo {author} {\bibfnamefont
  {G.}~\bibnamefont {Aeppli}},\ }\bibfield  {title} {\enquote {\bibinfo {title}
  {Quantum annealing of a disordered magnet},}\ }\href {\doibase
  10.1126/science.284.5415.779} {\bibfield  {journal} {\bibinfo  {journal}
  {Science}\ }\textbf {\bibinfo {volume} {284}},\ \bibinfo {pages} {779--781}
  (\bibinfo {year} {1999})}\BibitemShut {NoStop}%
\bibitem [{\citenamefont {Preskill}(2018)}]{JBL:nisq}%
  \BibitemOpen
  \bibfield  {author} {\bibinfo {author} {\bibfnamefont {John}\ \bibnamefont
  {Preskill}},\ }\bibfield  {title} {\enquote {\bibinfo {title} {Quantum
  {C}omputing in the {NISQ} era and beyond},}\ }\href {\doibase
  10.22331/q-2018-08-06-79} {\bibfield  {journal} {\bibinfo  {journal}
  {{Quantum}}\ }\textbf {\bibinfo {volume} {2}},\ \bibinfo {pages} {79}
  (\bibinfo {year} {2018})}\BibitemShut {NoStop}%
\bibitem [{\citenamefont {McClean}\ \emph {et~al.}(2016)\citenamefont
  {McClean}, \citenamefont {Romero}, \citenamefont {Babbush},\ and\
  \citenamefont {Aspuru-Guzik}}]{mcclean2016theory}%
  \BibitemOpen
  \bibfield  {author} {\bibinfo {author} {\bibfnamefont {Jarrod~R}\
  \bibnamefont {McClean}}, \bibinfo {author} {\bibfnamefont {Jonathan}\
  \bibnamefont {Romero}}, \bibinfo {author} {\bibfnamefont {Ryan}\ \bibnamefont
  {Babbush}}, \ and\ \bibinfo {author} {\bibfnamefont {Alán}\ \bibnamefont
  {Aspuru-Guzik}},\ }\bibfield  {title} {\enquote {\bibinfo {title} {The theory
  of variational hybrid quantum-classical algorithms},}\ }\href {\doibase
  10.1088/1367-2630/18/2/023023} {\bibfield  {journal} {\bibinfo  {journal}
  {New Journal of Physics}\ }\textbf {\bibinfo {volume} {18}},\ \bibinfo
  {pages} {023023} (\bibinfo {year} {2016})}\BibitemShut {NoStop}%
\bibitem [{\citenamefont {Bravo-Prieto}\ \emph {et~al.}(2023)\citenamefont
  {Bravo-Prieto}, \citenamefont {LaRose}, \citenamefont {Cerezo}, \citenamefont
  {Subasi}, \citenamefont {Cincio},\ and\ \citenamefont
  {Coles}}]{bravo2019variational}%
  \BibitemOpen
  \bibfield  {author} {\bibinfo {author} {\bibfnamefont {Carlos}\ \bibnamefont
  {Bravo-Prieto}}, \bibinfo {author} {\bibfnamefont {Ryan}\ \bibnamefont
  {LaRose}}, \bibinfo {author} {\bibfnamefont {M.}~\bibnamefont {Cerezo}},
  \bibinfo {author} {\bibfnamefont {Yigit}\ \bibnamefont {Subasi}}, \bibinfo
  {author} {\bibfnamefont {Lukasz}\ \bibnamefont {Cincio}}, \ and\ \bibinfo
  {author} {\bibfnamefont {Patrick~J.}\ \bibnamefont {Coles}},\ }\bibfield
  {title} {\enquote {\bibinfo {title} {Variational {Q}uantum {L}inear
  {S}olver},}\ }\href {\doibase 10.22331/q-2023-11-22-1188} {\bibfield
  {journal} {\bibinfo  {journal} {{Quantum}}\ }\textbf {\bibinfo {volume}
  {7}},\ \bibinfo {pages} {1188} (\bibinfo {year} {2023})}\BibitemShut
  {NoStop}%
\bibitem [{\citenamefont {Grimsley}\ \emph {et~al.}(2019)\citenamefont
  {Grimsley}, \citenamefont {Economou}, \citenamefont {Barnes},\ and\
  \citenamefont {Mayhall}}]{grimsley2019adaptive}%
  \BibitemOpen
  \bibfield  {author} {\bibinfo {author} {\bibfnamefont {Harper~R.}\
  \bibnamefont {Grimsley}}, \bibinfo {author} {\bibfnamefont {Sophia~E.}\
  \bibnamefont {Economou}}, \bibinfo {author} {\bibfnamefont {Edwin}\
  \bibnamefont {Barnes}}, \ and\ \bibinfo {author} {\bibfnamefont
  {Nicholas~J.}\ \bibnamefont {Mayhall}},\ }\bibfield  {title} {\enquote
  {\bibinfo {title} {An adaptive variational algorithm for exact molecular
  simulations on a quantum computer},}\ }\href {\doibase
  10.1038/s41467-019-10988-2} {\bibfield  {journal} {\bibinfo  {journal}
  {Nature Communications}\ }\textbf {\bibinfo {volume} {10}},\ \bibinfo {pages}
  {3007} (\bibinfo {year} {2019})}\BibitemShut {NoStop}%
\bibitem [{\citenamefont {Cerezo}\ \emph {et~al.}(2021)\citenamefont {Cerezo},
  \citenamefont {Arrasmith}, \citenamefont {Babbush}, \citenamefont {Benjamin},
  \citenamefont {Endo}, \citenamefont {Fujii}, \citenamefont {McClean},
  \citenamefont {Mitarai}, \citenamefont {Yuan}, \citenamefont {Cincio},\ and\
  \citenamefont {Coles}}]{cerezo2021variational}%
  \BibitemOpen
  \bibfield  {author} {\bibinfo {author} {\bibfnamefont {M.}~\bibnamefont
  {Cerezo}}, \bibinfo {author} {\bibfnamefont {Andrew}\ \bibnamefont
  {Arrasmith}}, \bibinfo {author} {\bibfnamefont {Ryan}\ \bibnamefont
  {Babbush}}, \bibinfo {author} {\bibfnamefont {Simon~C.}\ \bibnamefont
  {Benjamin}}, \bibinfo {author} {\bibfnamefont {Suguru}\ \bibnamefont {Endo}},
  \bibinfo {author} {\bibfnamefont {Keisuke}\ \bibnamefont {Fujii}}, \bibinfo
  {author} {\bibfnamefont {Jarrod~R.}\ \bibnamefont {McClean}}, \bibinfo
  {author} {\bibfnamefont {Kosuke}\ \bibnamefont {Mitarai}}, \bibinfo {author}
  {\bibfnamefont {Xiao}\ \bibnamefont {Yuan}}, \bibinfo {author} {\bibfnamefont
  {Lukasz}\ \bibnamefont {Cincio}}, \ and\ \bibinfo {author} {\bibfnamefont
  {Patrick~J.}\ \bibnamefont {Coles}},\ }\bibfield  {title} {\enquote {\bibinfo
  {title} {Variational quantum algorithms},}\ }\href {\doibase
  10.1038/s42254-021-00348-9} {\bibfield  {journal} {\bibinfo  {journal}
  {Nature Reviews Physics}\ }\textbf {\bibinfo {volume} {3}},\ \bibinfo {pages}
  {625--644} (\bibinfo {year} {2021})}\BibitemShut {NoStop}%
\bibitem [{\citenamefont {Magann}\ \emph {et~al.}(2021)\citenamefont {Magann},
  \citenamefont {Arenz}, \citenamefont {Grace}, \citenamefont {Ho},
  \citenamefont {Kosut}, \citenamefont {McClean}, \citenamefont {Rabitz},\ and\
  \citenamefont {Sarovar}}]{PRXQuantum.2.010101}%
  \BibitemOpen
  \bibfield  {author} {\bibinfo {author} {\bibfnamefont {Alicia~B.}\
  \bibnamefont {Magann}}, \bibinfo {author} {\bibfnamefont {Christian}\
  \bibnamefont {Arenz}}, \bibinfo {author} {\bibfnamefont {Matthew~D.}\
  \bibnamefont {Grace}}, \bibinfo {author} {\bibfnamefont {Tak-San}\
  \bibnamefont {Ho}}, \bibinfo {author} {\bibfnamefont {Robert~L.}\
  \bibnamefont {Kosut}}, \bibinfo {author} {\bibfnamefont {Jarrod~R.}\
  \bibnamefont {McClean}}, \bibinfo {author} {\bibfnamefont {Herschel~A.}\
  \bibnamefont {Rabitz}}, \ and\ \bibinfo {author} {\bibfnamefont {Mohan}\
  \bibnamefont {Sarovar}},\ }\bibfield  {title} {\enquote {\bibinfo {title}
  {From pulses to circuits and back again: A quantum optimal control
  perspective on variational quantum algorithms},}\ }\href {\doibase
  10.1103/PRXQuantum.2.010101} {\bibfield  {journal} {\bibinfo  {journal} {PRX
  Quantum}\ }\textbf {\bibinfo {volume} {2}},\ \bibinfo {pages} {010101}
  (\bibinfo {year} {2021})}\BibitemShut {NoStop}%
\bibitem [{\citenamefont {Bharti}\ \emph {et~al.}(2022)\citenamefont {Bharti},
  \citenamefont {Cervera-Lierta}, \citenamefont {Kyaw}, \citenamefont {Haug},
  \citenamefont {Alperin-Lea}, \citenamefont {Anand}, \citenamefont {Degroote},
  \citenamefont {Heimonen}, \citenamefont {Kottmann}, \citenamefont {Menke},
  \citenamefont {Mok}, \citenamefont {Sim}, \citenamefont {Kwek},\ and\
  \citenamefont {Aspuru-Guzik}}]{RMDKBharti2022}%
  \BibitemOpen
  \bibfield  {author} {\bibinfo {author} {\bibfnamefont {Kishor}\ \bibnamefont
  {Bharti}}, \bibinfo {author} {\bibfnamefont {Alba}\ \bibnamefont
  {Cervera-Lierta}}, \bibinfo {author} {\bibfnamefont {Thi~Ha}\ \bibnamefont
  {Kyaw}}, \bibinfo {author} {\bibfnamefont {Tobias}\ \bibnamefont {Haug}},
  \bibinfo {author} {\bibfnamefont {Sumner}\ \bibnamefont {Alperin-Lea}},
  \bibinfo {author} {\bibfnamefont {Abhinav}\ \bibnamefont {Anand}}, \bibinfo
  {author} {\bibfnamefont {Matthias}\ \bibnamefont {Degroote}}, \bibinfo
  {author} {\bibfnamefont {Hermanni}\ \bibnamefont {Heimonen}}, \bibinfo
  {author} {\bibfnamefont {Jakob~S.}\ \bibnamefont {Kottmann}}, \bibinfo
  {author} {\bibfnamefont {Tim}\ \bibnamefont {Menke}}, \bibinfo {author}
  {\bibfnamefont {Wai-Keong}\ \bibnamefont {Mok}}, \bibinfo {author}
  {\bibfnamefont {Sukin}\ \bibnamefont {Sim}}, \bibinfo {author} {\bibfnamefont
  {Leong-Chuan}\ \bibnamefont {Kwek}}, \ and\ \bibinfo {author} {\bibfnamefont
  {Al\'an}\ \bibnamefont {Aspuru-Guzik}},\ }\bibfield  {title} {\enquote
  {\bibinfo {title} {Noisy intermediate-scale quantum algorithms},}\ }\href
  {\doibase 10.1103/RevModPhys.94.015004} {\bibfield  {journal} {\bibinfo
  {journal} {Rev. Mod. Phys.}\ }\textbf {\bibinfo {volume} {94}},\ \bibinfo
  {pages} {015004} (\bibinfo {year} {2022})}\BibitemShut {NoStop}%
\bibitem [{\citenamefont {Farhi}\ \emph {et~al.}(2014)\citenamefont {Farhi},
  \citenamefont {Goldstone},\ and\ \citenamefont {Gutmann}}]{QAOA}%
  \BibitemOpen
  \bibfield  {author} {\bibinfo {author} {\bibfnamefont {Edward}\ \bibnamefont
  {Farhi}}, \bibinfo {author} {\bibfnamefont {Jeffrey}\ \bibnamefont
  {Goldstone}}, \ and\ \bibinfo {author} {\bibfnamefont {Sam}\ \bibnamefont
  {Gutmann}},\ }\bibfield  {title} {\enquote {\bibinfo {title} {A quantum
  approximate optimization algorithm},}\ }\href@noop {} {\  (\bibinfo {year}
  {2014})},\ \Eprint {http://arxiv.org/abs/1411.4028} {arXiv:1411.4028
  [quant-ph]} \BibitemShut {NoStop}%
\bibitem [{\citenamefont {Lloyd}(2018)}]{lloyd2018quantum}%
  \BibitemOpen
  \bibfield  {author} {\bibinfo {author} {\bibfnamefont {Seth}\ \bibnamefont
  {Lloyd}},\ }\href@noop {} {\enquote {\bibinfo {title} {Quantum approximate
  optimization is computationally universal},}\ } (\bibinfo {year} {2018}),\
  \Eprint {http://arxiv.org/abs/1812.11075} {arXiv:1812.11075 [quant-ph]}
  \BibitemShut {NoStop}%
\bibitem [{\citenamefont {Dalzell}\ \emph {et~al.}(2020)\citenamefont
  {Dalzell}, \citenamefont {Harrow}, \citenamefont {Koh},\ and\ \citenamefont
  {La~Placa}}]{dalzell2020many}%
  \BibitemOpen
  \bibfield  {author} {\bibinfo {author} {\bibfnamefont {Alexander~M.}\
  \bibnamefont {Dalzell}}, \bibinfo {author} {\bibfnamefont {Aram~W.}\
  \bibnamefont {Harrow}}, \bibinfo {author} {\bibfnamefont {Dax~Enshan}\
  \bibnamefont {Koh}}, \ and\ \bibinfo {author} {\bibfnamefont {Rolando~L.}\
  \bibnamefont {La~Placa}},\ }\bibfield  {title} {\enquote {\bibinfo {title}
  {How many qubits are needed for quantum computational supremacy?}}\ }\href
  {\doibase 10.22331/q-2020-05-11-264} {\bibfield  {journal} {\bibinfo
  {journal} {{Quantum}}\ }\textbf {\bibinfo {volume} {4}},\ \bibinfo {pages}
  {264} (\bibinfo {year} {2020})}\BibitemShut {NoStop}%
\bibitem [{\citenamefont {Peruzzo}\ \emph {et~al.}(2014)\citenamefont
  {Peruzzo}, \citenamefont {McClean}, \citenamefont {Shadbolt}, \citenamefont
  {Yung}, \citenamefont {Zhou}, \citenamefont {Love}, \citenamefont
  {Aspuru-Guzik},\ and\ \citenamefont {O'Brien}}]{peruzzo2014variational}%
  \BibitemOpen
  \bibfield  {author} {\bibinfo {author} {\bibfnamefont {Alberto}\ \bibnamefont
  {Peruzzo}}, \bibinfo {author} {\bibfnamefont {Jarrod}\ \bibnamefont
  {McClean}}, \bibinfo {author} {\bibfnamefont {Peter}\ \bibnamefont
  {Shadbolt}}, \bibinfo {author} {\bibfnamefont {Man-Hong}\ \bibnamefont
  {Yung}}, \bibinfo {author} {\bibfnamefont {Xiao-Qi}\ \bibnamefont {Zhou}},
  \bibinfo {author} {\bibfnamefont {Peter~J.}\ \bibnamefont {Love}}, \bibinfo
  {author} {\bibfnamefont {Al{\'a}n}\ \bibnamefont {Aspuru-Guzik}}, \ and\
  \bibinfo {author} {\bibfnamefont {Jeremy~L.}\ \bibnamefont {O'Brien}},\
  }\bibfield  {title} {\enquote {\bibinfo {title} {A variational eigenvalue
  solver on a photonic quantum processor},}\ }\href {\doibase
  10.1038/ncomms5213} {\bibfield  {journal} {\bibinfo  {journal} {Nature
  Communications}\ }\textbf {\bibinfo {volume} {5}},\ \bibinfo {pages} {4213}
  (\bibinfo {year} {2014})}\BibitemShut {NoStop}%
\bibitem [{\citenamefont {Kandala}\ \emph {et~al.}(2017)\citenamefont
  {Kandala}, \citenamefont {Mezzacapo}, \citenamefont {Temme}, \citenamefont
  {Takita}, \citenamefont {Brink}, \citenamefont {Chow},\ and\ \citenamefont
  {Gambetta}}]{kandala2017hardware}%
  \BibitemOpen
  \bibfield  {author} {\bibinfo {author} {\bibfnamefont {Abhinav}\ \bibnamefont
  {Kandala}}, \bibinfo {author} {\bibfnamefont {Antonio}\ \bibnamefont
  {Mezzacapo}}, \bibinfo {author} {\bibfnamefont {Kristan}\ \bibnamefont
  {Temme}}, \bibinfo {author} {\bibfnamefont {Maika}\ \bibnamefont {Takita}},
  \bibinfo {author} {\bibfnamefont {Markus}\ \bibnamefont {Brink}}, \bibinfo
  {author} {\bibfnamefont {Jerry~M.}\ \bibnamefont {Chow}}, \ and\ \bibinfo
  {author} {\bibfnamefont {Jay~M.}\ \bibnamefont {Gambetta}},\ }\bibfield
  {title} {\enquote {\bibinfo {title} {Hardware-efficient variational quantum
  eigensolver for small molecules and quantum magnets},}\ }\href {\doibase
  10.1038/nature23879} {\bibfield  {journal} {\bibinfo  {journal} {Nature}\
  }\textbf {\bibinfo {volume} {549}},\ \bibinfo {pages} {242--246} (\bibinfo
  {year} {2017})}\BibitemShut {NoStop}%
\bibitem [{\citenamefont {Colless}\ \emph {et~al.}(2018)\citenamefont
  {Colless}, \citenamefont {Ramasesh}, \citenamefont {Dahlen}, \citenamefont
  {Blok}, \citenamefont {Kimchi-Schwartz}, \citenamefont {McClean},
  \citenamefont {Carter}, \citenamefont {de~Jong},\ and\ \citenamefont
  {Siddiqi}}]{colless2018computation}%
  \BibitemOpen
  \bibfield  {author} {\bibinfo {author} {\bibfnamefont {J.~I.}\ \bibnamefont
  {Colless}}, \bibinfo {author} {\bibfnamefont {V.~V.}\ \bibnamefont
  {Ramasesh}}, \bibinfo {author} {\bibfnamefont {D.}~\bibnamefont {Dahlen}},
  \bibinfo {author} {\bibfnamefont {M.~S.}\ \bibnamefont {Blok}}, \bibinfo
  {author} {\bibfnamefont {M.~E.}\ \bibnamefont {Kimchi-Schwartz}}, \bibinfo
  {author} {\bibfnamefont {J.~R.}\ \bibnamefont {McClean}}, \bibinfo {author}
  {\bibfnamefont {J.}~\bibnamefont {Carter}}, \bibinfo {author} {\bibfnamefont
  {W.~A.}\ \bibnamefont {de~Jong}}, \ and\ \bibinfo {author} {\bibfnamefont
  {I.}~\bibnamefont {Siddiqi}},\ }\bibfield  {title} {\enquote {\bibinfo
  {title} {Computation of molecular spectra on a quantum processor with an
  error-resilient algorithm},}\ }\href {\doibase 10.1103/PhysRevX.8.011021}
  {\bibfield  {journal} {\bibinfo  {journal} {Phys. Rev. X}\ }\textbf {\bibinfo
  {volume} {8}},\ \bibinfo {pages} {011021} (\bibinfo {year}
  {2018})}\BibitemShut {NoStop}%
\bibitem [{\citenamefont {Tilly}\ \emph {et~al.}(2022)\citenamefont {Tilly},
  \citenamefont {Chen}, \citenamefont {Cao}, \citenamefont {Picozzi},
  \citenamefont {Setia}, \citenamefont {Li}, \citenamefont {Grant},
  \citenamefont {Wossnig}, \citenamefont {Rungger}, \citenamefont {Booth},\
  and\ \citenamefont {Tennyson}}]{tilly2022variational}%
  \BibitemOpen
  \bibfield  {author} {\bibinfo {author} {\bibfnamefont {Jules}\ \bibnamefont
  {Tilly}}, \bibinfo {author} {\bibfnamefont {Hongxiang}\ \bibnamefont {Chen}},
  \bibinfo {author} {\bibfnamefont {Shuxiang}\ \bibnamefont {Cao}}, \bibinfo
  {author} {\bibfnamefont {Dario}\ \bibnamefont {Picozzi}}, \bibinfo {author}
  {\bibfnamefont {Kanav}\ \bibnamefont {Setia}}, \bibinfo {author}
  {\bibfnamefont {Ying}\ \bibnamefont {Li}}, \bibinfo {author} {\bibfnamefont
  {Edward}\ \bibnamefont {Grant}}, \bibinfo {author} {\bibfnamefont {Leonard}\
  \bibnamefont {Wossnig}}, \bibinfo {author} {\bibfnamefont {Ivan}\
  \bibnamefont {Rungger}}, \bibinfo {author} {\bibfnamefont {George~H.}\
  \bibnamefont {Booth}}, \ and\ \bibinfo {author} {\bibfnamefont {Jonathan}\
  \bibnamefont {Tennyson}},\ }\bibfield  {title} {\enquote {\bibinfo {title}
  {The variational quantum eigensolver: A review of methods and best
  practices},}\ }\href {\doibase https://doi.org/10.1016/j.physrep.2022.08.003}
  {\bibfield  {journal} {\bibinfo  {journal} {Physics Reports}\ }\textbf
  {\bibinfo {volume} {986}},\ \bibinfo {pages} {1--128} (\bibinfo {year}
  {2022})}\BibitemShut {NoStop}%
\bibitem [{\citenamefont {Zhou}\ \emph {et~al.}(2020)\citenamefont {Zhou},
  \citenamefont {Wang}, \citenamefont {Choi}, \citenamefont {Pichler},\ and\
  \citenamefont {Lukin}}]{zhou2020quantum}%
  \BibitemOpen
  \bibfield  {author} {\bibinfo {author} {\bibfnamefont {Leo}\ \bibnamefont
  {Zhou}}, \bibinfo {author} {\bibfnamefont {Sheng-Tao}\ \bibnamefont {Wang}},
  \bibinfo {author} {\bibfnamefont {Soonwon}\ \bibnamefont {Choi}}, \bibinfo
  {author} {\bibfnamefont {Hannes}\ \bibnamefont {Pichler}}, \ and\ \bibinfo
  {author} {\bibfnamefont {Mikhail~D.}\ \bibnamefont {Lukin}},\ }\bibfield
  {title} {\enquote {\bibinfo {title} {Quantum approximate optimization
  algorithm: Performance, mechanism, and implementation on near-term
  devices},}\ }\href {\doibase 10.1103/PhysRevX.10.021067} {\bibfield
  {journal} {\bibinfo  {journal} {Phys. Rev. X}\ }\textbf {\bibinfo {volume}
  {10}},\ \bibinfo {pages} {021067} (\bibinfo {year} {2020})}\BibitemShut
  {NoStop}%
\bibitem [{\citenamefont {Liu}\ \emph {et~al.}(2022)\citenamefont {Liu},
  \citenamefont {Angone}, \citenamefont {Shaydulin}, \citenamefont {Safro},
  \citenamefont {Alexeev},\ and\ \citenamefont {Cincio}}]{liu2022layer}%
  \BibitemOpen
  \bibfield  {author} {\bibinfo {author} {\bibfnamefont {Xiaoyuan}\
  \bibnamefont {Liu}}, \bibinfo {author} {\bibfnamefont {Anthony}\ \bibnamefont
  {Angone}}, \bibinfo {author} {\bibfnamefont {Ruslan}\ \bibnamefont
  {Shaydulin}}, \bibinfo {author} {\bibfnamefont {Ilya}\ \bibnamefont {Safro}},
  \bibinfo {author} {\bibfnamefont {Yuri}\ \bibnamefont {Alexeev}}, \ and\
  \bibinfo {author} {\bibfnamefont {Lukasz}\ \bibnamefont {Cincio}},\
  }\bibfield  {title} {\enquote {\bibinfo {title} {Layer vqe: A variational
  approach for combinatorial optimization on noisy quantum computers},}\ }\href
  {\doibase 10.1109/TQE.2021.3140190} {\bibfield  {journal} {\bibinfo
  {journal} {IEEE Transactions on Quantum Engineering}\ }\textbf {\bibinfo
  {volume} {3}},\ \bibinfo {pages} {1--20} (\bibinfo {year}
  {2022})}\BibitemShut {NoStop}%
\bibitem [{\citenamefont {Cao}\ \emph {et~al.}(2019)\citenamefont {Cao},
  \citenamefont {Romero}, \citenamefont {Olson}, \citenamefont {Degroote},
  \citenamefont {Johnson}, \citenamefont {Kieferová}, \citenamefont
  {Kivlichan}, \citenamefont {Menke}, \citenamefont {Peropadre}, \citenamefont
  {Sawaya}, \citenamefont {Sim}, \citenamefont {Veis},\ and\ \citenamefont
  {Aspuru-Guzik}}]{cao2019quantum}%
  \BibitemOpen
  \bibfield  {author} {\bibinfo {author} {\bibfnamefont {Yudong}\ \bibnamefont
  {Cao}}, \bibinfo {author} {\bibfnamefont {Jonathan}\ \bibnamefont {Romero}},
  \bibinfo {author} {\bibfnamefont {Jonathan~P.}\ \bibnamefont {Olson}},
  \bibinfo {author} {\bibfnamefont {Matthias}\ \bibnamefont {Degroote}},
  \bibinfo {author} {\bibfnamefont {Peter~D.}\ \bibnamefont {Johnson}},
  \bibinfo {author} {\bibfnamefont {Mária}\ \bibnamefont {Kieferová}},
  \bibinfo {author} {\bibfnamefont {Ian~D.}\ \bibnamefont {Kivlichan}},
  \bibinfo {author} {\bibfnamefont {Tim}\ \bibnamefont {Menke}}, \bibinfo
  {author} {\bibfnamefont {Borja}\ \bibnamefont {Peropadre}}, \bibinfo {author}
  {\bibfnamefont {Nicolas P.~D.}\ \bibnamefont {Sawaya}}, \bibinfo {author}
  {\bibfnamefont {Sukin}\ \bibnamefont {Sim}}, \bibinfo {author} {\bibfnamefont
  {Libor}\ \bibnamefont {Veis}}, \ and\ \bibinfo {author} {\bibfnamefont
  {Alán}\ \bibnamefont {Aspuru-Guzik}},\ }\bibfield  {title} {\enquote
  {\bibinfo {title} {Quantum chemistry in the age of quantum computing},}\
  }\href {\doibase 10.1021/acs.chemrev.8b00803} {\bibfield  {journal} {\bibinfo
   {journal} {Chemical Reviews}\ }\textbf {\bibinfo {volume} {119}},\ \bibinfo
  {pages} {10856--10915} (\bibinfo {year} {2019})}\BibitemShut {NoStop}%
\bibitem [{\citenamefont {Chakrabarti}\ and\ \citenamefont
  {Rabitz}(2007)}]{chakrabarti2007quantum}%
  \BibitemOpen
  \bibfield  {author} {\bibinfo {author} {\bibfnamefont {Raj}\ \bibnamefont
  {Chakrabarti}}\ and\ \bibinfo {author} {\bibfnamefont {Herschel}\
  \bibnamefont {Rabitz}},\ }\bibfield  {title} {\enquote {\bibinfo {title}
  {Quantum control landscapes},}\ }\href {\doibase 10.1080/01442350701633300}
  {\bibfield  {journal} {\bibinfo  {journal} {International Reviews in Physical
  Chemistry}\ }\textbf {\bibinfo {volume} {26}},\ \bibinfo {pages} {671--735}
  (\bibinfo {year} {2007})}\BibitemShut {NoStop}%
\bibitem [{\citenamefont {Russell}\ \emph {et~al.}(2017)\citenamefont
  {Russell}, \citenamefont {Rabitz},\ and\ \citenamefont
  {Wu}}]{russell2017control}%
  \BibitemOpen
  \bibfield  {author} {\bibinfo {author} {\bibfnamefont {Benjamin}\
  \bibnamefont {Russell}}, \bibinfo {author} {\bibfnamefont {Herschel}\
  \bibnamefont {Rabitz}}, \ and\ \bibinfo {author} {\bibfnamefont {Re-Bing}\
  \bibnamefont {Wu}},\ }\bibfield  {title} {\enquote {\bibinfo {title} {Control
  landscapes are almost always trap free: a geometric assessment},}\ }\href
  {\doibase 10.1088/1751-8121/aa6b77} {\bibfield  {journal} {\bibinfo
  {journal} {Journal of Physics A: Mathematical and Theoretical}\ }\textbf
  {\bibinfo {volume} {50}},\ \bibinfo {pages} {205302} (\bibinfo {year}
  {2017})}\BibitemShut {NoStop}%
\bibitem [{\citenamefont {McClean}\ \emph {et~al.}(2018)\citenamefont
  {McClean}, \citenamefont {Boixo}, \citenamefont {Smelyanskiy}, \citenamefont
  {Babbush},\ and\ \citenamefont {Neven}}]{mcclean2018barren}%
  \BibitemOpen
  \bibfield  {author} {\bibinfo {author} {\bibfnamefont {Jarrod~R.}\
  \bibnamefont {McClean}}, \bibinfo {author} {\bibfnamefont {Sergio}\
  \bibnamefont {Boixo}}, \bibinfo {author} {\bibfnamefont {Vadim~N.}\
  \bibnamefont {Smelyanskiy}}, \bibinfo {author} {\bibfnamefont {Ryan}\
  \bibnamefont {Babbush}}, \ and\ \bibinfo {author} {\bibfnamefont {Hartmut}\
  \bibnamefont {Neven}},\ }\bibfield  {title} {\enquote {\bibinfo {title}
  {Barren plateaus in quantum neural network training landscapes},}\ }\href
  {\doibase 10.1038/s41467-018-07090-4} {\bibfield  {journal} {\bibinfo
  {journal} {Nature Communications}\ }\textbf {\bibinfo {volume} {9}},\
  \bibinfo {pages} {4812} (\bibinfo {year} {2018})}\BibitemShut {NoStop}%
\bibitem [{\citenamefont {Wiersema}\ \emph {et~al.}(2020)\citenamefont
  {Wiersema}, \citenamefont {Zhou}, \citenamefont {de~Sereville}, \citenamefont
  {Carrasquilla}, \citenamefont {Kim},\ and\ \citenamefont
  {Yuen}}]{wiersema2020exploring}%
  \BibitemOpen
  \bibfield  {author} {\bibinfo {author} {\bibfnamefont {Roeland}\ \bibnamefont
  {Wiersema}}, \bibinfo {author} {\bibfnamefont {Cunlu}\ \bibnamefont {Zhou}},
  \bibinfo {author} {\bibfnamefont {Yvette}\ \bibnamefont {de~Sereville}},
  \bibinfo {author} {\bibfnamefont {Juan~Felipe}\ \bibnamefont {Carrasquilla}},
  \bibinfo {author} {\bibfnamefont {Yong~Baek}\ \bibnamefont {Kim}}, \ and\
  \bibinfo {author} {\bibfnamefont {Henry}\ \bibnamefont {Yuen}},\ }\bibfield
  {title} {\enquote {\bibinfo {title} {Exploring entanglement and optimization
  within the hamiltonian variational ansatz},}\ }\href {\doibase
  10.1103/PRXQuantum.1.020319} {\bibfield  {journal} {\bibinfo  {journal} {PRX
  Quantum}\ }\textbf {\bibinfo {volume} {1}},\ \bibinfo {pages} {020319}
  (\bibinfo {year} {2020})}\BibitemShut {NoStop}%
\bibitem [{\citenamefont {Bittel}\ and\ \citenamefont
  {Kliesch}(2021)}]{bittel2021training}%
  \BibitemOpen
  \bibfield  {author} {\bibinfo {author} {\bibfnamefont {Lennart}\ \bibnamefont
  {Bittel}}\ and\ \bibinfo {author} {\bibfnamefont {Martin}\ \bibnamefont
  {Kliesch}},\ }\bibfield  {title} {\enquote {\bibinfo {title} {Training
  variational quantum algorithms is np-hard},}\ }\href {\doibase
  10.1103/PhysRevLett.127.120502} {\bibfield  {journal} {\bibinfo  {journal}
  {Phys. Rev. Lett.}\ }\textbf {\bibinfo {volume} {127}},\ \bibinfo {pages}
  {120502} (\bibinfo {year} {2021})}\BibitemShut {NoStop}%
\bibitem [{\citenamefont {Larocca}\ \emph {et~al.}(2022)\citenamefont
  {Larocca}, \citenamefont {Czarnik}, \citenamefont {Sharma}, \citenamefont
  {Muraleedharan}, \citenamefont {Coles},\ and\ \citenamefont
  {Cerezo}}]{larocca2022diagnosing}%
  \BibitemOpen
  \bibfield  {author} {\bibinfo {author} {\bibfnamefont {Martin}\ \bibnamefont
  {Larocca}}, \bibinfo {author} {\bibfnamefont {Piotr}\ \bibnamefont
  {Czarnik}}, \bibinfo {author} {\bibfnamefont {Kunal}\ \bibnamefont {Sharma}},
  \bibinfo {author} {\bibfnamefont {Gopikrishnan}\ \bibnamefont
  {Muraleedharan}}, \bibinfo {author} {\bibfnamefont {Patrick~J.}\ \bibnamefont
  {Coles}}, \ and\ \bibinfo {author} {\bibfnamefont {M.}~\bibnamefont
  {Cerezo}},\ }\bibfield  {title} {\enquote {\bibinfo {title} {Diagnosing
  {B}arren {P}lateaus with {T}ools from {Q}uantum {O}ptimal {C}ontrol},}\
  }\href {\doibase 10.22331/q-2022-09-29-824} {\bibfield  {journal} {\bibinfo
  {journal} {{Quantum}}\ }\textbf {\bibinfo {volume} {6}},\ \bibinfo {pages}
  {824} (\bibinfo {year} {2022})}\BibitemShut {NoStop}%
\bibitem [{\citenamefont {Magann}\ \emph
  {et~al.}(2022{\natexlab{a}})\citenamefont {Magann}, \citenamefont {Rudinger},
  \citenamefont {Grace},\ and\ \citenamefont {Sarovar}}]{FeedbackPRL}%
  \BibitemOpen
  \bibfield  {author} {\bibinfo {author} {\bibfnamefont {Alicia~B.}\
  \bibnamefont {Magann}}, \bibinfo {author} {\bibfnamefont {Kenneth~M.}\
  \bibnamefont {Rudinger}}, \bibinfo {author} {\bibfnamefont {Matthew~D.}\
  \bibnamefont {Grace}}, \ and\ \bibinfo {author} {\bibfnamefont {Mohan}\
  \bibnamefont {Sarovar}},\ }\bibfield  {title} {\enquote {\bibinfo {title}
  {Feedback-based quantum optimization},}\ }\href {\doibase
  10.1103/PhysRevLett.129.250502} {\bibfield  {journal} {\bibinfo  {journal}
  {Phys. Rev. Lett.}\ }\textbf {\bibinfo {volume} {129}},\ \bibinfo {pages}
  {250502} (\bibinfo {year} {2022}{\natexlab{a}})}\BibitemShut {NoStop}%
\bibitem [{\citenamefont {Magann}\ \emph
  {et~al.}(2022{\natexlab{b}})\citenamefont {Magann}, \citenamefont {Rudinger},
  \citenamefont {Grace},\ and\ \citenamefont {Sarovar}}]{FeedbackPRA}%
  \BibitemOpen
  \bibfield  {author} {\bibinfo {author} {\bibfnamefont {Alicia~B.}\
  \bibnamefont {Magann}}, \bibinfo {author} {\bibfnamefont {Kenneth~M.}\
  \bibnamefont {Rudinger}}, \bibinfo {author} {\bibfnamefont {Matthew~D.}\
  \bibnamefont {Grace}}, \ and\ \bibinfo {author} {\bibfnamefont {Mohan}\
  \bibnamefont {Sarovar}},\ }\bibfield  {title} {\enquote {\bibinfo {title}
  {Lyapunov-control-inspired strategies for quantum combinatorial
  optimization},}\ }\href {\doibase 10.1103/PhysRevA.106.062414} {\bibfield
  {journal} {\bibinfo  {journal} {Phys. Rev. A}\ }\textbf {\bibinfo {volume}
  {106}},\ \bibinfo {pages} {062414} (\bibinfo {year}
  {2022}{\natexlab{b}})}\BibitemShut {NoStop}%
\bibitem [{\citenamefont {Grivopoulos}\ and\ \citenamefont
  {Bamieh}(2003)}]{grivopoulos2003lyapunov}%
  \BibitemOpen
  \bibfield  {author} {\bibinfo {author} {\bibfnamefont {S.}~\bibnamefont
  {Grivopoulos}}\ and\ \bibinfo {author} {\bibfnamefont {B.}~\bibnamefont
  {Bamieh}},\ }\bibfield  {title} {\enquote {\bibinfo {title} {Lyapunov-based
  control of quantum systems},}\ }in\ \href {\doibase 10.1109/CDC.2003.1272601}
  {\emph {\bibinfo {booktitle} {42nd IEEE International Conference on Decision
  and Control (IEEE Cat. No.03CH37475)}}},\ Vol.~\bibinfo {volume} {1}\
  (\bibinfo {year} {2003})\ pp.\ \bibinfo {pages} {434--438 Vol.1}\BibitemShut
  {NoStop}%
\bibitem [{\citenamefont {Wang}\ \emph {et~al.}(2013)\citenamefont {Wang},
  \citenamefont {Jiang}, \citenamefont {Chen}, \citenamefont {Dong},
  \citenamefont {Cong},\ and\ \citenamefont {Meng}}]{QLC_Wang_2013}%
  \BibitemOpen
  \bibfield  {author} {\bibinfo {author} {\bibfnamefont {L.~C.}\ \bibnamefont
  {Wang}}, \bibinfo {author} {\bibfnamefont {M.}~\bibnamefont {Jiang}},
  \bibinfo {author} {\bibfnamefont {C.}~\bibnamefont {Chen}}, \bibinfo {author}
  {\bibfnamefont {D.}~\bibnamefont {Dong}}, \bibinfo {author} {\bibfnamefont
  {Shuang}\ \bibnamefont {Cong}}, \ and\ \bibinfo {author} {\bibfnamefont
  {Fangfang}\ \bibnamefont {Meng}},\ }\bibfield  {title} {\enquote {\bibinfo
  {title} {A survey of quantum lyapunov control methods},}\ }\href {\doibase
  10.1155/2013/967529} {\bibfield  {journal} {\bibinfo  {journal} {The
  Scientific World Journal}\ }\textbf {\bibinfo {volume} {2013}},\ \bibinfo
  {pages} {967529} (\bibinfo {year} {2013})}\BibitemShut {NoStop}%
\bibitem [{\citenamefont {Demirplak}\ and\ \citenamefont
  {Rice}(2003)}]{demirplak2003adiabatic}%
  \BibitemOpen
  \bibfield  {author} {\bibinfo {author} {\bibfnamefont {Mustafa}\ \bibnamefont
  {Demirplak}}\ and\ \bibinfo {author} {\bibfnamefont {Stuart~A.}\ \bibnamefont
  {Rice}},\ }\bibfield  {title} {\enquote {\bibinfo {title} {Adiabatic
  population transfer with control fields},}\ }\href {\doibase
  10.1021/jp030708a} {\bibfield  {journal} {\bibinfo  {journal} {The Journal of
  Physical Chemistry A}\ }\textbf {\bibinfo {volume} {107}},\ \bibinfo {pages}
  {9937--9945} (\bibinfo {year} {2003})}\BibitemShut {NoStop}%
\bibitem [{\citenamefont {Yao}\ \emph {et~al.}(2021)\citenamefont {Yao},
  \citenamefont {Lin},\ and\ \citenamefont {Bukov}}]{yao2021reinforcement}%
  \BibitemOpen
  \bibfield  {author} {\bibinfo {author} {\bibfnamefont {Jiahao}\ \bibnamefont
  {Yao}}, \bibinfo {author} {\bibfnamefont {Lin}\ \bibnamefont {Lin}}, \ and\
  \bibinfo {author} {\bibfnamefont {Marin}\ \bibnamefont {Bukov}},\ }\bibfield
  {title} {\enquote {\bibinfo {title} {Reinforcement learning for many-body
  ground-state preparation inspired by counterdiabatic driving},}\ }\href
  {\doibase 10.1103/PhysRevX.11.031070} {\bibfield  {journal} {\bibinfo
  {journal} {Phys. Rev. X}\ }\textbf {\bibinfo {volume} {11}},\ \bibinfo
  {pages} {031070} (\bibinfo {year} {2021})}\BibitemShut {NoStop}%
\bibitem [{\citenamefont {Chandarana}\ \emph {et~al.}(2022)\citenamefont
  {Chandarana}, \citenamefont {Hegade}, \citenamefont {Paul}, \citenamefont
  {Albarr\'an-Arriagada}, \citenamefont {Solano}, \citenamefont {del Campo},\
  and\ \citenamefont {Chen}}]{chandarana2022digitized}%
  \BibitemOpen
  \bibfield  {author} {\bibinfo {author} {\bibfnamefont {P.}~\bibnamefont
  {Chandarana}}, \bibinfo {author} {\bibfnamefont {N.~N.}\ \bibnamefont
  {Hegade}}, \bibinfo {author} {\bibfnamefont {K.}~\bibnamefont {Paul}},
  \bibinfo {author} {\bibfnamefont {F.}~\bibnamefont {Albarr\'an-Arriagada}},
  \bibinfo {author} {\bibfnamefont {E.}~\bibnamefont {Solano}}, \bibinfo
  {author} {\bibfnamefont {A.}~\bibnamefont {del Campo}}, \ and\ \bibinfo
  {author} {\bibfnamefont {Xi}~\bibnamefont {Chen}},\ }\bibfield  {title}
  {\enquote {\bibinfo {title} {Digitized-counterdiabatic quantum approximate
  optimization algorithm},}\ }\href {\doibase 10.1103/PhysRevResearch.4.013141}
  {\bibfield  {journal} {\bibinfo  {journal} {Phys. Rev. Res.}\ }\textbf
  {\bibinfo {volume} {4}},\ \bibinfo {pages} {013141} (\bibinfo {year}
  {2022})}\BibitemShut {NoStop}%
\bibitem [{\citenamefont {Keever}\ and\ \citenamefont
  {Lubasch}(2023)}]{keever2023adiabatic}%
  \BibitemOpen
  \bibfield  {author} {\bibinfo {author} {\bibfnamefont {Conor~Mc}\
  \bibnamefont {Keever}}\ and\ \bibinfo {author} {\bibfnamefont {Michael}\
  \bibnamefont {Lubasch}},\ }\href@noop {} {\enquote {\bibinfo {title} {Towards
  adiabatic quantum computing using compressed quantum circuits},}\ } (\bibinfo
  {year} {2023}),\ \Eprint {http://arxiv.org/abs/2311.05544} {arXiv:2311.05544
  [quant-ph]} \BibitemShut {NoStop}%
\bibitem [{\citenamefont {Beauchard}\ \emph {et~al.}(2007)\citenamefont
  {Beauchard}, \citenamefont {Coron}, \citenamefont {Mirrahimi},\ and\
  \citenamefont {Rouchon}}]{QLC_Beauchard_2007}%
  \BibitemOpen
  \bibfield  {author} {\bibinfo {author} {\bibfnamefont {Karine}\ \bibnamefont
  {Beauchard}}, \bibinfo {author} {\bibfnamefont {Jean~Michel}\ \bibnamefont
  {Coron}}, \bibinfo {author} {\bibfnamefont {Mazyar}\ \bibnamefont
  {Mirrahimi}}, \ and\ \bibinfo {author} {\bibfnamefont {Pierre}\ \bibnamefont
  {Rouchon}},\ }\bibfield  {title} {\enquote {\bibinfo {title} {Implicit
  lyapunov control of finite dimensional schr{\"o}dinger equations},}\ }\href
  {\doibase https://doi.org/10.1016/j.sysconle.2006.10.024} {\bibfield
  {journal} {\bibinfo  {journal} {Systems and Control Letters}\ }\textbf
  {\bibinfo {volume} {56}},\ \bibinfo {pages} {388--395} (\bibinfo {year}
  {2007})}\BibitemShut {NoStop}%
\bibitem [{\citenamefont {Zhao}\ \emph {et~al.}(2012)\citenamefont {Zhao},
  \citenamefont {Lin}, \citenamefont {Sun},\ and\ \citenamefont
  {Xue}}]{QLC_Zhao_2012}%
  \BibitemOpen
  \bibfield  {author} {\bibinfo {author} {\bibfnamefont {Shouwei}\ \bibnamefont
  {Zhao}}, \bibinfo {author} {\bibfnamefont {Hai}\ \bibnamefont {Lin}},
  \bibinfo {author} {\bibfnamefont {Jitao}\ \bibnamefont {Sun}}, \ and\
  \bibinfo {author} {\bibfnamefont {Zhengui}\ \bibnamefont {Xue}},\ }\bibfield
  {title} {\enquote {\bibinfo {title} {An implicit lyapunov control for
  finite-dimensional closed quantum systems},}\ }\href {\doibase
  https://doi.org/10.1002/rnc.1748} {\bibfield  {journal} {\bibinfo  {journal}
  {International Journal of Robust and Nonlinear Control}\ }\textbf {\bibinfo
  {volume} {22}},\ \bibinfo {pages} {1212--1228} (\bibinfo {year}
  {2012})}\BibitemShut {NoStop}%
\bibitem [{\citenamefont {La~Salle}(1976)}]{la1976stability}%
  \BibitemOpen
  \bibfield  {author} {\bibinfo {author} {\bibfnamefont {J.~P.}\ \bibnamefont
  {La~Salle}},\ }\href {\doibase 10.1137/1.9781611970432} {\emph {\bibinfo
  {title} {The Stability of Dynamical Systems}}}\ (\bibinfo  {publisher}
  {Society for Industrial and Applied Mathematics},\ \bibinfo {year}
  {1976})\BibitemShut {NoStop}%
\bibitem [{\citenamefont {Sels}\ and\ \citenamefont
  {Polkovnikov}(2017)}]{sels2017minimizing}%
  \BibitemOpen
  \bibfield  {author} {\bibinfo {author} {\bibfnamefont {Dries}\ \bibnamefont
  {Sels}}\ and\ \bibinfo {author} {\bibfnamefont {Anatoli}\ \bibnamefont
  {Polkovnikov}},\ }\bibfield  {title} {\enquote {\bibinfo {title} {Minimizing
  irreversible losses in quantum systems by local counterdiabatic driving},}\
  }\href {\doibase 10.1073/pnas.1619826114} {\bibfield  {journal} {\bibinfo
  {journal} {Proceedings of the National Academy of Sciences}\ }\textbf
  {\bibinfo {volume} {114}},\ \bibinfo {pages} {E3909--E3916} (\bibinfo {year}
  {2017})}\BibitemShut {NoStop}%
\bibitem [{\citenamefont {Chen}\ \emph {et~al.}(2010)\citenamefont {Chen},
  \citenamefont {Ruschhaupt}, \citenamefont {Schmidt}, \citenamefont {del
  Campo}, \citenamefont {Gu\'ery-Odelin},\ and\ \citenamefont
  {Muga}}]{chen2010fast}%
  \BibitemOpen
  \bibfield  {author} {\bibinfo {author} {\bibfnamefont {Xi}~\bibnamefont
  {Chen}}, \bibinfo {author} {\bibfnamefont {A.}~\bibnamefont {Ruschhaupt}},
  \bibinfo {author} {\bibfnamefont {S.}~\bibnamefont {Schmidt}}, \bibinfo
  {author} {\bibfnamefont {A.}~\bibnamefont {del Campo}}, \bibinfo {author}
  {\bibfnamefont {D.}~\bibnamefont {Gu\'ery-Odelin}}, \ and\ \bibinfo {author}
  {\bibfnamefont {J.~G.}\ \bibnamefont {Muga}},\ }\bibfield  {title} {\enquote
  {\bibinfo {title} {Fast optimal frictionless atom cooling in harmonic traps:
  Shortcut to adiabaticity},}\ }\href {\doibase 10.1103/PhysRevLett.104.063002}
  {\bibfield  {journal} {\bibinfo  {journal} {Phys. Rev. Lett.}\ }\textbf
  {\bibinfo {volume} {104}},\ \bibinfo {pages} {063002} (\bibinfo {year}
  {2010})}\BibitemShut {NoStop}%
\bibitem [{\citenamefont {del Campo}(2013)}]{del2013shortcuts}%
  \BibitemOpen
  \bibfield  {author} {\bibinfo {author} {\bibfnamefont {Adolfo}\ \bibnamefont
  {del Campo}},\ }\bibfield  {title} {\enquote {\bibinfo {title} {Shortcuts to
  adiabaticity by counterdiabatic driving},}\ }\href {\doibase
  10.1103/PhysRevLett.111.100502} {\bibfield  {journal} {\bibinfo  {journal}
  {Phys. Rev. Lett.}\ }\textbf {\bibinfo {volume} {111}},\ \bibinfo {pages}
  {100502} (\bibinfo {year} {2013})}\BibitemShut {NoStop}%
\bibitem [{\citenamefont {Gu\'ery-Odelin}\ \emph {et~al.}(2019)\citenamefont
  {Gu\'ery-Odelin}, \citenamefont {Ruschhaupt}, \citenamefont {Kiely},
  \citenamefont {Torrontegui}, \citenamefont {Mart\'{\i}nez-Garaot},\ and\
  \citenamefont {Muga}}]{guery2019shortcuts}%
  \BibitemOpen
  \bibfield  {author} {\bibinfo {author} {\bibfnamefont {D.}~\bibnamefont
  {Gu\'ery-Odelin}}, \bibinfo {author} {\bibfnamefont {A.}~\bibnamefont
  {Ruschhaupt}}, \bibinfo {author} {\bibfnamefont {A.}~\bibnamefont {Kiely}},
  \bibinfo {author} {\bibfnamefont {E.}~\bibnamefont {Torrontegui}}, \bibinfo
  {author} {\bibfnamefont {S.}~\bibnamefont {Mart\'{\i}nez-Garaot}}, \ and\
  \bibinfo {author} {\bibfnamefont {J.~G.}\ \bibnamefont {Muga}},\ }\bibfield
  {title} {\enquote {\bibinfo {title} {Shortcuts to adiabaticity: Concepts,
  methods, and applications},}\ }\href {\doibase 10.1103/RevModPhys.91.045001}
  {\bibfield  {journal} {\bibinfo  {journal} {Rev. Mod. Phys.}\ }\textbf
  {\bibinfo {volume} {91}},\ \bibinfo {pages} {045001} (\bibinfo {year}
  {2019})}\BibitemShut {NoStop}%
\bibitem [{\citenamefont {Takahashi}(2019)}]{takahashi2019hamiltonian}%
  \BibitemOpen
  \bibfield  {author} {\bibinfo {author} {\bibfnamefont {Kazutaka}\
  \bibnamefont {Takahashi}},\ }\bibfield  {title} {\enquote {\bibinfo {title}
  {Hamiltonian engineering for adiabatic quantum computation: Lessons from
  shortcuts to adiabaticity},}\ }\href {\doibase 10.7566/JPSJ.88.061002}
  {\bibfield  {journal} {\bibinfo  {journal} {Journal of the Physical Society
  of Japan}\ }\textbf {\bibinfo {volume} {88}},\ \bibinfo {pages} {061002}
  (\bibinfo {year} {2019})}\BibitemShut {NoStop}%
\bibitem [{\citenamefont {Claeys}\ \emph {et~al.}(2019)\citenamefont {Claeys},
  \citenamefont {Pandey}, \citenamefont {Sels},\ and\ \citenamefont
  {Polkovnikov}}]{PolkovnikovPRL2019}%
  \BibitemOpen
  \bibfield  {author} {\bibinfo {author} {\bibfnamefont {Pieter~W.}\
  \bibnamefont {Claeys}}, \bibinfo {author} {\bibfnamefont {Mohit}\
  \bibnamefont {Pandey}}, \bibinfo {author} {\bibfnamefont {Dries}\
  \bibnamefont {Sels}}, \ and\ \bibinfo {author} {\bibfnamefont {Anatoli}\
  \bibnamefont {Polkovnikov}},\ }\bibfield  {title} {\enquote {\bibinfo {title}
  {Floquet-engineering counterdiabatic protocols in quantum many-body
  systems},}\ }\href {\doibase 10.1103/PhysRevLett.123.090602} {\bibfield
  {journal} {\bibinfo  {journal} {Phys. Rev. Lett.}\ }\textbf {\bibinfo
  {volume} {123}},\ \bibinfo {pages} {090602} (\bibinfo {year}
  {2019})}\BibitemShut {NoStop}%
\bibitem [{\citenamefont {\ifmmode \check{C}\else
  \v{C}\fi{}epait\ifmmode~\dot{e}\else \.{e}\fi{}}\ \emph
  {et~al.}(2023)\citenamefont {\ifmmode \check{C}\else
  \v{C}\fi{}epait\ifmmode~\dot{e}\else \.{e}\fi{}}, \citenamefont
  {Polkovnikov}, \citenamefont {Daley},\ and\ \citenamefont
  {Duncan}}]{vcepaite2023counterdiabatic}%
  \BibitemOpen
  \bibfield  {author} {\bibinfo {author} {\bibfnamefont {Ieva}\ \bibnamefont
  {\ifmmode \check{C}\else \v{C}\fi{}epait\ifmmode~\dot{e}\else \.{e}\fi{}}},
  \bibinfo {author} {\bibfnamefont {Anatoli}\ \bibnamefont {Polkovnikov}},
  \bibinfo {author} {\bibfnamefont {Andrew~J.}\ \bibnamefont {Daley}}, \ and\
  \bibinfo {author} {\bibfnamefont {Callum~W.}\ \bibnamefont {Duncan}},\
  }\bibfield  {title} {\enquote {\bibinfo {title} {Counterdiabatic optimized
  local driving},}\ }\href {\doibase 10.1103/PRXQuantum.4.010312} {\bibfield
  {journal} {\bibinfo  {journal} {PRX Quantum}\ }\textbf {\bibinfo {volume}
  {4}},\ \bibinfo {pages} {010312} (\bibinfo {year} {2023})}\BibitemShut
  {NoStop}%
\bibitem [{\citenamefont {Gokhale}\ \emph {et~al.}(2019)\citenamefont
  {Gokhale}, \citenamefont {Angiuli}, \citenamefont {Ding}, \citenamefont
  {Gui}, \citenamefont {Tomesh}, \citenamefont {Suchara}, \citenamefont
  {Martonosi},\ and\ \citenamefont {Chong}}]{gokhale2019minimizing}%
  \BibitemOpen
  \bibfield  {author} {\bibinfo {author} {\bibfnamefont {Pranav}\ \bibnamefont
  {Gokhale}}, \bibinfo {author} {\bibfnamefont {Olivia}\ \bibnamefont
  {Angiuli}}, \bibinfo {author} {\bibfnamefont {Yongshan}\ \bibnamefont
  {Ding}}, \bibinfo {author} {\bibfnamefont {Kaiwen}\ \bibnamefont {Gui}},
  \bibinfo {author} {\bibfnamefont {Teague}\ \bibnamefont {Tomesh}}, \bibinfo
  {author} {\bibfnamefont {Martin}\ \bibnamefont {Suchara}}, \bibinfo {author}
  {\bibfnamefont {Margaret}\ \bibnamefont {Martonosi}}, \ and\ \bibinfo
  {author} {\bibfnamefont {Frederic~T.}\ \bibnamefont {Chong}},\ }\bibfield
  {title} {\enquote {\bibinfo {title} {Minimizing state preparations in
  variational quantum eigensolver by partitioning into commuting families},}\
  }\href@noop {} {\  (\bibinfo {year} {2019})},\ \Eprint
  {http://arxiv.org/abs/1907.13623} {arXiv:1907.13623 [quant-ph]} \BibitemShut
  {NoStop}%
\bibitem [{\citenamefont {Verteletskyi}\ \emph {et~al.}(2020)\citenamefont
  {Verteletskyi}, \citenamefont {Yen},\ and\ \citenamefont
  {Izmaylov}}]{verteletskyi2020measurement}%
  \BibitemOpen
  \bibfield  {author} {\bibinfo {author} {\bibfnamefont {Vladyslav}\
  \bibnamefont {Verteletskyi}}, \bibinfo {author} {\bibfnamefont {Tzu-Ching}\
  \bibnamefont {Yen}}, \ and\ \bibinfo {author} {\bibfnamefont {Artur~F.}\
  \bibnamefont {Izmaylov}},\ }\bibfield  {title} {\enquote {\bibinfo {title}
  {{Measurement optimization in the variational quantum eigensolver using a
  minimum clique cover}},}\ }\href {\doibase 10.1063/1.5141458} {\bibfield
  {journal} {\bibinfo  {journal} {The Journal of Chemical Physics}\ }\textbf
  {\bibinfo {volume} {152}},\ \bibinfo {pages} {124114} (\bibinfo {year}
  {2020})}\BibitemShut {NoStop}%
\bibitem [{\citenamefont {Reggio}\ \emph {et~al.}(2023)\citenamefont {Reggio},
  \citenamefont {Butt}, \citenamefont {Lytle},\ and\ \citenamefont
  {Draper}}]{reggio2023fast}%
  \BibitemOpen
  \bibfield  {author} {\bibinfo {author} {\bibfnamefont {Ben}\ \bibnamefont
  {Reggio}}, \bibinfo {author} {\bibfnamefont {Nouman}\ \bibnamefont {Butt}},
  \bibinfo {author} {\bibfnamefont {Andrew}\ \bibnamefont {Lytle}}, \ and\
  \bibinfo {author} {\bibfnamefont {Patrick}\ \bibnamefont {Draper}},\
  }\bibfield  {title} {\enquote {\bibinfo {title} {Fast partitioning of pauli
  strings into commuting families for optimal expectation value measurements of
  dense operators},}\ }\href@noop {} {\  (\bibinfo {year} {2023})},\ \Eprint
  {http://arxiv.org/abs/2305.11847} {arXiv:2305.11847 [quant-ph]} \BibitemShut
  {NoStop}%
\bibitem [{\citenamefont {Anastasiou}\ \emph {et~al.}(2023)\citenamefont
  {Anastasiou}, \citenamefont {Mayhall}, \citenamefont {Barnes},\ and\
  \citenamefont {Economou}}]{anastasiou2023really}%
  \BibitemOpen
  \bibfield  {author} {\bibinfo {author} {\bibfnamefont {Panagiotis~G.}\
  \bibnamefont {Anastasiou}}, \bibinfo {author} {\bibfnamefont {Nicholas~J.}\
  \bibnamefont {Mayhall}}, \bibinfo {author} {\bibfnamefont {Edwin}\
  \bibnamefont {Barnes}}, \ and\ \bibinfo {author} {\bibfnamefont {Sophia~E.}\
  \bibnamefont {Economou}},\ }\bibfield  {title} {\enquote {\bibinfo {title}
  {How to really measure operator gradients in adapt-vqe},}\ }\href@noop {} {\
  (\bibinfo {year} {2023})},\ \Eprint {http://arxiv.org/abs/2306.03227}
  {arXiv:2306.03227 [quant-ph]} \BibitemShut {NoStop}%
\bibitem [{\citenamefont {Zhu}\ \emph {et~al.}(2023)\citenamefont {Zhu},
  \citenamefont {Liang}, \citenamefont {Yang},\ and\ \citenamefont
  {Li}}]{zhu2023optimizing}%
  \BibitemOpen
  \bibfield  {author} {\bibinfo {author} {\bibfnamefont {Linghua}\ \bibnamefont
  {Zhu}}, \bibinfo {author} {\bibfnamefont {Senwei}\ \bibnamefont {Liang}},
  \bibinfo {author} {\bibfnamefont {Chao}\ \bibnamefont {Yang}}, \ and\
  \bibinfo {author} {\bibfnamefont {Xiaosong}\ \bibnamefont {Li}},\ }\bibfield
  {title} {\enquote {\bibinfo {title} {Optimizing shot assignment in
  variational quantum eigensolver measurement},}\ }\href@noop {} {\  (\bibinfo
  {year} {2023})},\ \Eprint {http://arxiv.org/abs/2307.06504} {arXiv:2307.06504
  [quant-ph]} \BibitemShut {NoStop}%
\bibitem [{\citenamefont {Hegade}\ \emph {et~al.}(2022)\citenamefont {Hegade},
  \citenamefont {Chen},\ and\ \citenamefont {Solano}}]{Chen2022}%
  \BibitemOpen
  \bibfield  {author} {\bibinfo {author} {\bibfnamefont {Narendra~N.}\
  \bibnamefont {Hegade}}, \bibinfo {author} {\bibfnamefont {Xi}~\bibnamefont
  {Chen}}, \ and\ \bibinfo {author} {\bibfnamefont {Enrique}\ \bibnamefont
  {Solano}},\ }\bibfield  {title} {\enquote {\bibinfo {title} {Digitized
  counterdiabatic quantum optimization},}\ }\href {\doibase
  10.1103/PhysRevResearch.4.L042030} {\bibfield  {journal} {\bibinfo  {journal}
  {Phys. Rev. Res.}\ }\textbf {\bibinfo {volume} {4}},\ \bibinfo {pages}
  {L042030} (\bibinfo {year} {2022})}\BibitemShut {NoStop}%
\bibitem [{\citenamefont {Bravyi}\ \emph {et~al.}(2006)\citenamefont {Bravyi},
  \citenamefont {Hastings},\ and\ \citenamefont {Verstraete}}]{bravyi2006lieb}%
  \BibitemOpen
  \bibfield  {author} {\bibinfo {author} {\bibfnamefont {Sergey}\ \bibnamefont
  {Bravyi}}, \bibinfo {author} {\bibfnamefont {Matthew~B}\ \bibnamefont
  {Hastings}}, \ and\ \bibinfo {author} {\bibfnamefont {Frank}\ \bibnamefont
  {Verstraete}},\ }\bibfield  {title} {\enquote {\bibinfo {title}
  {Lieb-robinson bounds and the generation of correlations and topological
  quantum order},}\ }\href {\doibase 10.1103/PhysRevLett.97.050401} {\bibfield
  {journal} {\bibinfo  {journal} {\prl}\ }\textbf {\bibinfo {volume} {97}},\
  \bibinfo {pages} {050401} (\bibinfo {year} {2006})}\BibitemShut {NoStop}%
\bibitem [{\citenamefont {Yu}\ and\ \citenamefont
  {Wei}(2023)}]{yu2023learning}%
  \BibitemOpen
  \bibfield  {author} {\bibinfo {author} {\bibfnamefont {Nengkun}\ \bibnamefont
  {Yu}}\ and\ \bibinfo {author} {\bibfnamefont {Tzu-Chieh}\ \bibnamefont
  {Wei}},\ }\bibfield  {title} {\enquote {\bibinfo {title} {Learning marginals
  suffices!}}\ }\href@noop {} {\  (\bibinfo {year} {2023})},\ \Eprint
  {http://arxiv.org/abs/2303.08938} {arXiv:2303.08938 [quant-ph]} \BibitemShut
  {NoStop}%
\bibitem [{\citenamefont {Temme}\ \emph {et~al.}(2017)\citenamefont {Temme},
  \citenamefont {Bravyi},\ and\ \citenamefont {Gambetta}}]{temme2017error}%
  \BibitemOpen
  \bibfield  {author} {\bibinfo {author} {\bibfnamefont {Kristan}\ \bibnamefont
  {Temme}}, \bibinfo {author} {\bibfnamefont {Sergey}\ \bibnamefont {Bravyi}},
  \ and\ \bibinfo {author} {\bibfnamefont {Jay~M}\ \bibnamefont {Gambetta}},\
  }\bibfield  {title} {\enquote {\bibinfo {title} {Error mitigation for
  short-depth quantum circuits},}\ }\href@noop {} {\bibfield  {journal}
  {\bibinfo  {journal} {Physical review letters}\ }\textbf {\bibinfo {volume}
  {119}},\ \bibinfo {pages} {180509} (\bibinfo {year} {2017})}\BibitemShut
  {NoStop}%
\bibitem [{\citenamefont {Giurgica-Tiron}\ \emph {et~al.}(2020)\citenamefont
  {Giurgica-Tiron}, \citenamefont {Hindy}, \citenamefont {LaRose},
  \citenamefont {Mari},\ and\ \citenamefont {Zeng}}]{giurgica2020digital}%
  \BibitemOpen
  \bibfield  {author} {\bibinfo {author} {\bibfnamefont {Tudor}\ \bibnamefont
  {Giurgica-Tiron}}, \bibinfo {author} {\bibfnamefont {Yousef}\ \bibnamefont
  {Hindy}}, \bibinfo {author} {\bibfnamefont {Ryan}\ \bibnamefont {LaRose}},
  \bibinfo {author} {\bibfnamefont {Andrea}\ \bibnamefont {Mari}}, \ and\
  \bibinfo {author} {\bibfnamefont {William~J}\ \bibnamefont {Zeng}},\
  }\bibfield  {title} {\enquote {\bibinfo {title} {Digital zero noise
  extrapolation for quantum error mitigation},}\ }in\ \href@noop {} {\emph
  {\bibinfo {booktitle} {2020 IEEE International Conference on Quantum
  Computing and Engineering (QCE)}}}\ (\bibinfo {organization} {IEEE},\
  \bibinfo {year} {2020})\ pp.\ \bibinfo {pages} {306--316}\BibitemShut
  {NoStop}%
\bibitem [{\citenamefont {Malla}(2024)}]{malla2024cdfqa}%
  \BibitemOpen
  \bibfield  {author} {\bibinfo {author} {\bibfnamefont {Rajesh~K.}\
  \bibnamefont {Malla}},\ }\href@noop {} {\enquote {\bibinfo {title} {Github
  repository for cdfqa},}\ }\bibinfo {howpublished}
  {\url{https://github.com/rajeshkmalla/CDFQA}} (\bibinfo {year} {2024}),\
  \bibinfo {note} {accessed: 2024-10-02}\BibitemShut {NoStop}%
\bibitem [{\citenamefont {Bergholm}\ \emph {et~al.}(2022)\citenamefont
  {Bergholm}, \citenamefont {Izaac}, \citenamefont {Schuld}, \citenamefont
  {Gogolin}, \citenamefont {Ahmed}, \citenamefont {Ajith}, \citenamefont
  {Alam}, \citenamefont {Alonso-Linaje}, \citenamefont {AkashNarayanan},
  \citenamefont {Asadi}, \citenamefont {Arrazola}, \citenamefont {Azad},
  \citenamefont {Banning}, \citenamefont {Blank}, \citenamefont {Bromley},
  \citenamefont {Cordier}, \citenamefont {Ceroni}, \citenamefont {Delgado},
  \citenamefont {Matteo}, \citenamefont {Dusko}, \citenamefont {Garg},
  \citenamefont {Guala}, \citenamefont {Hayes}, \citenamefont {Hill},
  \citenamefont {Ijaz}, \citenamefont {Isacsson}, \citenamefont {Ittah},
  \citenamefont {Jahangiri}, \citenamefont {Jain}, \citenamefont {Jiang},
  \citenamefont {Khandelwal}, \citenamefont {Kottmann}, \citenamefont {Lang},
  \citenamefont {Lee}, \citenamefont {Loke}, \citenamefont {Lowe},
  \citenamefont {McKiernan}, \citenamefont {Meyer}, \citenamefont
  {Montañez-Barrera}, \citenamefont {Moyard}, \citenamefont {Niu},
  \citenamefont {O'Riordan}, \citenamefont {Oud}, \citenamefont {Panigrahi},
  \citenamefont {Park}, \citenamefont {Polatajko}, \citenamefont {Quesada},
  \citenamefont {Roberts}, \citenamefont {Sá}, \citenamefont {Schoch},
  \citenamefont {Shi}, \citenamefont {Shu}, \citenamefont {Sim}, \citenamefont
  {Singh}, \citenamefont {Strandberg}, \citenamefont {Soni}, \citenamefont
  {Száva}, \citenamefont {Thabet}, \citenamefont {Vargas-Hernández},
  \citenamefont {Vincent}, \citenamefont {Vitucci}, \citenamefont {Weber},
  \citenamefont {Wierichs}, \citenamefont {Wiersema}, \citenamefont {Willmann},
  \citenamefont {Wong}, \citenamefont {Zhang},\ and\ \citenamefont
  {Killoran}}]{bergholm2018pennylane}%
  \BibitemOpen
  \bibfield  {author} {\bibinfo {author} {\bibfnamefont {Ville}\ \bibnamefont
  {Bergholm}}, \bibinfo {author} {\bibfnamefont {Josh}\ \bibnamefont {Izaac}},
  \bibinfo {author} {\bibfnamefont {Maria}\ \bibnamefont {Schuld}}, \bibinfo
  {author} {\bibfnamefont {Christian}\ \bibnamefont {Gogolin}}, \bibinfo
  {author} {\bibfnamefont {Shahnawaz}\ \bibnamefont {Ahmed}}, \bibinfo {author}
  {\bibfnamefont {Vishnu}\ \bibnamefont {Ajith}}, \bibinfo {author}
  {\bibfnamefont {M.~Sohaib}\ \bibnamefont {Alam}}, \bibinfo {author}
  {\bibfnamefont {Guillermo}\ \bibnamefont {Alonso-Linaje}}, \bibinfo {author}
  {\bibfnamefont {B.}~\bibnamefont {AkashNarayanan}}, \bibinfo {author}
  {\bibfnamefont {Ali}\ \bibnamefont {Asadi}}, \bibinfo {author} {\bibfnamefont
  {Juan~Miguel}\ \bibnamefont {Arrazola}}, \bibinfo {author} {\bibfnamefont
  {Utkarsh}\ \bibnamefont {Azad}}, \bibinfo {author} {\bibfnamefont {Sam}\
  \bibnamefont {Banning}}, \bibinfo {author} {\bibfnamefont {Carsten}\
  \bibnamefont {Blank}}, \bibinfo {author} {\bibfnamefont {Thomas~R}\
  \bibnamefont {Bromley}}, \bibinfo {author} {\bibfnamefont {Benjamin~A.}\
  \bibnamefont {Cordier}}, \bibinfo {author} {\bibfnamefont {Jack}\
  \bibnamefont {Ceroni}}, \bibinfo {author} {\bibfnamefont {Alain}\
  \bibnamefont {Delgado}}, \bibinfo {author} {\bibfnamefont {Olivia~Di}\
  \bibnamefont {Matteo}}, \bibinfo {author} {\bibfnamefont {Amintor}\
  \bibnamefont {Dusko}}, \bibinfo {author} {\bibfnamefont {Tanya}\ \bibnamefont
  {Garg}}, \bibinfo {author} {\bibfnamefont {Diego}\ \bibnamefont {Guala}},
  \bibinfo {author} {\bibfnamefont {Anthony}\ \bibnamefont {Hayes}}, \bibinfo
  {author} {\bibfnamefont {Ryan}\ \bibnamefont {Hill}}, \bibinfo {author}
  {\bibfnamefont {Aroosa}\ \bibnamefont {Ijaz}}, \bibinfo {author}
  {\bibfnamefont {Theodor}\ \bibnamefont {Isacsson}}, \bibinfo {author}
  {\bibfnamefont {David}\ \bibnamefont {Ittah}}, \bibinfo {author}
  {\bibfnamefont {Soran}\ \bibnamefont {Jahangiri}}, \bibinfo {author}
  {\bibfnamefont {Prateek}\ \bibnamefont {Jain}}, \bibinfo {author}
  {\bibfnamefont {Edward}\ \bibnamefont {Jiang}}, \bibinfo {author}
  {\bibfnamefont {Ankit}\ \bibnamefont {Khandelwal}}, \bibinfo {author}
  {\bibfnamefont {Korbinian}\ \bibnamefont {Kottmann}}, \bibinfo {author}
  {\bibfnamefont {Robert~A.}\ \bibnamefont {Lang}}, \bibinfo {author}
  {\bibfnamefont {Christina}\ \bibnamefont {Lee}}, \bibinfo {author}
  {\bibfnamefont {Thomas}\ \bibnamefont {Loke}}, \bibinfo {author}
  {\bibfnamefont {Angus}\ \bibnamefont {Lowe}}, \bibinfo {author}
  {\bibfnamefont {Keri}\ \bibnamefont {McKiernan}}, \bibinfo {author}
  {\bibfnamefont {Johannes~Jakob}\ \bibnamefont {Meyer}}, \bibinfo {author}
  {\bibfnamefont {J.~A.}\ \bibnamefont {Montañez-Barrera}}, \bibinfo {author}
  {\bibfnamefont {Romain}\ \bibnamefont {Moyard}}, \bibinfo {author}
  {\bibfnamefont {Zeyue}\ \bibnamefont {Niu}}, \bibinfo {author} {\bibfnamefont
  {Lee~James}\ \bibnamefont {O'Riordan}}, \bibinfo {author} {\bibfnamefont
  {Steven}\ \bibnamefont {Oud}}, \bibinfo {author} {\bibfnamefont {Ashish}\
  \bibnamefont {Panigrahi}}, \bibinfo {author} {\bibfnamefont {Chae-Yeun}\
  \bibnamefont {Park}}, \bibinfo {author} {\bibfnamefont {Daniel}\ \bibnamefont
  {Polatajko}}, \bibinfo {author} {\bibfnamefont {Nicolás}\ \bibnamefont
  {Quesada}}, \bibinfo {author} {\bibfnamefont {Chase}\ \bibnamefont
  {Roberts}}, \bibinfo {author} {\bibfnamefont {Nahum}\ \bibnamefont {Sá}},
  \bibinfo {author} {\bibfnamefont {Isidor}\ \bibnamefont {Schoch}}, \bibinfo
  {author} {\bibfnamefont {Borun}\ \bibnamefont {Shi}}, \bibinfo {author}
  {\bibfnamefont {Shuli}\ \bibnamefont {Shu}}, \bibinfo {author} {\bibfnamefont
  {Sukin}\ \bibnamefont {Sim}}, \bibinfo {author} {\bibfnamefont {Arshpreet}\
  \bibnamefont {Singh}}, \bibinfo {author} {\bibfnamefont {Ingrid}\
  \bibnamefont {Strandberg}}, \bibinfo {author} {\bibfnamefont {Jay}\
  \bibnamefont {Soni}}, \bibinfo {author} {\bibfnamefont {Antal}\ \bibnamefont
  {Száva}}, \bibinfo {author} {\bibfnamefont {Slimane}\ \bibnamefont
  {Thabet}}, \bibinfo {author} {\bibfnamefont {Rodrigo~A.}\ \bibnamefont
  {Vargas-Hernández}}, \bibinfo {author} {\bibfnamefont {Trevor}\ \bibnamefont
  {Vincent}}, \bibinfo {author} {\bibfnamefont {Nicola}\ \bibnamefont
  {Vitucci}}, \bibinfo {author} {\bibfnamefont {Maurice}\ \bibnamefont
  {Weber}}, \bibinfo {author} {\bibfnamefont {David}\ \bibnamefont {Wierichs}},
  \bibinfo {author} {\bibfnamefont {Roeland}\ \bibnamefont {Wiersema}},
  \bibinfo {author} {\bibfnamefont {Moritz}\ \bibnamefont {Willmann}}, \bibinfo
  {author} {\bibfnamefont {Vincent}\ \bibnamefont {Wong}}, \bibinfo {author}
  {\bibfnamefont {Shaoming}\ \bibnamefont {Zhang}}, \ and\ \bibinfo {author}
  {\bibfnamefont {Nathan}\ \bibnamefont {Killoran}},\ }\bibfield  {title}
  {\enquote {\bibinfo {title} {Pennylane: Automatic differentiation of hybrid
  quantum-classical computations},}\ }\href@noop {} {\  (\bibinfo {year}
  {2022})},\ \Eprint {http://arxiv.org/abs/1811.04968} {arXiv:1811.04968
  [quant-ph]} \BibitemShut {NoStop}%
\end{thebibliography}%
